\documentstyle[12pt,psfig]{article}
\newcommand{\mysection}{\setcounter{equation}{0}\section}

\begin{document}


\hfill{ITP-SB-93-72}
\vskip 1cm
\centerline{\large\bf Exclusive $W^+ + \gamma$ production in}
\centerline{\large\bf proton-antiproton collisions I: general formalism}
\vskip 0.5cm
\centerline{\bf S. Mendoza and J. Smith}
\centerline{\small \sl Institute for Theoretical Physics,}
\centerline{\small \sl State University of New York at Stony Brook,}
\centerline{\small \sl New York 11794-3840}
\vskip 1cm
\centerline{February 1994}
\vskip 0.5cm
\centerline{\bf Abstract}

We present a detailed computation of the fully exclusive cross section of $ p +
\bar p \rightarrow W^+ + \gamma + X$ with $X=0$ and $1$ jet in the framework of
the factorization theorem and dimensional regularization.
Order $\alpha_S$ and photon bremsstrahlung contributions are discussed in
the $\overline{MS}$ mass factorization scheme. The resulting expressions are
ready to be implemented numerically using Monte Carlo techniques to compute
single and double differential cross
sections and correlations between outgoing pairs of particles.
\newpage

\pagestyle{myheadings}
\mysection{\bf INTRODUCTION.}

Ever since Mikaelian's discovery of a zero in the amplitude of the partonic
subprocess $ q + \bar{q}\rightarrow W + \gamma $ \cite{M}, radiative
production of\ W bosons has been discussed as a way of testing the validity of
the
Electroweak Theory. The study of differential distributions in
$ p +\bar p  \rightarrow W +\gamma + X $ may be the best way to place bounds on
the
magnitude of the magnetic dipole and electric quadrupole moments of the W
boson.
Deviations from the Standard Model could show up as a shift of the photon
distributions near the dip that is a reflection of the partonic zero.
The QCD corrections to the reaction $ p +\bar p  \rightarrow W +\gamma + X $
and
its deviations from the Standard Model have been studied in
\cite{STN},\cite{MSN},\cite{O1},\cite{BHO} and other references therein. These
papers have been
mainly devoted to the analysis of single photon distributions, photon-$W$ boson
pair
mass correlations and charged lepton-photon pseudorapidity correlations and
they either neglect or
approximate the photon bremsstrahlung contributions.

When computing the photon inclusive process in \cite{STN} and \cite{MSN} all
the singularities
associated with a jet emitted in a collinear or a soft region of
phase space were regularized by analytically performing all the integrals
associated with the jet and the $W$ boson in $n$ space-time dimensions.
Although the
numerical advantage of this procedure is obvious -one is left with only
photon variables to be integrated over numerically- the predictive power of the
whole computation is limited by the fact that one looses information about the
energies and angles of the jet and the W boson.

In the present work we redo the exact first order calculation reported in
\cite{STN} in an
exclusive fashion. We present analytical results for the integrands
needed in the computation of physical observables related to any of the
outgoing particles in the reactions
$p +\bar p  \rightarrow W^+ +\gamma$ and $p +\bar p  \rightarrow W^+ +\gamma +
\rm{jet}$. Using these results we will extend the studies of the Electroweak
and QCD sectors of the Standard Model by providing a complete set of single and
double differential distributions and correlations including the $W$ boson and,
when applicable, the jet. Deviations of the experimental data from the
theoretical predictions could not only mean new physics in the Electroweak
sector, but would also probe the QCD behavior and the underlying photon
bremsstrahlung processes. In particular, an inadequate photon bremsstrahlung
approximation would also result in deviations from the predicted photon single
and double differential distributions and correlations.

The method that we employ for computing exclusive cross sections is based on
the one used by Mele et. al.
in the context of $Z^0$ pair production and production of heavy quarks
\cite{MNR}.
This method allows for control of all soft and initial (final) state collinear
singular
regions of phase space in the framework of dimensional regularization and the
factorization theorem.

We consider three different scenarios: (1) the 2-body inclusive production of
$W^+$ and $\gamma$,
(2) the exclusive production of $W^+$, $\gamma$ and $1$ jet and
(3) the exclusive production of $W^+$ and $\gamma$ accompanied by $0$ jet.
In all three cases we take into account exact $O(\alpha_S)$ QCD contributions
in the
$\overline{MS}$ mass factorization scheme.
Contributions arising from photon-quark and photon-gluon fragmentation
functions (generically called ``photon bremsstrahlung contributions'') are also
included in our discussion.

In Section II we present a detailed review of the process $p +\bar p
\rightarrow W^+ +\gamma + X$
for $X=0,1$ jet in the framework of the parton model and the factorization
theorem.

In Section III we show how the cancellation of singularities is performed in an
exclusive fashion in each of the hard scattering channels of our process in
the framework of $n$ dimensional regularization.

Section IV is devoted to the definition of the three experimental scenarios
and their corresponding cuts.

We end our study in Section V with a discussion of the numerical implementation
of the several expressions
for the cross sections, together with a listing of the relevant formulae.

Results for total, single and double differential cross sections and
correlations between pairs of outgoing particles are given in a separate paper
\cite{MS}.

\newpage

\pagestyle{myheadings}
\mysection{\bf THE PROCESS $p + \bar p \rightarrow W^+ + \gamma\ +X$, THE
PARTON MODEL AND THE FACTORIZATION THEOREM.}

\subsection{\sc Introduction.}
We are considering the hadronic processes given by
\begin{eqnarray}
p + \bar p \rightarrow W^+ + \gamma\,,
\end{eqnarray}
and
\begin{eqnarray}
p + \bar p \rightarrow W^+ + \gamma + jet\,.
\end{eqnarray}

In what follows we will omit the charge index ``$^+$'' when referring to the
$W^+$ boson. In the framework of the parton model we can formally write the
hadronic 2-body inclusive differential cross section in the following way:
\begin{eqnarray}
\lefteqn{\sum_{X} \frac{D^2\sigma^H}{DQ_1DQ_2}\left[ p(P_1)\ \bar p(P_2)
\rightarrow W(Q_1)\ \gamma(Q_2) \ X \right] = } \nonumber \\
& & \sum_{i,j,k}\left( \int_{0}^{1}du_1\int_{0}^{1}du_2\int_{0}^{1}du_3\
D_{ip}(u_1)D_{j\bar p}(u_2)D_{\gamma k}(u_3) \right. \nonumber \\
& &\hspace{-0.15in}\left. \times\sum_{x} \frac{D^2\sigma^P}{DQ_1DQ_2}\left[
i(u_1P_1)\ j(u_2P_2) \rightarrow W(Q_1)\ k(\frac{Q_2}{u_3})\ x\right]
\right)\,,
\end{eqnarray}
where \( \sum_{X} \) is a sum over sets $X$ of physical particles integrated
over their phase space. \( \sum_{i,j,k} \)
denotes sums over partons $i,j,k$ (by partons we mean quarks, antiquarks,
gluons and photons.)
\( \sum_{x} \) is a sum over sets $x$ of outgoing partons integrated over their
phase space.
In $n$ dimensional space-time $DQ_1$ and $DQ_2$ are generically given by:
\begin{eqnarray}
DQ_i = \frac{d^{n-1}\vec{Q}_i}{(2\pi)^{n-1}2Q_{i,0}} \,,
\end{eqnarray}
where $Q_i$ is an $n$-momentum vector with space components $\vec{Q}_i$ and
time component $Q_{i,0}$.
\( \sum_{x} D^2\sigma^P/DQ_1DQ_2 \) formally denotes the ``bare'' partonic
2-body inclusive differential cross section, which can be directly computed
using perturbation theory in the Standard Model.
$D_{iA}(u)$ for $A = p,\bar p$ are the bare partonic densities.
The bare fragmentation function $D_{\gamma k}(u)$ gives the photon momentum
fraction density
when a parton of type $k$ and momentum $q$ fragments into a photon of momentum
$uq$ and any number of hadrons.
After renormalization has been performed the bare partonic cross sections in
(2.3) contain soft and collinear singularities coming from virtual corrections
as well as integration over phase space of non-empty sets of outgoing partons
$x$. Cancellation of soft singularities will occur after addition of the soft
pole terms in the virtual corrections
with the soft pole terms in the corresponding emission processes. By virtue of
the factorization theorem \cite{CSS} the bare partonic densities
and the bare fragmentation function are defined to contain singularities that
cancel against the remaining collinear singularities in the bare partonic cross
sections so that the hadronic
cross section on the LHS of  (2.3) is a finite quantity. This procedure is
implemented at a specific mass factorization scale $M$. We will define this
scale to equal the renormalization scale $\mu$. To avoid complicating the
subsequent formulae this scale dependence is not explicitly written, except
where necessary for the discussion. According to the factorization theorem we
can rewrite the singular
bare partonic 2-body inclusive cross section in terms of non-singular ``hard
scattering cross sections'':
\begin{eqnarray}
\label{(2.5)}
\lefteqn{\sum_{x} \frac{D^2\sigma^P}{DQ_1DQ_2}\left[ i(p_1)\ j(p_2) \rightarrow
W(Q_1)\ k(Q_2) \ x \right] = } \nonumber \\
& &\hspace{-0.1in} \sum_{a,b,c,x_a,x_b,x_c} \left(
\int_{0}^{1}dv_1\int_{0}^{1}dv_2\int_{0}^{1}dv_3\
d_{ai}^{x_a}(v_1)d_{bj}^{x_b}(v_2)d_{kc}^{x_c}(v_3)
 \right. \nonumber \\
& & \left. \times\sum_{y} \frac{D^2\sigma}{DQ_1DQ_2}\left[ a(v_1p_1)\ b(v_2p_2)
\rightarrow W(Q_1)\ c(\frac{Q_2}{v_3})\ y\right] \right)\,,
\end{eqnarray}
where \( \sum_{y} D^2\sigma/DQ_1DQ_2 \) denotes a 2-body inclusive hard
scattering differential cross section.

$d_{ai}^{x_a}(v)$ denotes the splitting function of parton $i$ into a parton
$a$ and a set of partons $x_a$ with $a$
carrying a momentum fraction $v$ of its parent parton $i$. These splitting
functions factorize the collinear singularities contained in the
bare partonic cross section $\sum_{x} D^2\sigma^P/DQ_1DQ_2$ and they can be
exactly computed order by order in perturbation
theory. In this way  (2.5) is solved perturbatively
for the hard scattering cross sections \( \sum_{y} D^2\sigma/DQ_1DQ_2 \).

Using (2.5) in (2.3) we can rewrite the hadronic 2-body inclusive differential
cross section in terms of only non-singular quantities:
\begin{eqnarray}
\lefteqn{\sum_{X} \frac{D^2\sigma^H}{DQ_1DQ_2}\left[ p(P_1)\ \bar p(P_2)
\rightarrow W(Q_1)\ \gamma(Q_2) \ X \right] = } \nonumber \\
& & \sum_{a,b,c}\left(
\int_{0}^{1}d\tau_1\int_{0}^{1}d\tau_2\int_{0}^{1}d\tau_3\
f_{ap}(\tau_1)f_{b\bar p}(\tau_2)f_{\gamma c}(\tau_3) \right. \nonumber \\
& &\hspace{-0.15in}\left. \times \sum_{x} \frac{D^2\sigma}{DQ_1DQ_2}\left[
a(\tau_1P_1)\ b(\tau_2P_2) \rightarrow W(Q_1)\ c(\frac{Q_2}{\tau_3})\ x\right]
\right)\,.
\end{eqnarray}

The parton densities $f_{ap}(\tau), f_{b\bar p}(\tau)$ and the fragmentation
function $f_{\gamma c}(\tau)$ are defined by:
\begin{eqnarray}
f_{ap}(\tau) &=& \sum_{x_a,i} \int_{\tau}^{1}du\
\frac{1}{u}d_{ai}^{x_a}(\frac{\tau}{u})D_{ip}(u) \nonumber \\
f_{b\bar p}(\tau) &=& \sum_{x_b,j} \int_{\tau}^{1}du\
\frac{1}{u}d_{bj}^{x_b}(\frac{\tau}{u})D_{j\bar p}(u) \nonumber \\
f_{\gamma c}(\tau) &=& \sum_{x_c,k} \int_{\tau}^{1}du\ \frac{1}{u}D_{\gamma
k}(u)d_{kc}^{x_c}(\frac{\tau}{u})\,,
\end{eqnarray}
and are obtained by fitting data of deep inelastic scattering to results of
perturbation theory at a given mass
factorization scale $M$. At present there is not enough data available to fit
the photon fragmentation functions so one has to rely upon an approximation,
for example, the so called leading-log-approximation \cite{DO},\cite{O}.

In the computation of hadronic quantities we use  (2.6) with \\
\( a,b\ \epsilon \left\{q,\bar q,g \right\} \) and\ \( c\ \epsilon
\left\{q,\bar q,g,\gamma \right\} \) where
$q, \bar q$ and $g$ denote quark, antiquark and gluon respectively. The photon
$(\gamma)$ is
treated in a dual way: it is a hadron, i.e. an observable final state particle,
and it is also a parton of our Lagrangian.
In our 1-body inclusive computation in \cite{MSN} we only considered
contributions from $f_{\gamma \gamma}$
and neglected $f_{\gamma c}$ for \( c\ \epsilon \left\{q,\bar q,g \right\} \).
In the present work we include the four contributions keeping terms up to
$O(\alpha_S\alpha\alpha_W)$, where $\alpha_S, \alpha$ and $\alpha_W$ are the
strong, electromagnetic and electroweak fine structure constants.

\subsection{\sc Contributions from $f_{\gamma \gamma}$.}

The leading photon-photon splitting function is given by $d_{\gamma
\gamma}^{\{\ \}}(\tau/u) = \delta(1-\tau/u)$,
i.e. when no partons are emitted from the photon. This leading order splitting
function can be identified with the bare fragmentation function
$D_{\gamma \gamma}(u)$ when zero hadrons are fragmented from the photon.
Using this in  (2.7) we obtain the leading contribution to the photon-photon
fragmentation function
\begin{eqnarray}
f_{\gamma \gamma}(\tau) = \delta(1-\tau)\,.
\end{eqnarray}
Setting $c=\gamma$ in  (2.6) and keeping hard scattering contributions up to
$O(\alpha_S\alpha\alpha_W)$ we obtain
\begin{eqnarray}
\lefteqn{ \sum_{X} \left( \frac{D^2\sigma^H}{DQ_1DQ_2} \right)^{\gamma
\gamma}\left[ p(P_1)\ \bar p(P_2) \rightarrow W(Q_1)\ \gamma(Q_2) \ X \right] =
} \nonumber \\
& &\int_{0}^{1}d\tau_1\int_{0}^{1}d\tau_2 \left\{ f_{qp}(\tau_1)f_{\bar q\bar
p}(\tau_2) \left( \frac{D^2\sigma}{DQ_1DQ_2}\left[ q(\tau_1P_1)\ \bar
q(\tau_2P_2)\rightarrow W(Q_1)\ \gamma(Q_2) \right] \right. \right.\nonumber \\
& &\hspace{2.in} \left.+ \frac{D^2\sigma}{DQ_1DQ_2}\left[ q(\tau_1P_1)\ \bar
q(\tau_2P_2)\rightarrow W(Q_1)\ \gamma(Q_2) \ g \right] \right)\nonumber \\
& &\hspace{1.in}+ f_{qp}(\tau_1)f_{g \bar p}(\tau_2)\
\frac{D^2\sigma}{DQ_1DQ_2}\left[ q(\tau_1P_1)\ g(\tau_2P_2)\rightarrow W(Q_1)\
\gamma(Q_2) \ q\right]\nonumber \\
& &\hspace{1.in}+ f_{gp}(\tau_1)f_{\bar q \bar p}(\tau_2)\
\frac{D^2\sigma}{DQ_1DQ_2}\left[ g(\tau_1P_1)\ \bar q(\tau_2P_2)\rightarrow
W(Q_1)\ \gamma(Q_2)\ \bar q\right]\nonumber \\
\nonumber \\
& &\hspace{1.in} + \left( q \longleftrightarrow \bar q \right) \bigg\} \,.
\nonumber \\
\,
\end{eqnarray}
Note that the lowest order hard scattering cross section is always equal to the
corresponding lowest order bare
partonic cross section, as we will verify in subsections F,G and H.
%
\subsection{\sc Contributions from $f_{\gamma q}$ and $f_{\gamma \bar q}$.}

Setting $c=q$ and $c=\bar q$ in  (2.6) we obtain for these contributions:
\begin{eqnarray}
\sum_{X} \left( \frac{D^2\sigma^H}{DQ_1DQ_2} \right)^{\gamma q}\left[ p(P_1)\
\bar p(P_2) \rightarrow W(Q_1)\ \gamma(Q_2) \ X \right] &=&
\int_{0}^{1}d\tau_1\int_{0}^{1}d\tau_2\int_{0}^{1}d\tau_3\nonumber \\
& &\hspace{-3.in}\times\left\{ f_{qp}(\tau_1)f_{g \bar p}(\tau_2)f_{\gamma
q}(\tau_3) \frac{D^2\sigma}{DQ_1DQ_2}\left[ q(\tau_1P_1)\
g(\tau_2P_2)\rightarrow W(Q_1)\ q(\frac{Q_2}{\tau_3}) \right] \right.\nonumber
\\
& &\hspace{-2.9in}\left.+ f_{gp}(\tau_1)f_{q \bar p}(\tau_2)f_{\gamma
q}(\tau_3) \frac{D^2\sigma}{DQ_1DQ_2}\left[ g(\tau_1P_1)\
q(\tau_2P_2)\rightarrow W(Q_1)\ q(\frac{Q_2}{\tau_3}) \right] \right\}
\nonumber \\
\,
\end{eqnarray}
and
\begin{eqnarray}
\sum_{X} \left( \frac{D^2\sigma^H}{DQ_1DQ_2} \right)^{\gamma \bar q}\left[
p(P_1)\ \bar p(P_2) \rightarrow W(Q_1)\ \gamma(Q_2) \ X \right] &=&
\int_{0}^{1}d\tau_1\int_{0}^{1}d\tau_2\int_{0}^{1}d\tau_3 \nonumber \\
& &\hspace{-3.in}\times\left\{ f_{gp}(\tau_1)f_{\bar q \bar p}(\tau_2)f_{\gamma
\bar q}(\tau_3) \frac{D^2\sigma}{DQ_1DQ_2}\left[ g(\tau_1P_1)\ \bar
q(\tau_2P_2)\rightarrow W(Q_1)\ \bar q(\frac{Q_2}{\tau_3}) \right]
\right.\nonumber \\
& &\hspace{-2.9in}\left.+ f_{\bar q p}(\tau_1)f_{g \bar p}(\tau_2)f_{\gamma
\bar q}(\tau_3) \frac{D^2\sigma}{DQ_1DQ_2}\left[ \bar q(\tau_1P_1)\
g(\tau_2P_2)\rightarrow W(Q_1)\ \bar q(\frac{Q_2}{\tau_3}) \right] \right\}\,.
\nonumber \\
\,
\end{eqnarray}
\\

\subsection{\sc Contributions from $f_{\gamma g}$.}

Setting $c=g$ in  (2.6) we have:
\begin{eqnarray}
\sum_{X} \left( \frac{D^2\sigma^H}{DQ_1DQ_2} \right)^{\gamma g}\left[ p(P_1)\
\bar p(P_2) \rightarrow W(Q_1)\ \gamma(Q_2) \ X \right] &=&
\int_{0}^{1}d\tau_1\int_{0}^{1}d\tau_2\int_{0}^{1}d\tau_3 \nonumber \\
& &\hspace{-3.in}\times\left\{ f_{qp}(\tau_1)f_{\bar q \bar p}(\tau_2)f_{\gamma
g}(\tau_3) \frac{D^2\sigma}{DQ_1DQ_2}\left[ q(\tau_1P_1)\ \bar
q(\tau_2P_2)\rightarrow W(Q_1)\ g(\frac{Q_2}{\tau_3}) \right] \right.\nonumber
\\
\nonumber \\
& &\hspace{-2.8in} + \left( q \longleftrightarrow \bar q \right) \bigg\} \,.
\nonumber \\
\,
\end{eqnarray}
\\
\\

In {\sc B, C} and {\sc D} sums over flavors of quarks $q$ and antiquarks $\bar
q$ satisfying the electric charge
conservation for production of $W^+$ are implicit. For a more detailed
discussion on the way we treat this issue see section IV of \cite{MSN}.
The contributions from {\sc C} and {\sc D} will be referred to as ``photon
bremsstrahlung contributions''.\\

\subsection{\sc The incoming hard scattering channels.}

We can now regroup all terms in {\sc B, C} and {\sc D} according to three
partonic channels in the incoming hard scattering state:

\begin{eqnarray}
\label{2.13}
\sum_{X} \left( \frac{D^2\sigma^H}{DQ_1DQ_2} \right)^{q \bar q}\left[ p(P_1)\
\bar p(P_2) \rightarrow W(Q_1)\ \gamma(Q_2) \ X \right]
&=&\int_{0}^{1}d\tau_1\int_{0}^{1}d\tau_2 \nonumber \\
& &\hspace{-3.in}\times\left\{ f_{qp}(\tau_1)f_{\bar q\bar p}(\tau_2) \left(
\frac{D^2\sigma}{DQ_1DQ_2}\left[ q(\tau_1P_1)\ \bar q(\tau_2P_2)\rightarrow
W(Q_1)\ \gamma(Q_2) \right] \right. \right.\nonumber \\
& &\hspace{-1.8in}+ \frac{D^2\sigma}{DQ_1DQ_2}\left[ q(\tau_1P_1)\ \bar
q(\tau_2P_2)\rightarrow W(Q_1)\ \gamma(Q_2) \ g \right]\nonumber \\
& &\hspace{-2.8in} \left. + \int_{0}^{1}d\tau_3\ f_{\gamma
g}(\tau_3)\frac{D^2\sigma}{DQ_1DQ_2}\left[ q(\tau_1P_1)\ \bar
q(\tau_2P_2)\rightarrow W(Q_1)\ g(\frac{Q_2}{\tau_3}) \right] \right)\nonumber
\\
& &\hspace{-2.8in} + \left( q \longleftrightarrow \bar q \right) \bigg\}
\nonumber \\
\,
\end{eqnarray}
\begin{eqnarray}
\label{2.14}
\sum_{X} \left( \frac{D^2\sigma^H}{DQ_1DQ_2} \right)^{q g}\left[ p(P_1)\ \bar
p(P_2) \rightarrow W(Q_1)\ \gamma(Q_2) \ X \right] &=&
\int_{0}^{1}d\tau_1\int_{0}^{1}d\tau_2 \nonumber \\
& &\hspace{-3.in}\times\left\{ f_{qp}(\tau_1)f_{g \bar p}(\tau_2)\
\frac{D^2\sigma}{DQ_1DQ_2}\left[ q(\tau_1P_1)\ g(\tau_2P_2)\rightarrow W(Q_1)\
\gamma(Q_2) \ q\right] \right.\nonumber \\
& &\hspace{-2.8in}\left.+ f_{gp}(\tau_1)f_{q \bar p}(\tau_2)\
\frac{D^2\sigma}{DQ_1DQ_2}\left[ g(\tau_1P_1)\ q(\tau_2P_2)\rightarrow W(Q_1)\
\gamma(Q_2) \ q\right] \right\}\nonumber \\
& &\hspace{-3.in} +
\int_{0}^{1}d\tau_1\int_{0}^{1}d\tau_2\int_{0}^{1}d\tau_3\nonumber \\
& &\hspace{-3.in}\times \left\{ f_{qp}(\tau_1)f_{g \bar p}(\tau_2)f_{\gamma
q}(\tau_3) \frac{D^2\sigma}{DQ_1DQ_2}\left[ q(\tau_1P_1)\
g(\tau_2P_2)\rightarrow W(Q_1)\ q(\frac{Q_2}{\tau_3}) \right] \right.\nonumber
\\
& &\hspace{-2.9in}\left.+ f_{gp}(\tau_1)f_{q \bar p}(\tau_2)f_{\gamma
q}(\tau_3) \frac{D^2\sigma}{DQ_1DQ_2}\left[ g(\tau_1P_1)\
q(\tau_2P_2)\rightarrow W(Q_1)\ q(\frac{Q_2}{\tau_3}) \right] \right\}
\nonumber \\
\,
\end{eqnarray}
\begin{eqnarray}
\label{2.15}
\sum_{X} \left( \frac{D^2\sigma^H}{DQ_1DQ_2} \right)^{g \bar q}\left[ p(P_1)\
\bar p(P_2) \rightarrow W(Q_1)\ \gamma(Q_2) \ X \right] &=&
\int_{0}^{1}d\tau_1\int_{0}^{1}d\tau_2 \nonumber \\
& &\hspace{-3.in}\times\left\{ f_{gp}(\tau_1)f_{\bar q \bar p}(\tau_2)\
\frac{D^2\sigma}{DQ_1DQ_2}\left[ g(\tau_1P_1)\ \bar q(\tau_2P_2)\rightarrow
W(Q_1)\ \gamma(Q_2)\ \bar q \right] \right.\nonumber \\
& &\hspace{-2.8in}\left. + f_{\bar qp}(\tau_1)f_{g \bar p}(\tau_2)\
\frac{D^2\sigma}{DQ_1DQ_2}\left[ \bar q(\tau_1P_1)\ g(\tau_2P_2)\rightarrow
W(Q_1)\ \gamma(Q_2) \ \bar q\right] \right\}\nonumber \\
& &\hspace{-3.in}+
\int_{0}^{1}d\tau_1\int_{0}^{1}d\tau_2\int_{0}^{1}d\tau_3\nonumber \\
& &\hspace{-3.in}\times\left\{ f_{gp}(\tau_1)f_{\bar q \bar p}(\tau_2)f_{\gamma
\bar q}(\tau_3) \frac{D^2\sigma}{DQ_1DQ_2}\left[ g(\tau_1P_1)\ \bar
q(\tau_2P_2)\rightarrow W(Q_1)\ \bar q(\frac{Q_2}{\tau_3}) \right]
\right.\nonumber \\
& &\hspace{-2.9in}\left. + f_{\bar qp}(\tau_1)f_{g \bar p}(\tau_2)f_{\gamma
\bar q}(\tau_3) \frac{D^2\sigma}{DQ_1DQ_2}\left[ \bar q(\tau_1P_1)\
g(\tau_2P_2)\rightarrow W(Q_1)\ \bar q(\frac{Q_2}{\tau_3}) \right] \right\}\,.
\nonumber \\
\,
\end{eqnarray}

\subsection{\sc Factorization in the $q \bar q$ channel.}

Let us write (2.5) for $i=q, j=\bar q$ and $k=\gamma$:
\begin{eqnarray}
\label{2.16}
\sum_{x} \frac{D^2\sigma^P}{DQ_1DQ_2}\left[ q(p_1)\ \bar q(p_2) \rightarrow
W(Q_1)\ \gamma(Q_2) \ x \right] &=& \nonumber \\
& &\hspace{-1.6in}\sum_{a,b,c,x_a,x_b,x_c} \left(
\int_{0}^{1}dv_1\int_{0}^{1}dv_2\int_{0}^{1}dv_3\ d_{aq}^{x_a}(v_1)d_{b\bar
q}^{x_b}(v_2)d_{\gamma c}^{x_c}(v_3)
 \right. \nonumber \\
& &\hspace{-1.5in}\left. \times\sum_{y} \frac{D^2\sigma}{DQ_1DQ_2}\left[
a(v_1p_1)\ b(v_2p_2) \rightarrow W(Q_1)\ c(\frac{Q_2}{v_3})\ y\right]
\right)\,. \nonumber \\
\,
\end{eqnarray}
We will solve (\ref{2.16}) at $O(\alpha\alpha_W)$ and
$O(\alpha_S\alpha\alpha_W)$ so only $x=\{\ \}$ and $x=\{g\}$ contribute on the
LHS. By constraining the sums on the RHS so that $x_a \cup x_b \cup x_c \cup y
\subseteq x$ we obtain at $O(\alpha_S^0)$:
\begin{eqnarray}
\label{2.17}
\frac{D^2\sigma^{(0)}}{DQ_1DQ_2}\left[ q(p_1)\ \bar q(p_2) \rightarrow W(Q_1)\
\gamma(Q_2) \right] &=&
\frac{D^2\sigma^{P(0)}}{DQ_1DQ_2}\left[ q(p_1)\ \bar q(p_2) \rightarrow W(Q_1)\
\gamma(Q_2)\right] \nonumber \\
\,
\end{eqnarray}
and at $O(\alpha_S)$:
\begin{eqnarray}
\label{2.18}
\lefteqn{\frac{D^2\sigma^{(1)}}{DQ_1DQ_2}\left[ q(p_1)\ \bar q(p_2) \rightarrow
W(Q_1)\ \gamma(Q_2)\right]}\nonumber \\
& &+ \frac{D^2\sigma^{(1)}}{DQ_1DQ_2}\left[ q(p_1)\ \bar q(p_2) \rightarrow
W(Q_1)\ \gamma(Q_2)\ g \right] = \nonumber \\
& &\hspace{1.in}\frac{D^2\sigma^{P(1)}}{DQ_1DQ_2}\left[ q(p_1)\ \bar q(p_2)
\rightarrow W(Q_1)\ \gamma(Q_2)\right]\nonumber \\
& &\hspace{0.9in}+ \frac{D^2\sigma^{P(1)}}{DQ_1DQ_2}\left[ q(p_1)\ \bar q(p_2)
\rightarrow W(Q_1)\ \gamma(Q_2) \ g \right]\nonumber \\
& &\hspace{0.9in}+ \frac{\alpha_S}{2\pi\overline{\epsilon}}\int_{0}^{1}dv\
\overline{P}_{qq}(v)\frac{D^2\sigma^{(0)}}{DQ_1DQ_2}\left[ q(vp_1)\ \bar q(p_2)
\rightarrow W(Q_1)\ \gamma(Q_2) \right]\nonumber \\
& &\hspace{0.9in}+ \frac{\alpha_S}{2\pi\overline{\epsilon}}\int_{0}^{1}dv\
\overline{P}_{\bar q \bar q}(v)\frac{D^2\sigma^{(0)}}{DQ_1DQ_2}\left[ q(p_1)\
\bar q(vp_2) \rightarrow W(Q_1)\ \gamma(Q_2) \right]\,, \nonumber \\
\,
\end{eqnarray}
and the corresponding equations for $p_1 \longleftrightarrow p_2$. In deriving
(\ref{2.17}) and (\ref{2.18}) we have used the following splitting functions:
\begin{eqnarray}
d_{ai}^{\{\ \}}(v) &=& \delta_{ai}\delta(1-v) \nonumber \\
d_{aq}^{\{g\}}(v) &=&
-\frac{\alpha_S}{2\pi\overline{\epsilon}}\overline{P}_{qq}(v)\delta_{aq}
\nonumber \\
d_{b \bar q}^{\{g\}}(v) &=&
-\frac{\alpha_S}{2\pi\overline{\epsilon}}\overline{P}_{\bar q \bar
q}(v)\delta_{b \bar q} \,
\end{eqnarray}
and the definition:
\begin{eqnarray}
\overline{P}_{ij}(v) \equiv P_{ij}(v) - \overline{\epsilon}K_{ij}(v) \,
\end{eqnarray}
with
\begin{eqnarray}
\frac{1}{\overline{\epsilon}} \equiv \frac{1}{\epsilon} - \gamma_E + \ln(4\pi)
\,
\end{eqnarray}
and $\epsilon\equiv (4-n)/2$. The running strong fine structure constant is
$\alpha_S=g^2_S(\mu)/4\pi$. In the $\overline{MS}$ mass factorization scheme
$K_{ij}(v)=0$ for all relevant $i,j$ in this and the following two
subsections.\\

For $i = j = q(\bar q)$ we have:
\begin{eqnarray}
P_{qq}(v) = P_{\bar q \bar q}(v) &=& C_F\left[ (1+v^2)\left\{ \frac{1}{1-v}
\right\}_{0} + \frac{3}{2}\delta(v-1) \right] \nonumber \\
&=& C_F\left[ (1+v^2)\left\{ \frac{1}{1-v} \right\}_{v_0} + \left( \frac{3}{2}
+ 2\ln\left(1-v_0\right) \right)\delta(1-v) \right] \nonumber \\
\,
\end{eqnarray}
with
\begin{eqnarray}
\int_{0}^{1}dv\left\{ \frac{1}{1-v} \right\}_{v_0}f(v) \equiv \int_{0}^{v_0}dv
\frac{f(v)}{1-v}\ +\
\int_{v_0}^{1}dv \frac{f(v)-f(1)}{1-v} \,
\end{eqnarray}
where $0 \leq v_0 < 1$ and the color factor is given by $C_F = 4/3$.

In an analogous way, setting $i=q, j=\bar q, k=g$ in (\ref{(2.5)}) we obtain:
\begin{eqnarray}
\label{2.24}
\frac{D^2\sigma^{(1)}}{DQ_1DQ_2}\left[ q(p_1)\ \bar q(p_2) \rightarrow W(Q_1)\
g(Q_2) \right] &=&
\frac{D^2\sigma^{P(1)}}{DQ_1DQ_2}\left[ q(p_1)\ \bar q(p_2) \rightarrow W(Q_1)\
g(Q_2)\right]\,. \nonumber \\
\,
\end{eqnarray}

\subsection{\sc Factorization in the $qg$ channel.}

Setting $i=q, j=g$ and $k=q$ in (\ref{(2.5)}) and again keeping terms up to
$O(\alpha_S\alpha\alpha_W)$
we can only have $x=\{\ \}$, thus constraining the sums on the RHS
to $x_a \cup x_b \cup x_c \cup y = \{\ \}$. We obtain at $O(\alpha_S)$:
\begin{eqnarray}
\label{2.25}
\frac{D^2\sigma^{(1)}}{DQ_1DQ_2}\left[ q(p_1)\ g(p_2) \rightarrow W(Q_1)\
q(Q_2) \right] &=&
\frac{D^2\sigma^{P(1)}}{DQ_1DQ_2}\left[ q(p_1)\ g(p_2) \rightarrow W(Q_1)\
q(Q_2)\right]\,. \nonumber \\
\,
\end{eqnarray}
Resetting $k=\gamma$ in (\ref{(2.5)}) we can now only have $x=\{q\}$, thus
constraining the sums on
the RHS to $x_a \cup x_b \cup x_c \cup y = \{q\}$. We obtain at $O(\alpha_S)$:
\begin{eqnarray}
\label{2.26}
\lefteqn{ \frac{D^2\sigma^{(1)}}{DQ_1DQ_2}\left[ q(p_1)\ g(p_2) \rightarrow
W(Q_1)\ \gamma(Q_2)\ q \right] = } \nonumber \\
& &\frac{D^2\sigma^{P(1)}}{DQ_1DQ_2}\left[ q(p_1)\ g(p_2) \rightarrow W(Q_1)\
\gamma(Q_2) \ q \right]\nonumber \\
& &\hspace{-0.1in}+ \frac{\alpha_S}{2\pi\overline{\epsilon}}\int_{0}^{1}dv\
\overline{P}_{\bar q g}(v)\frac{D^2\sigma^{(0)}}{DQ_1DQ_2}\left[ q(p_1)\ \bar
q(vp_2) \rightarrow W(Q_1)\ \gamma(Q_2) \right]\nonumber \\
& &\hspace{-0.1in}+ \frac{\alpha}{2\pi\overline{\epsilon}}\int_{0}^{1}dv\
\overline{P}_{\gamma q}(v)\frac{D^2\sigma^{(1)}}{DQ_1DQ_2}\left[ q(p_1)\ g(p_2)
\rightarrow W(Q_1)\ q(\frac{Q_2}{v}) \right] \nonumber \\
\,
\end{eqnarray}
and the corresponding equations for $p_1 \longleftrightarrow p_2$.
The following splitting functions have been used when deriving  (\ref{2.26}):
\begin{eqnarray}
d_{bg}^{\{q\}}(v) &=&
-\frac{\alpha_S}{2\pi\overline{\epsilon}}\overline{P}_{\bar qg}(v)\delta_{b\bar
q}  \nonumber \\
d_{\gamma c}^{\{q\}}(v) &=&
-\frac{\alpha}{2\pi\overline{\epsilon}}\overline{P}_{\gamma q}(v)\delta_{cq} \,
\end{eqnarray}
with
\begin{eqnarray}
P_{\bar q g}(v) &=& \frac{v^2 + (1-v)^2}{2} \nonumber \\
P_{\gamma q}(v) &=& \left(\hat{e}_q\right)^2\frac{1 + (1-v)^2}{v} \,
\end{eqnarray}
where $\hat{e}_q = -1/3$ is the charge of the outgoing quark $q$ on the LHS of
(\ref{2.26}), in units of $e$ and
$\alpha = e^2(\mu)/4\pi$ is the running electromagnetic fine structure
constant.

\subsection{\sc Factorization in the $g\bar q$ channel.}

Analogously to the previous case, setting $i=g, j=\bar q$ and $k=\bar q,\gamma
$ in (\ref{(2.5)}) and keeping up
to $O(\alpha_S\alpha\alpha_W)$ we obtain:
\begin{eqnarray}
\label{2.29}
\frac{D^2\sigma^{(1)}}{DQ_1DQ_2}\left[ g(p_1)\ \bar q(p_2) \rightarrow W(Q_1)\
\bar q(Q_2) \right] &=&
\frac{D^2\sigma^{P(1)}}{DQ_1DQ_2}\left[ g(p_1)\ \bar q(p_2) \rightarrow W(Q_1)\
\bar q(Q_2)\right] \nonumber \\
\,
\end{eqnarray}
and
\begin{eqnarray}
\label{2.30}
\frac{D^2\sigma^{(1)}}{DQ_1DQ_2}\left[ g(p_1)\ \bar q(p_2) \rightarrow W(Q_1)\
\gamma(Q_2)\ \bar q \right] &=& \nonumber \\
& &\hspace{-3.in}\frac{D^2\sigma^{P(1)}}{DQ_1DQ_2}\left[ g(p_1)\ \bar q(p_2)
\rightarrow W(Q_1)\ \gamma(Q_2)\ \bar q \right]\nonumber \\
& &\hspace{-3.1in}+ \frac{\alpha_S}{2\pi\overline{\epsilon}}\int_{0}^{1}dv\
\overline{P}_{\bar q g}(v)\frac{D^2\sigma^{(0)}}{DQ_1DQ_2}\left[ q(vp_1)\ \bar
q(p_2) \rightarrow W(Q_1)\ \gamma(Q_2) \right]\nonumber \\
& &\hspace{-3.1in}+ \frac{\alpha}{2\pi\overline{\epsilon}}\int_{0}^{1}dv\
\overline{P}_{\gamma \bar q}(v)\frac{D^2\sigma^{(1)}}{DQ_1DQ_2}\left[ g(p_1)\
\bar q(p_2) \rightarrow W(Q_1)\ \bar q(\frac{Q_2}{v}) \right] \nonumber \\
\,
\end{eqnarray}
and the corresponding equations for $p_1 \longleftrightarrow p_2$.
In deriving  (\ref{2.30}) we have used the following splitting functions:
\begin{eqnarray}
\label{2.33}
d_{ag}^{\{\bar q\}}(v) &=&
-\frac{\alpha_S}{2\pi\overline{\epsilon}}\overline{P}_{qg}(v)\delta_{aq}
\nonumber \\
d_{\gamma c}^{\{\bar q\}}(v) &=&
-\frac{\alpha}{2\pi\overline{\epsilon}}\overline{P}_{\gamma \bar
q}(v)\delta_{c\bar q}\,
\end{eqnarray}
with
\begin{eqnarray}
P_{q g}(v) &=& \frac{v^2 + (1-v)^2}{2} \nonumber \\
P_{\gamma \bar q}(v) &=& \left(\hat{e}_{\bar q}\right)^2\frac{1 + (1-v)^2}{v}\,
\end{eqnarray}
where $\hat{e}_{\bar q} = -2/3$ is the charge fraction of the outgoing
antiquark $\bar q$ on the LHS of (\ref{2.30}), in units of $e$.

\newpage

\newcommand{\plusdist}[3]{ \left\{ \frac{#1}{#2} \right\}_{#3} }
\newcommand{\sla}[1]{ /\hspace{-0.1in}{#1} }
\mysection{\bf EXCLUSIVE CANCELLATION OF SINGULARITIES.}
\label{3}

\subsection{\sc Introduction.}

In order to compute the hard scattering cross sections required
in  (\ref{2.13})-(\ref{2.15}) using  (\ref{2.17}), (\ref{2.18}),
(\ref{2.24})-(\ref{2.26}), (\ref{2.29}) and (\ref{2.30})
we first need the bare partonic cross sections $D^2\sigma^P/DQ_1DQ_2$
on the RHS of these equations evaluated in $n = 4-2\epsilon$ space-time
dimensions.

When computing phase space integrations of outgoing massless particles
singularities
appear in regions of phase space where one of these particles is collinear to
any
other massless on-shell parton or where one of the outgoing massless gauge
bosons is soft.
Since we are tagging the outgoing photon, we don't have to worry about
singularities associated with integration over the photon's phase space.

With this in mind we will classify the $2$ to $3$ body Feynman amplitudes
according to the
way the outgoing massless particle $q, \bar q$ or $g$ which is integrated over
is attached to the
rest of the legs of the diagram.
In Fig. 1 we have decomposed the $2$ to $3$ body Feynman amplitude
for the partonic reaction $q\ +\ \bar q\rightarrow W\ +\ \gamma\ +\ g$
into three pieces: $M^{q\bar q \rightarrow W\gamma g} = M^{q\bar q}_{Ia} +
M^{q\bar q}_{Ib} + M^{q\bar q}_{III}$:\\
\hspace{1.in}
\par
\centerline{\hbox{
\psfig{figure=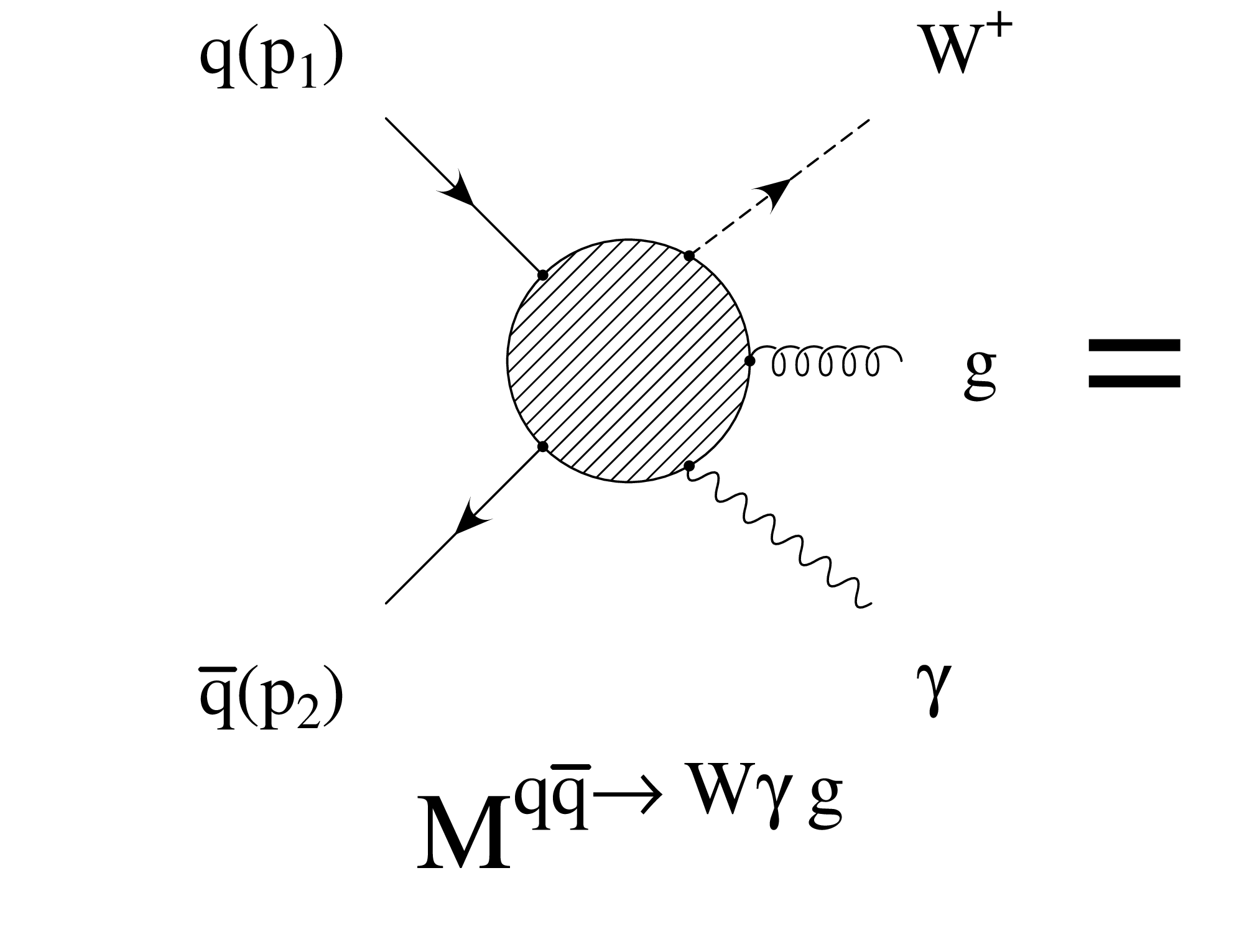,height=1.5in}
\psfig{figure=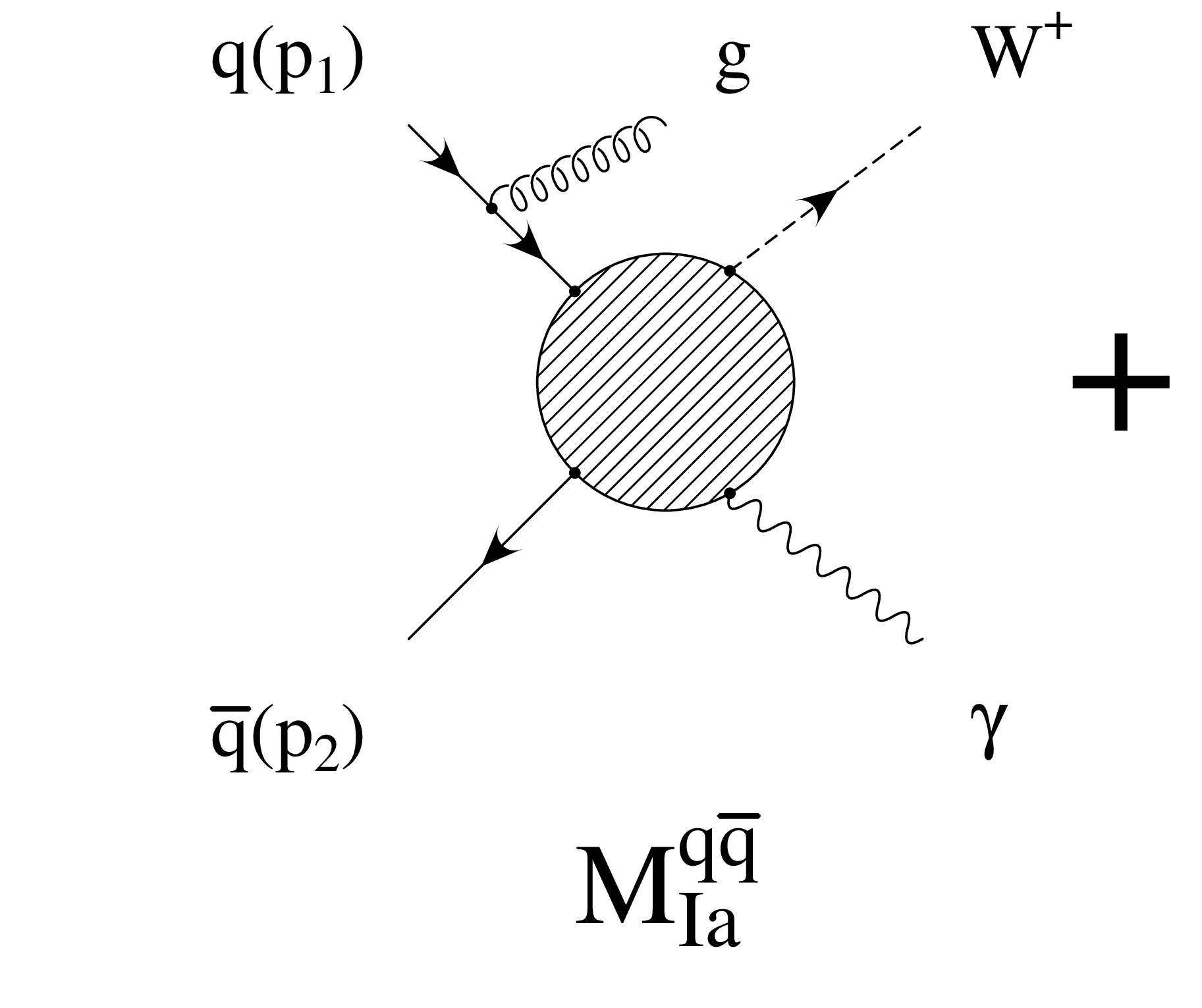,height=1.5in}
\psfig{figure=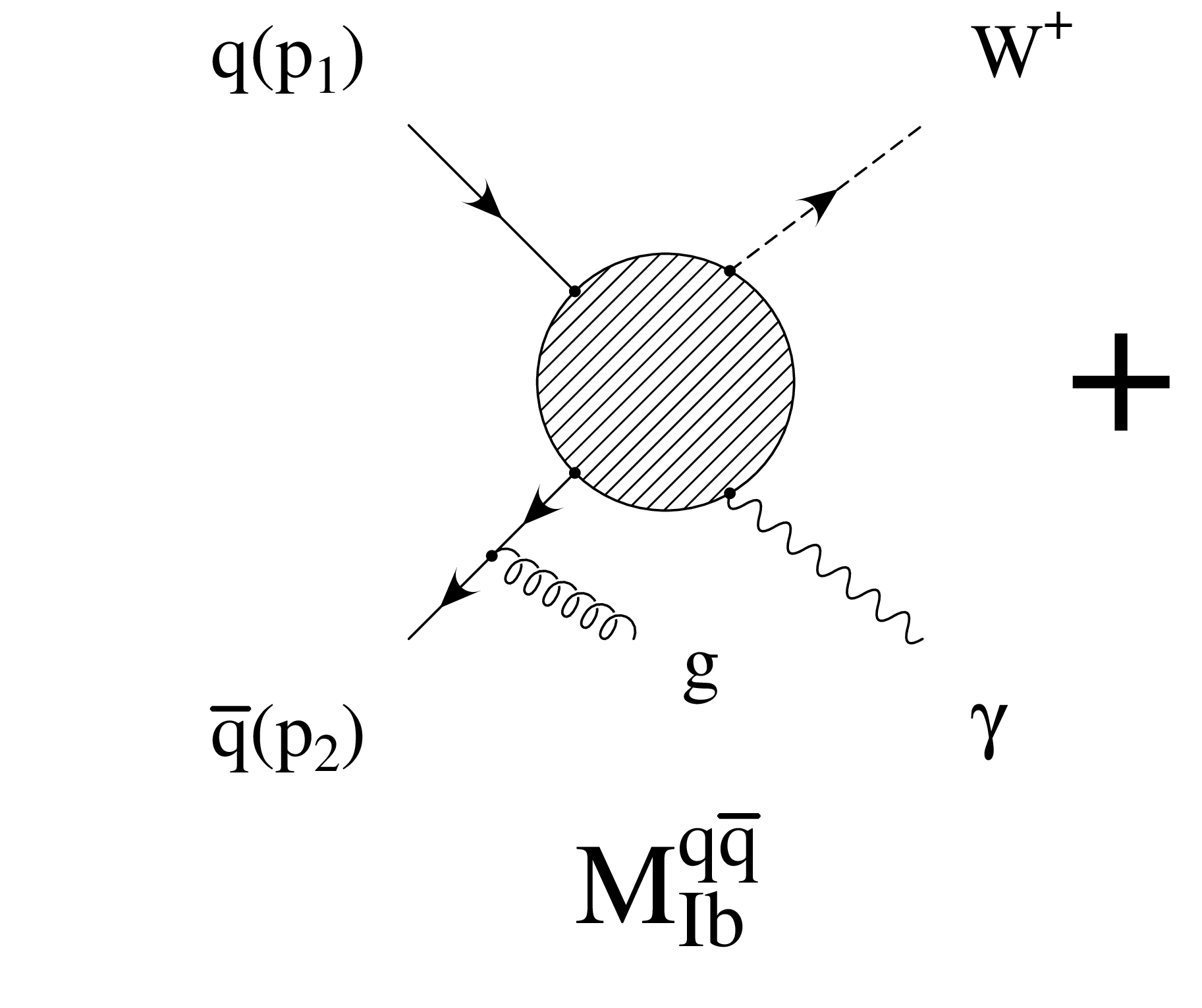,height=1.5in}
\psfig{figure=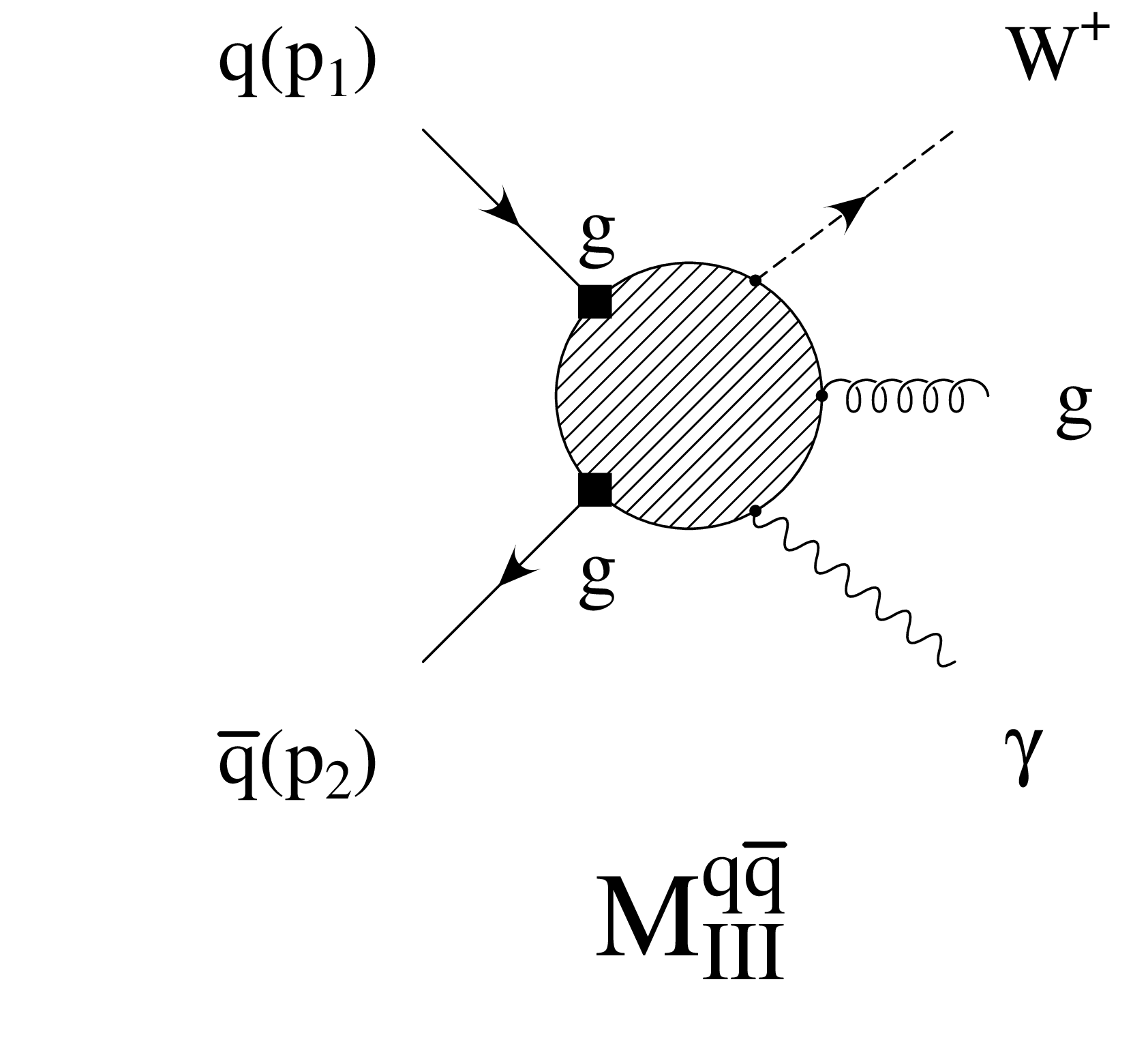,height=1.5in}
}}
\centerline{Figure 1. Decomposition of the 2 to 3 body Feynman amplitude in the
$q\bar q$
channel.\\}
\par
\hspace{1.in}

The labels ``g'' near the solid vertices at the end of the incoming quark and
antiquark
legs in the amplitude $M^{q\bar q}_{III}$ mean that these vertices do not
contain the outgoing gluon.
Arrows show the direction of the fermionic charge and the $W^+$ charge.
The momenta $p_1$ and $p_2$ are always incoming.
The shaded blobs denote the inclusion of all possible
Feynman diagrams (except for the mentioned constraints in $M^{q\bar q}_{III}$.)
If we integrate the squared matrix element over the phase space of the outgoing
gluon summed over physical
polarizations the $\sum\left|M^{q\bar q}_{Ia}\right|^2$ and
$\sum\left|M^{q\bar q}_{Ib}\right|^2$ pieces of the squared matrix element have
collinear and soft singularities
while the interference term $(\sum M^{q\bar q}_{Ia}M^{\star q\bar q}_{Ib} +
c.c)$ has only soft singularities.
Other pieces of the squared matrix element have no singularities.

For the partonic reaction $q\ +\ g\rightarrow W\ +\ \gamma\ +\ q$
we have $M^{qg\rightarrow W\gamma q} = M^{qg}_{Ib} + M^{qg}_{II} +
M^{qg}_{III}$, as shown in Fig. 2.\\
\hspace{1.in}
\par
\centerline{\hbox{
\psfig{figure=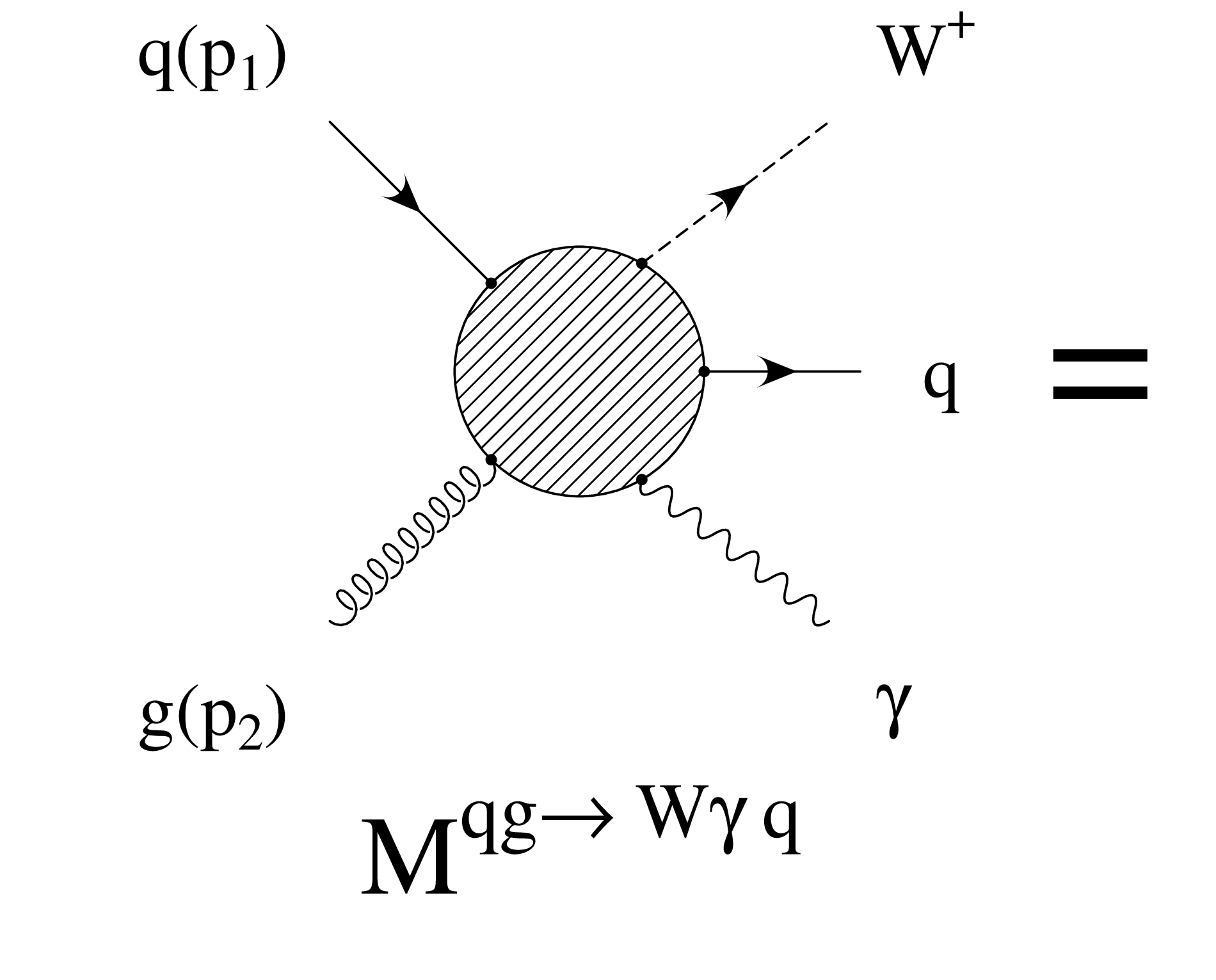,height=1.5in}
\psfig{figure=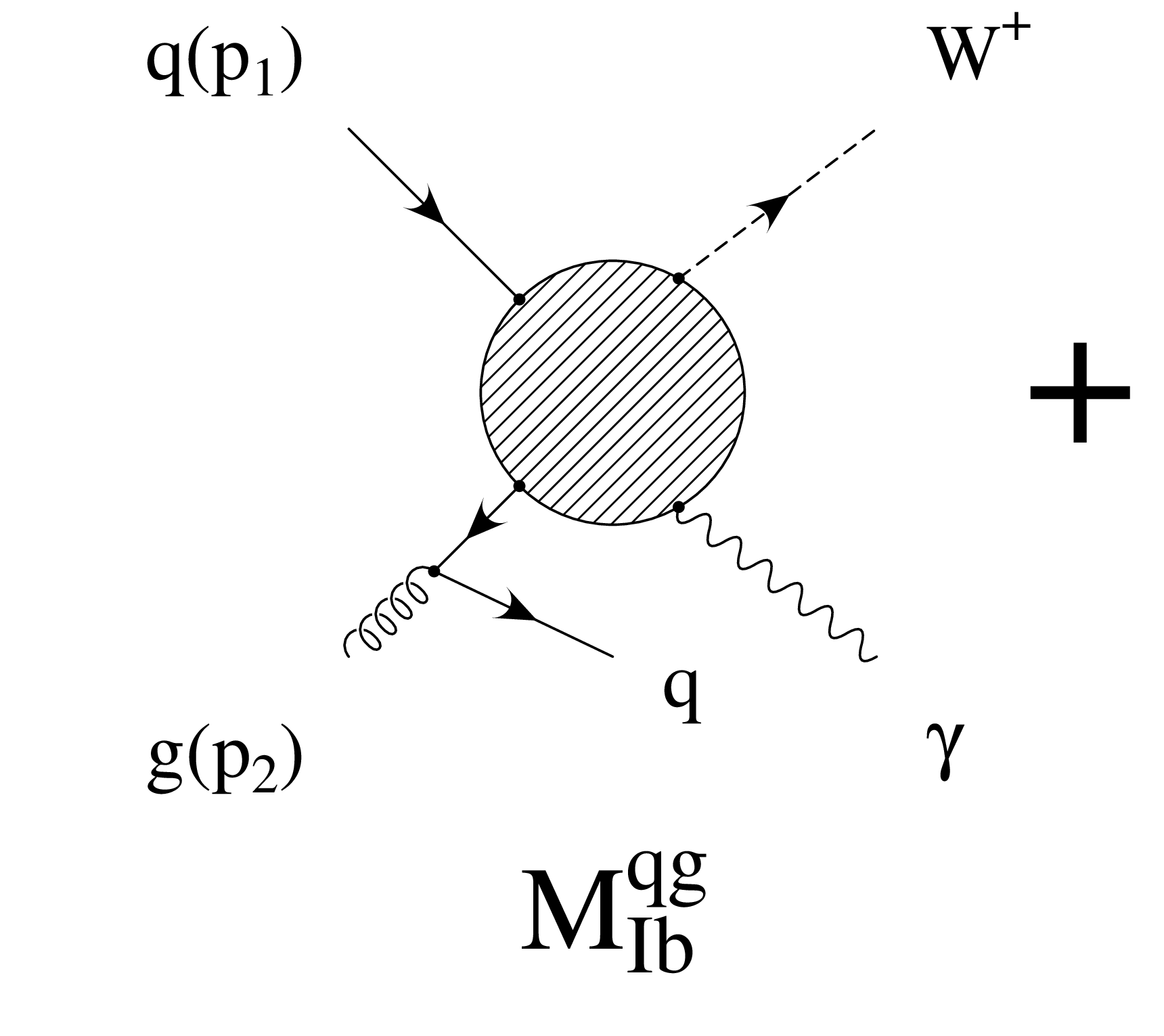,height=1.5in}
\psfig{figure=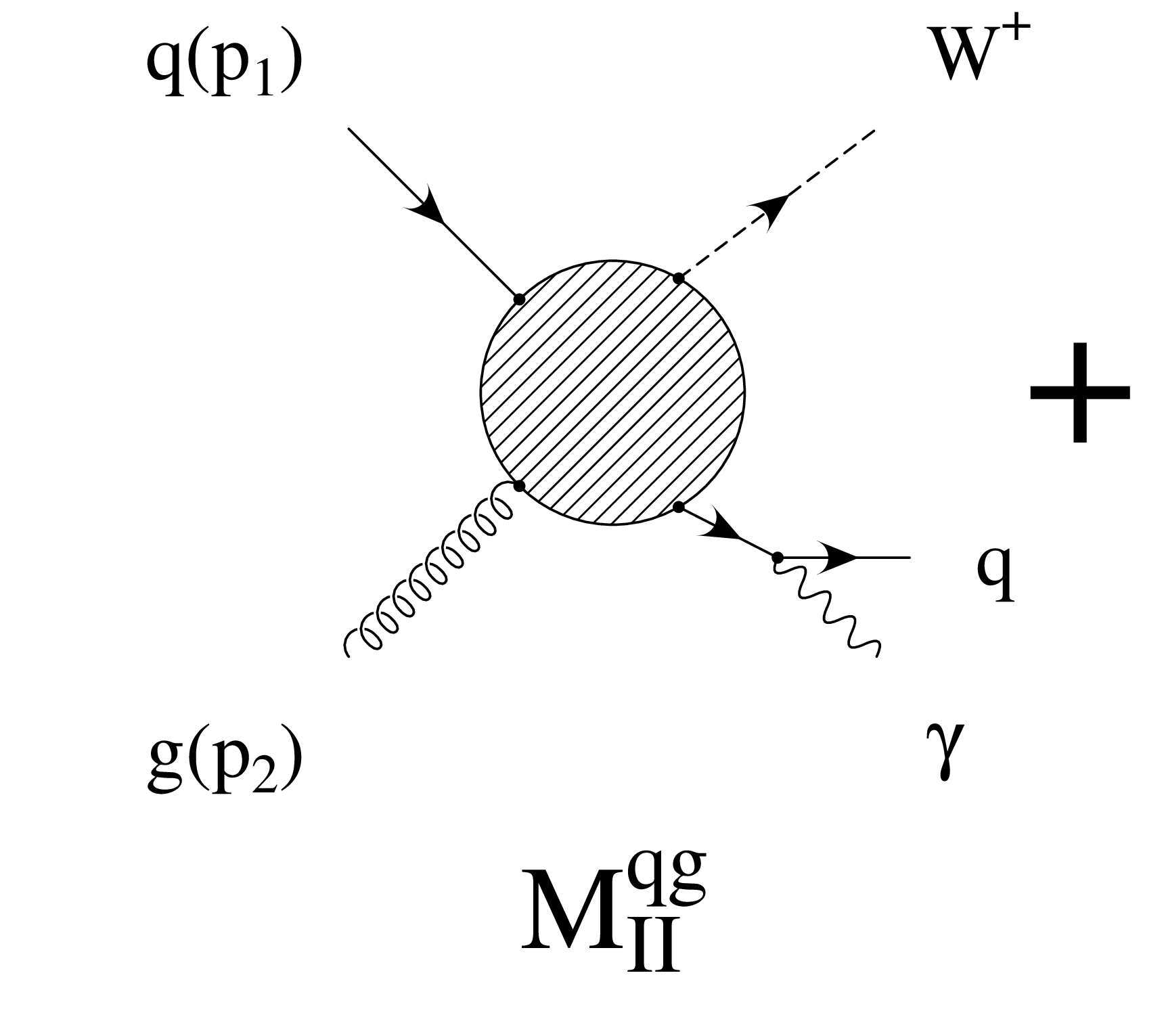,height=1.5in}
\psfig{figure=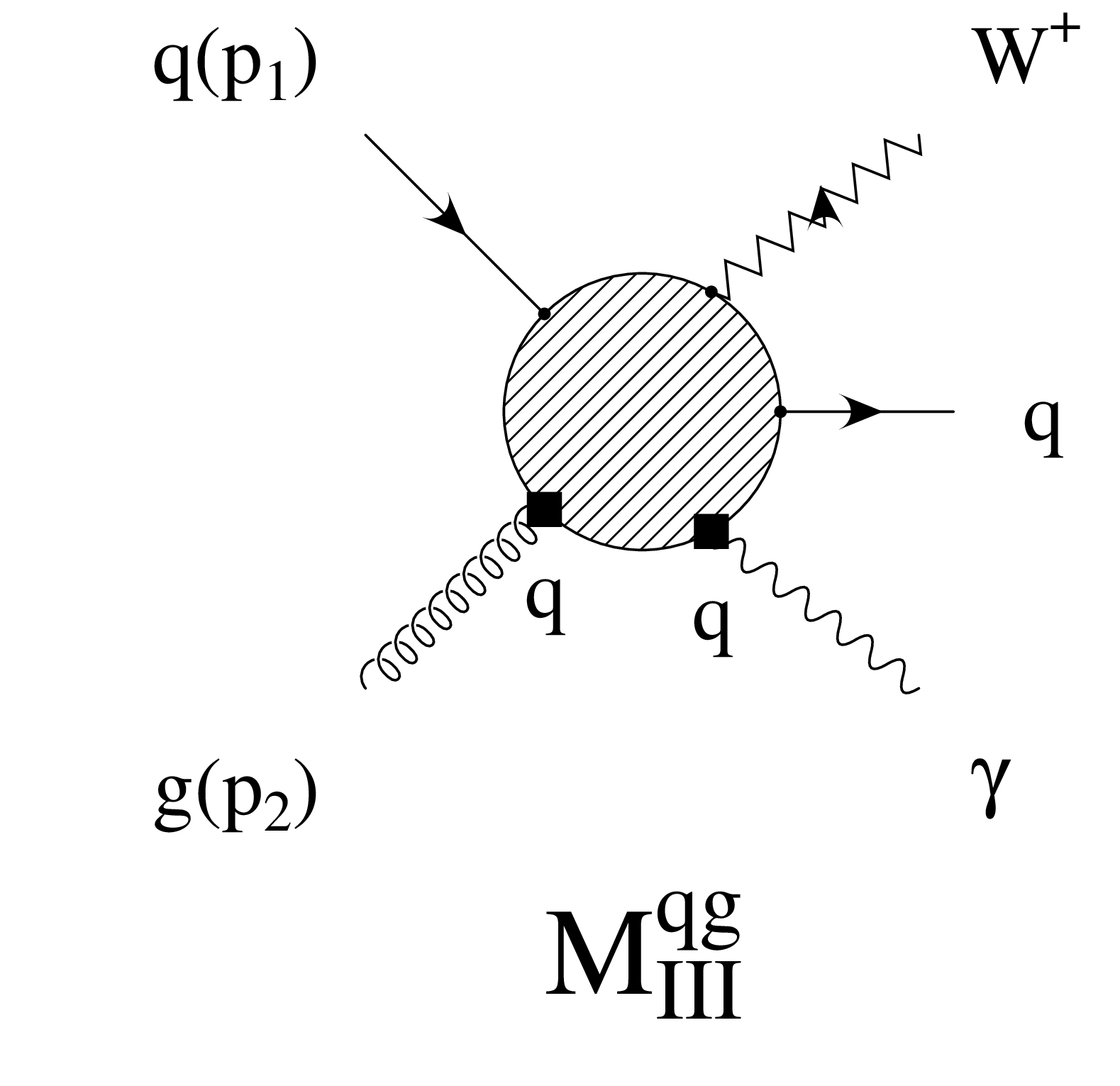,height=1.5in}
}}
\centerline{Figure 2. Decomposition of the 2 to 3 body Feynman amplitude in the
$q g$
channel.\\}
\hspace{1.in}

In this case, only the $\sum\left|M^{qg}_{Ib}\right|^2$ and
$\sum\left|M^{qg}_{II}\right|^2$ pieces of the squared matrix element have
collinear singularities, with no soft singularities in this channel.

In Fig.3 we show the analogous decomposition
for the partonic reaction $g\ +\ \bar q\rightarrow W\ +\ \gamma\ +\ \bar q$:
$M^{g\bar q\rightarrow W\gamma \bar q} = M^{g\bar q}_{Ia} + M^{g\bar q}_{II} +
M^{g\bar q}_{III}$. As in the previous case, only  the $\sum\left|M^{g\bar
q}_{Ia}\right|^2$ and $\sum\left|M^{g\bar q}_{II}\right|^2$ pieces of the
squared matrix element heve collinear singularities and no soft singularities
are present in this channel.\\
\hspace{1.in}
\par
\centerline{\hbox{
\psfig{figure=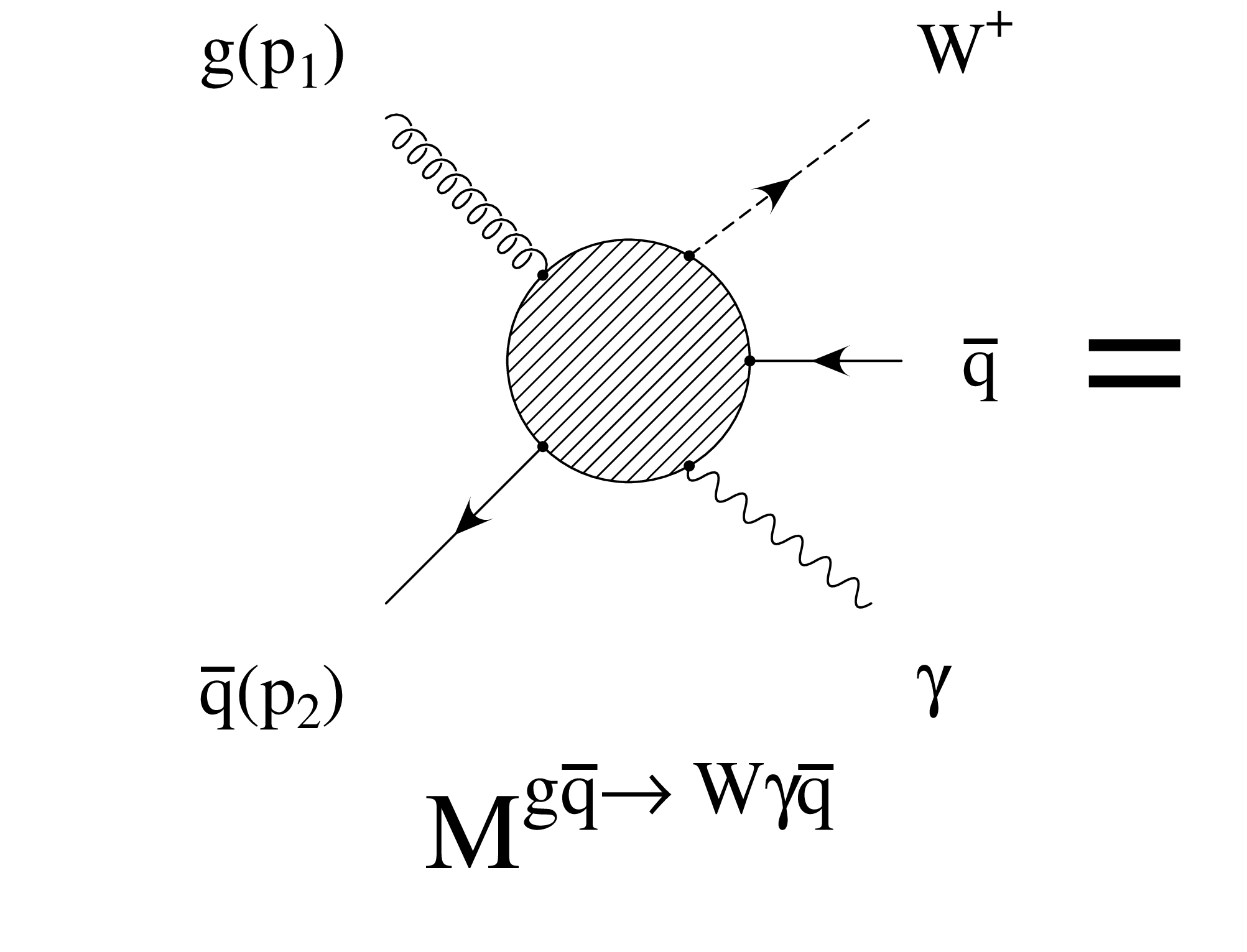,height=1.5in}
\psfig{figure=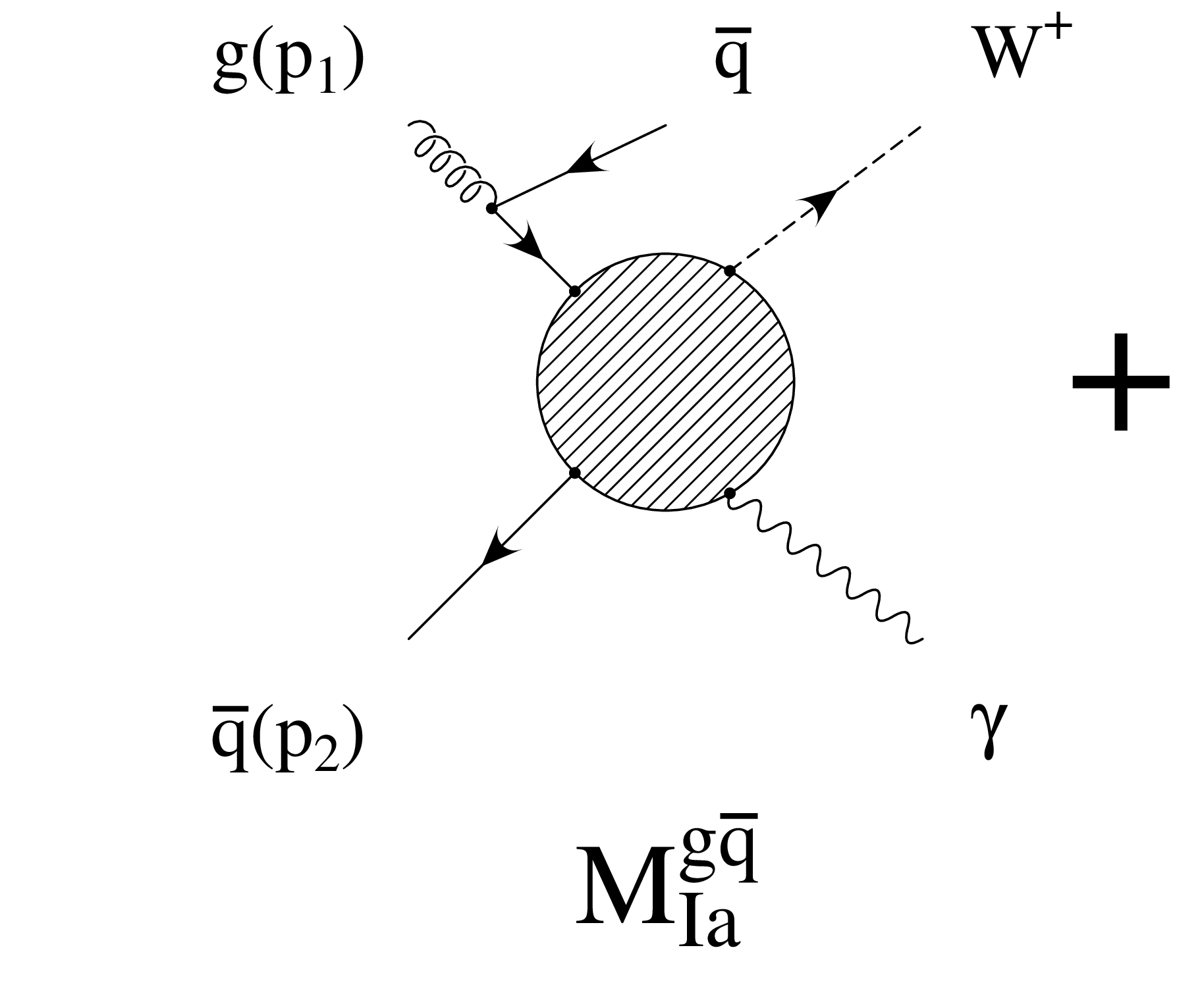,height=1.5in}
\psfig{figure=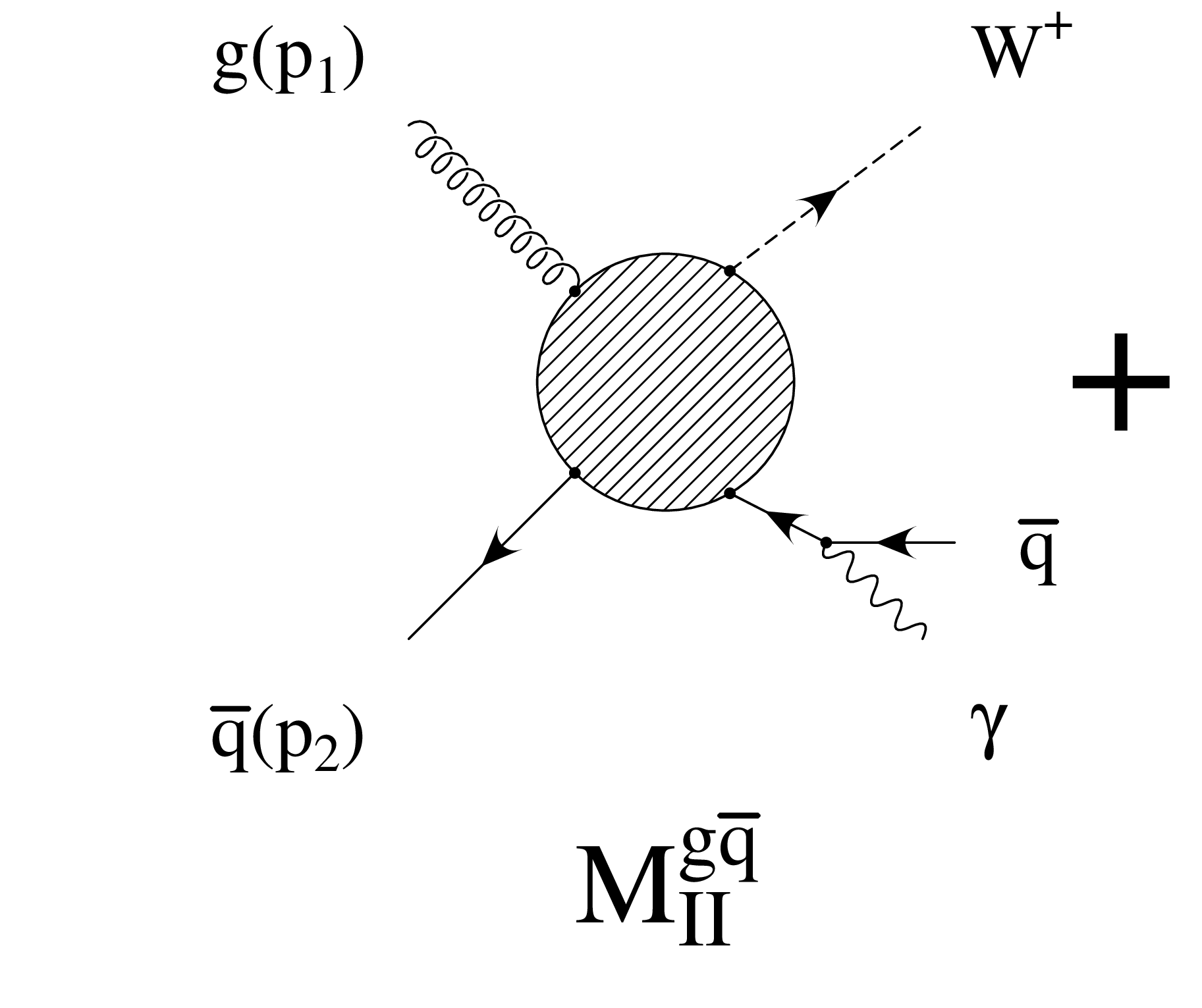,height=1.5in}
\psfig{figure=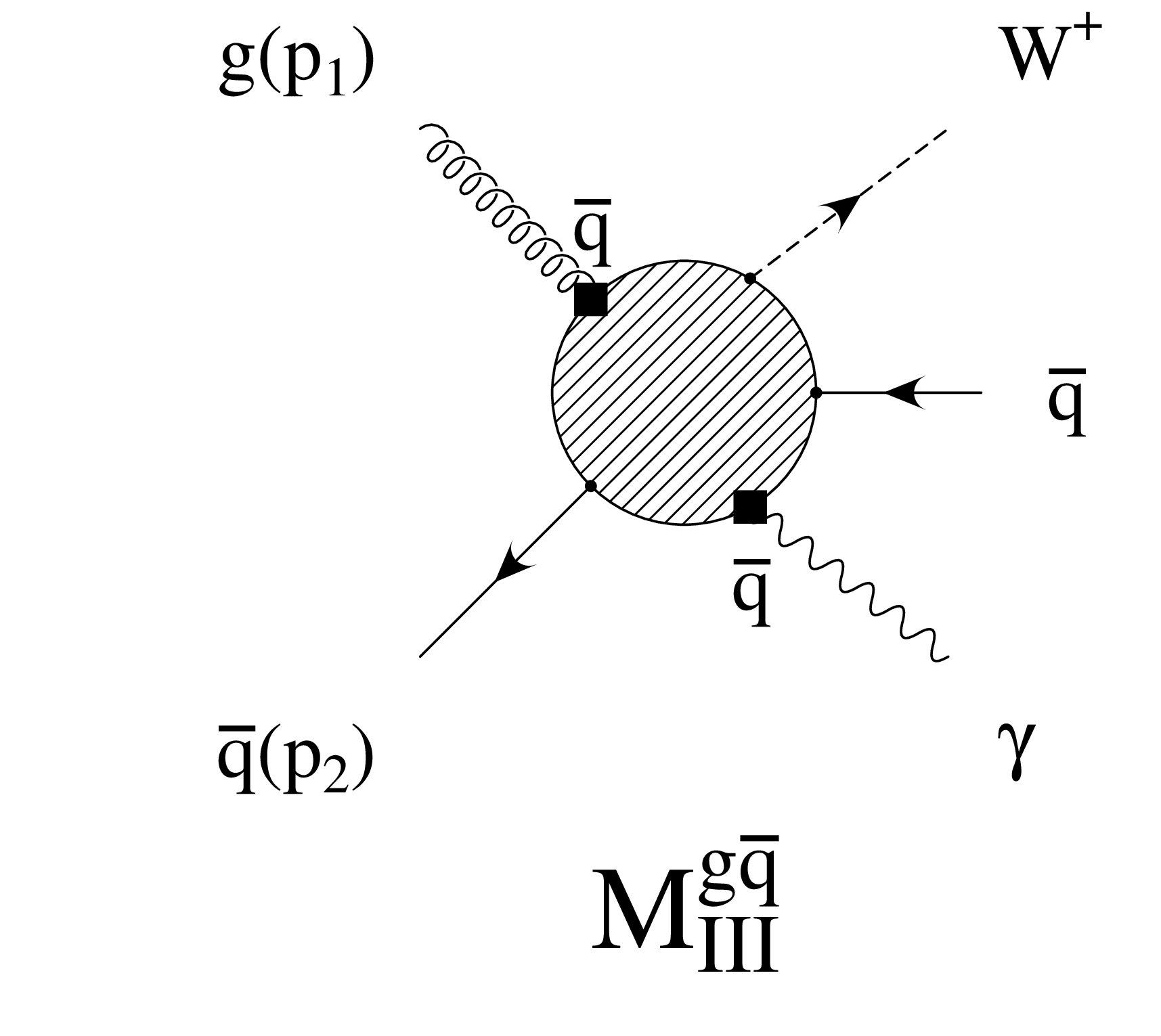,height=1.5in}
}}
\centerline{Figure 3. Decomposition of the 2 to 3 body Feynman amplitude in the
$g \bar q$ channel.\\}
\par
\hspace{1.in}

We will develop the above decompositions in more detail for each channel in
subsections III.B, C and D.

The $2$ to $(2+l)$ body partonic differential cross section for the
$W+\gamma+X$ production process
is defined in $n$ dimensions of phase space by:
\begin{eqnarray}
\label{3.1}
d\sigma^P_{(2+l)}&=&\frac{1}{N}\frac{1}{2s}DQ_1DQ_2Dk_1\cdots
Dk_l(2\pi)^n\delta^n(p_1+p_2-Q_1-Q_2-k_1\cdots -k_{l})\nonumber \\
& &\times\sum|M|^2\,, \nonumber \\
\,
\end{eqnarray}
where we have averaged over $N$ possible incoming states of different
polarizations and colors and
the sum on the RHS is over polarizations and colors of all particles.
$Q_1$ and $Q_2$ are the $n$-momenta of the $W^+$ boson and the photon
respectively.
The variables in (\ref{3.1}) are defined in the following way:
\begin{eqnarray}
\label{3.2}
s&=&2p_1\cdot p_2 \nonumber \\
DQ_1&=&\frac{1}{2(2\pi)^{n-1}}\frac{\left|\vec{Q}_1\right|^{n-2}}{\sqrt{\left|\vec{Q}_1\right|^2 + M^2_W}}
d\left|\vec{Q}_1\right|
d\Omega_{n-2}\left(\hat{Q}_1\left(\theta_1,\ldots,\theta_{n-2}\right)\right)
\nonumber \\
DQ_2&=&\frac{1}{2(2\pi)^{n-1}}\left|\vec{Q}_2\right|^{n-3}d\left|\vec{Q}_2\right| d\Omega_{n-2}\left(\hat{Q}_2\left(\phi_1,\ldots,\phi_{n-2}\right)\right) \nonumber \\
Dk_1&=&Dk=\frac{1}{2(2\pi)^{n-1}}\left|\vec{k}\right|^{n-3}d\left|\vec{k}\right|d\Omega_{n-2}\left(\hat{k}\left(\psi_1,\ldots,\psi_{n-2}\right)\right)\,. \nonumber \\
\,
\end{eqnarray}
The angle differentials in (\ref{3.2}) are generically given by:
\begin{eqnarray}
\label{3.3}
d\Omega_{n-2}\left(\hat{p}\left(\alpha_1,\ldots,\alpha_{n-2}\right)\right)&\equiv&d\cos\alpha_1\ \sin^{n-4}\alpha_1
\cdots d\cos\alpha_{n-3}\ \sin^{0}\alpha_{n-3}d\alpha_{n-2}\,. \nonumber \\
\,
\end{eqnarray}
To account for the experimental cuts on the outgoing particles that define the
experimental scenario under consideration
we have to include on the RHS of (\ref{3.1}) an extra factor of
$C(Q_1,Q_2,k_1,\ldots,k_l)$. These cuts may be expressed in a covariant way in
terms of $\Theta$ step functions as we will see in more detail in section IV.
In the rest of this chapter and the following sections we omit the charge index
``$^+$'' when referring to the $W^+$ boson.

\subsection{\sc The $q\bar q$ channel.}

The 2 to 2 body total partonic cross section for the reaction $q\bar
q\rightarrow W\gamma$ may be written in the following way:
\begin{eqnarray}
\label{3.4a}
\int DQ_1DQ_2\frac{D^2\sigma^P}{DQ_1DQ_2}\left[q(p_1)\bar q(p_2)\rightarrow
W(Q_1)\gamma(Q_2)\right] &=&\nonumber \\
&
&\hspace{-3.in}\frac{1}{4N^2_c}\frac{1}{2s}\Theta\left(\beta(s)\right)\Phi(s)\int_{-1}^{1}d\cos\theta_1\ \sin^{-2\epsilon}\theta_1\sum\left|M^{q\bar q\rightarrow W\gamma}(s,b)\right|^2\nonumber \\
\,
\end{eqnarray}
with
\begin{eqnarray}
\label{3.5}
N_c &=& 3 \nonumber \\
\Phi(s)& =
&\frac{2^{2\epsilon}}{\Gamma(1-\epsilon)}\left(\frac{4\pi}{s}\right)^{\epsilon}
\frac{1}{16\pi}\beta^{1-2\epsilon}(s)\ \nonumber \\
\beta(s) &=& 1-\rho(s) \nonumber \\
\rho(s) &\equiv& \frac{M^2_W}{s}\,.
\end{eqnarray}
The only independent invariants in the squared matrix elements in  (\ref{3.4a})
are $s$ and
\begin{eqnarray}
\label{3.6}
b &\equiv& 2p_1\cdot Q_2 = \frac{s}{2}\beta(s)(1+\cos\theta_1)\,.
\end{eqnarray}
The rest of the invariants may be expressed in terms of $b$ and $s$ as follows
\begin{eqnarray}
\label{3.7}
2p_1\cdot Q_1&=&s-b \nonumber \\
2p_2\cdot Q_1&=&b+M^2_W \nonumber \\
2p_2\cdot Q_2&=&s-M^2_W-b \nonumber \\
2Q_1\cdot Q_2&=&s-M^2_W\,.
\end{eqnarray}

In  (\ref{3.4a}) we chose the $(n-1)$-th axis to be the one pointing in the
direction of $\vec{p}_1$.
The integration over the angles $\theta_2,\ldots,\theta_{n-2}$ has been
performed
because there is no dependence on them in the 2 to 2 body squared matrix
element.
To account for the experimental cuts on the outgoing particles we implicitly
include an extra factor of
$C(Q_1,Q_2,0)$ on the RHS of (\ref{3.4a}).
In (\ref{3.5}) and subsequent equations the unprimed variables refer to
variables in the center of mass
system of the incoming partons.

To obtain the $2$ to $3$ body partonic cross section for the reaction $q\bar
q\rightarrow W\gamma g$ we define a primed
reference frame in the $W\gamma$ center of mass system such that the
$n$-momenta are given by:
\begin{eqnarray}
\label{3.8}
p'_1 &=&p'_{1,0}\left(1,0,\ldots,0,0,1\right) \nonumber \\
p'_2 &=&p'_{2,0}\left(1,0,\ldots,0,\sin\eta',\cos\eta'\right) \nonumber \\
k' &=&k'_{0}\left(1,0,\ldots,0,\sin\psi',\cos\psi'\right) \nonumber \\
Q'_1
&=&\left|\vec{Q}'_1\right|\left(\frac{Q'_{1,0}}{\left|\vec{Q}'_1\right|},\ldots,\sin\theta'_1\sin\theta'_2\cos\theta'_3,
\sin\theta'_1\cos\theta'_2,\cos\theta'_1\right) \nonumber \\
Q'_2
&=&\left|\vec{Q}'_1\right|\left(1,\ldots,-\sin\theta'_1\sin\theta'_2\cos\theta'_3,
-\sin\theta'_1\cos\theta'_2,-\cos\theta'_1\right)\,. \nonumber \\
\,
\end{eqnarray}
The 2 to 3 body total partonic cross section in the $q\bar q$ channel may thus
be written as follows:
\begin{eqnarray}
\label{3.9}
\int DQ_1DQ_2\frac{D^2\sigma^P}{DQ_1DQ_2}\left[q(p_1)\bar q(p_2)\rightarrow
W(Q_1)\gamma(Q_2)g\right] &=& \nonumber \\
&
&\hspace{-3.in}\frac{1}{4N^2_c}\frac{1}{2s}\left(4\pi\right)^{\epsilon-2}\frac{\Gamma(1-\epsilon)}{\Gamma(1-2\epsilon)}\frac{s^{1-\epsilon}}{2\pi}\int_{\rho(s)}^{1}dx\ \Phi(sx)(1-x)^{1-2\epsilon}\int_{-1}^{1}dy\left(1-y^2\right)^{-\epsilon}\nonumber \\
& &\hspace{-3.in}\times\int_{-1}^{1}d\cos\theta'_1\
\sin^{-2\epsilon}\theta'_1\int_{0}^{\pi}d\theta'_2\ \sin^{-2\epsilon}\theta'_2\
\sum\left|M^{q\bar q\rightarrow W\gamma g}(s,a,b,c,d)\right|^2 \nonumber \\
\,
\end{eqnarray}
where
\begin{eqnarray}
\label{3.10}
y&\equiv& \cos\psi \nonumber \\
k_0&=&(1-x)p_{1,0} = (1-x)\frac{\sqrt{s}}{2}\,.
\end{eqnarray}
We implicitly include on the RHS of (\ref{3.9}) the factor $C(Q_1,Q_2,k)$. We
have chosen as independent invariants $s$ and:
\begin{eqnarray}
\label{3.11}
a&\equiv& 2p_1\cdot k = \frac{s}{2}(1-x)(1-y)      \nonumber \\
b&\equiv& 2p_1\cdot Q_2 = 2p'_{1,0}\left|\vec{Q}'_1\right|\left(1 +
\cos\theta'_1\right)   \nonumber \\
c&\equiv& 2k\cdot Q_2 =
2k'_0\left|\vec{Q}'_1\right|\left(1+\sin\psi'\sin\theta'_1\cos\theta'_2+\cos\psi'\cos\theta'_1\right)\nonumber \\
d&\equiv& 2p_2\cdot Q_2 = 2p'_{2,0}\left|\vec{Q}'_1\right|\left(1 +
\sin\eta'\sin\theta'_1\cos\theta'_2 + \cos\eta'\cos\theta'_1\right)\nonumber \\
\,
\end{eqnarray}
so we have the following dependent invariants:
\begin{eqnarray}
\label{3.12}
2p_1\cdot Q_1 &=& s - a - b      \nonumber \\
2p_2\cdot Q_1 &=& M^2_W + a + b - c      \nonumber \\
2Q_1\cdot Q_2 &=& b + d - c = sx - M^2_W \nonumber \\
2Q_1\cdot k &=& s - M^2_W - b - d \nonumber \\
2p_2\cdot k &=& s - M^2_W + c - a - b - d = \frac{s}{2}(1-x)(1+y)\,. \nonumber
\\
\,
\end{eqnarray}
All invariants can now be expressed in terms of $s, x, y, \cos\theta'_1$ and
$\cos\theta'_2$ by solving
for all the necessary quantities in the primed reference frame:
\begin{eqnarray}
\label{3.13}
p'_{1,0} &=& \frac{\sqrt{s}}{4}\left(\frac{1+x+(1-x)y}{\sqrt{x}}\right)
\nonumber \\
p'_{2,0} &=& \frac{\sqrt{s}}{4}\left(\frac{1+x-(1-x)y}{\sqrt{x}}\right)
\nonumber \\
Q'_{1,0}&=&\frac{\sqrt{sx}}{2}\left(1 + \frac{M^2_W}{sx}\right) \nonumber \\
Q'_{2,0}&=&\left|\vec{Q}'_1\right|=\frac{\sqrt{sx}}{2}\left(1 -
\frac{M^2_W}{sx}\right) = \frac{\sqrt{sx}}{2}\beta(sx) \nonumber \\
\cos\eta'&=&1-\left(\frac{8x}{(1+x)^2 - (1-x)^2y^2}\right) \nonumber \\
\cos\psi'&=&\frac{1-x+y(1+x)}{1+x+y(1-x)} \nonumber \\
k'_0&=& \frac{\sqrt{s}}{2}\left(\frac{1-x}{\sqrt{x}}\right)\,. \nonumber \\
\,
\end{eqnarray}

Looking at  (\ref{3.9}) it is clear that the soft divergences will be present
in those pieces of squared
matrix element that contain a factor $\left(1-x\right)^{-2}$, while the
collinear divergences will be due
to factors of $\left(1-y\right)^{-1}$ or $\left(1+y\right)^{-1}$ in the squared
matrix element.
The squared matrix element for the $q\bar q$ channel can be written as:
\begin{eqnarray}
\label{3.14}
\sum\left|M^{q\bar q\rightarrow W\gamma g}\right|^2 = \sum\left|M^{q\bar
q}_{Ia}\right|^2 + \sum\left|M^{q\bar q}_{Ib}\right|^2 + \left(\sum M^{q\bar
q}_{Ia}M^{q\bar q \ast}_{Ib} + c.c. \right) + {\rm remain}\,. \nonumber \\
\,
\end{eqnarray}
The type $Ia$ matrix element (see Fig.1) can be written in the following way:
\begin{eqnarray}
\label{3.15}
\lefteqn{M_{Ia}\left[q(p_1,l_1,\lambda_1)\bar q(p_2,l_2) \rightarrow
W(Q_1)\gamma(Q_2)g(k,c,\lambda)\right] =}\nonumber \\
& &g_S T^c_{l'_1l_1}\frac{\left[\left(2p_1^{\rho} -
\sla{k}\gamma^{\rho}\right)u_{\lambda_1}(p_1)\right]_{\alpha}}{2p_1\cdot k}
\epsilon^{\lambda}_{\rho}(k)\ M\left[q\left(p_1-k,l'_1,\alpha\right)\bar
q(p_2,l_2) \rightarrow W(Q_1)\gamma(Q_2)\right]\nonumber \\
\,
\end{eqnarray}
where $l_1,l_2$ and $l'_1$ are quark color indices, $\lambda_1$ is the quark
polarization index,
$c$ and $\lambda$ are the outgoing gluon color and
polarization indices, $\alpha$ and $\rho$ are Lorentz indices and $g_S$ is the
renormalized
strong coupling constant. Other indices have been omitted because they are
not necessary in what follows.
The partial squared matrix element is given by:
\begin{eqnarray}
\label{3.16}
\sum\left|M^{q\bar q}_{Ia}\right|^2 &=& \frac{g^2_S}{(2p_1\cdot k)^2}C_F
R_{q\bar q,Ia}^{\alpha\alpha'}\sum M\left[q(p_1-k,\alpha)\bar q(p_2)
\rightarrow W(Q_1)\gamma(Q_2)\right]\nonumber \\
& &\hspace{1.2in}\times M^{\ast}\left[q\left(p_1-k,\alpha'\right)\bar q(p_2)
\rightarrow W(Q_1)\gamma(Q_2)\right]\,. \nonumber \\
\,
\end{eqnarray}
The $Ib$ partial squared matrix element may be written in a similar fashion:
\begin{eqnarray}
\label{3.16.b}
\sum\left|M^{q\bar q}_{Ib}\right|^2 &=& \frac{g^2_S}{(2p_2\cdot k)^2}C_F
R_{q\bar q,Ib}^{\beta\beta'}\sum M\left[q(p_1)\bar q(p_2-k,\beta) \rightarrow
W(Q_1)\gamma(Q_2)\right]\nonumber \\
& &\hspace{1.2in}\times M^{\ast}\left[q\left(p_1\right)\bar q(p_2-k,\beta')
\rightarrow W(Q_1)\gamma(Q_2)\right]\,. \nonumber \\
\,
\end{eqnarray}

The repeated indices on the RHS of  (\ref{3.16}) and (\ref{3.16.b}) are
contracted and the sum is over quark(antiquark) colors and polarizations as
well as
$W$ and $\gamma$ polarizations. In the $q\bar q$ center of mass frame the
tensors
$R_{q\bar q,Ia}^{\alpha\alpha'}$ and $R_{q\bar q,Ib}^{\beta\beta'}$ can be
written in the following way:
\begin{eqnarray}
\label{3.17}
R_{q\bar q,Ia}^{\alpha\alpha'} &=& -4p_1\cdot k \left[ \left\{ \left(
\frac{2-n}{2} + \frac{1+y}{2(1-x)} \right) \sla{k} + \left(1-\frac{1+y}{1-x}
\right) \sla{p}_1 - \frac{1-y}{2(1-x)}\sla{\bar{k}} \right\} \gamma_0 \
\right]^{\alpha\alpha'} \nonumber \\
R_{q\bar q,Ib}^{\beta\beta'} &=& -4p_2\cdot k \left[ \gamma_0\left\{ \left(
\frac{2-n}{2} + \frac{1-y}{2(1-x)} \right) \sla{k} + \left(1-\frac{1-y}{1-x}
\right) \sla{p}_2 - \frac{1+y}{2(1-x)}\sla{\bar{k}} \right\}
\right]^{\beta'\beta}\,. \nonumber \\
\,
\end{eqnarray}
In deriving  (\ref{3.17}) we have summed over physical gluon polarizations in
the covariant gauge where we may write:
\begin{eqnarray}
\label{3.18}
P_{\rho\rho'}(k)\equiv \sum_{\lambda =
1}^{n-2}\epsilon^{\lambda}_{\rho}(k)\epsilon^{\lambda}_{\rho'}(k)=-g_{\rho
\rho'} +
\frac{k_{\rho}\overline{k}_{\rho'} + k_{\rho'}\overline{k}_{\rho}}{k\cdot
\overline{k}} \,
\end{eqnarray}
where, if $k=\left(k_0,\vec k\right)$ then $\bar k\equiv\left(k_0,-\vec
k\right)$.
The factors $p_1\cdot k$ and $p_2\cdot k$ in front of the RHS of
 (\ref{3.17}) will cancel similar factors in the denominators of the RHS of
(\ref{3.16}) and (\ref{3.16.b}) respectively, leaving the type $Ia$ and $Ib$
squared matrix elements with singular terms proportional to
$(1-x)^{-2}(1-y)^{-1}$, $(1-x)^{-1}(1-y)^{-1}$
and $(1-x)^{-2}(1+y)^{-1}$ and $(1-x)^{-1}(1+y)^{-1}$ respectively, i.e. both
terms will contribute to the soft and collinear singularities.
In a similar way the ${\rm remain}$ in  (\ref{3.14}) can be shown to have no
collinear or soft singularities while the interference term
$M_{Ia}M^{\ast}_{Ib}$ has only a soft singularity. It is thus convenient to
define the non-singular function $F^{q\bar q}$:
\begin{eqnarray}
\label{3.19}
F^{q\bar q}\left(s,x,y,\cos\theta'_1,\theta'_2\right) &\equiv& 4(p_1\cdot
k)(p_2\cdot k)\sum\left|M^{q\bar q\rightarrow W\gamma g}\right|^2 \,
\end{eqnarray}
so that we can now rewrite the $2$ to $3$ body total partonic cross section in
(\ref{3.9}) as follows:
\begin{eqnarray}
\label{3.20}
\sigma^P\left[q(p_1)\bar q(p_2)\rightarrow W\gamma g\right] &=&
\frac{1}{4N^2_c}\frac{1}{2s}\left(4\pi\right)^{\epsilon-2}\frac{\Gamma(1-\epsilon)}{\Gamma(1-2\epsilon)}\frac{s^{-1-\epsilon}}{2\pi}\nonumber \\
& &\hspace{-0.5in}\times\int_{\rho(s)}^{1}dx\
\Phi(sx)(1-x)^{-1-2\epsilon}\int_{-1}^{1}dy\left(1-y^2\right)^{-1-\epsilon}\int_{-1}^{1}d\cos\theta'_1\ \sin^{-2\epsilon}\theta'_1\nonumber \\
& &\hspace{-0.5in}\times\int_{0}^{\pi}d\theta'_2\ \sin^{-2\epsilon}\theta'_2\
F^{q\bar q}\left(s,x,y,\cos\theta'_1,\theta'_2\right)\,. \nonumber \\
\,
\end{eqnarray}
Now that the singular factors in $x$ and $y$ have been isolated we can rewrite
them as distributions for \( \epsilon < 0 \):
\begin{eqnarray}
\label{3.21}
(1-x)^{-1-2\epsilon} &\sim& -
\frac{1}{2\epsilon}\left(1-x_0\right)^{-2\epsilon}\delta(1-x) +
\plusdist{1}{1-x}{x_0} - 2\epsilon\plusdist{\ln(1-x)}{1-x}{x_0}  +
O\left(\epsilon^2\right)\nonumber \\
\left(1-y^2\right)^{-1-\epsilon} &\sim& -
\frac{(2y_0)^{-\epsilon}}{2\epsilon}\left[ \delta(1-y) + \delta(1+y) \right] +
\frac{1}{2}\plusdist{1}{1-y}{y_0} +\frac{1}{2}\plusdist{1}{1+y}{y_0} +
O(\epsilon) \nonumber \\
\,
\end{eqnarray}
with $\zeta(2)=\pi^2/6$.
We have introduced the following definitions:
\begin{eqnarray}
\label{3.22}
\int_{\rho(s)}^{1}dx f(x)\plusdist{1}{1-x}{x_0} &\equiv& \int_{\rho(s)}^{x_0}dx
\frac{f(x)}{1-x}\ + \int_{x_0}^{1}dx \frac{f(x)-f(1)}{1-x}\nonumber \\
\int_{-1}^{1}dy f(y)\plusdist{1}{1-y}{y_0} &\equiv& \int_{-1}^{1-y_0}dy
\frac{f(y)}{1-y}\ + \int_{1-y_0}^{1}dy \frac{f(y)-f(1)}{1-y}\nonumber \\
\int_{-1}^{1}dy f(y)\plusdist{1}{1+y}{y_0} &\equiv& \int_{-1+y_0}^{1}dy
\frac{f(y)}{1+y}\ + \int_{-1}^{-1+y_0}dy \frac{f(y)-f(-1)}{1+y}\,. \nonumber \\
\,
\end{eqnarray}
The parameters $x_0$ and $y_0$ are arbitrary as long as they satisfy the
conditions $\rho(s) \leq x_0 < 1$
and $0 < y_0 \leq 2$. The symbol $\sim$ in  (\ref{3.21}) means that the
equality only holds under an integration over $x$ ranging from $\rho(s)$ to $1$
for the first expression in  (\ref{3.21}) and under an integration over $y$
ranging from $-1$ to $1$ for the second expression.
When $x$ and $y$ are not integrated over their whole range care has to be taken
when defining $x_0$ and $y_0$ so as not to introduce
unphysical dependences into the quantities we want to compute. We will discuss
this in more detail in section IV.

Using  (\ref{3.21}) and (\ref{3.22}) in  (\ref{3.20}) we write the
$O(\alpha_S)$ $2$ to $3$ body cross section as follows:
\begin{eqnarray}
\label{3.23}
\sigma^{P(1)}\left[q(p_1)\bar q(p_2)\rightarrow W\gamma g\right] &=&
\sigma^P_{q\bar q\ (finite)} + \sigma^P_{q\bar q\ (col+)} + \sigma^P_{q\bar q\
(col-)} + \sigma^P_{q\bar q\ (soft)} + O(\epsilon)\nonumber \\
\,
\end{eqnarray}
with
\begin{eqnarray}
\label{3.24}
\sigma^P_{q\bar q\ (finite)} &=&
\frac{1}{4N^2_c}\frac{1}{2s}2^{-10}\pi^{-4}s^{-1}\int_{\rho(s)}^{1}dx\
\beta(sx)\plusdist{1}{1-x}{x_0}\nonumber \\
& &\times\int_{-1}^{1}dy\left[\plusdist{1}{1-y}{y_0} +
\plusdist{1}{1+y}{y_0}\right]\nonumber \\
& &\times\int_{-1}^{1}d\cos\theta'_1\int_{0}^{\pi}d\theta'_2\ F^{q\bar
q}\left(s,x,y,\cos\theta'_1,\theta'_2\right) \nonumber \\
\sigma^P_{q\bar q\ (col\pm)} &=&
-\frac{1}{4N^2_c}\frac{1}{2s}(4\pi)^{\epsilon-2}\frac{1}{\Gamma(1-\epsilon)}\frac{s^{-1-\epsilon}}{2\pi}\frac{\pi}{2\epsilon}\left(\frac{2}{y_0}\right)^{\epsilon}\nonumber \\
& &\times\int_{\rho(s)}^{1}dx\ \Phi(sx)\left[\plusdist{1}{1-x}{x_0} -
2\epsilon\plusdist{\ln(1-x)}{1-x}{x_0}\right]\nonumber \\
& &\times\int_{-1}^{1}d\cos\theta'_1\ \sin^{-2\epsilon}\theta'_1\ F^{q\bar q\
(col\pm)}\left(s,x,\cos\theta'_1\right) \nonumber \\
\sigma^P_{q\bar q\ (soft)} &=&
\frac{1}{4N^2_c}\frac{1}{2s}\left(4\pi\right)^{\epsilon-2}\frac{\Gamma(1-\epsilon)}{\Gamma(1-2\epsilon)}\frac{s^{-1-\epsilon}}{2\pi}\frac{\pi}{2\epsilon^2}\left(1-x_0\right)^{-2\epsilon}\Phi(s)\nonumber \\
& &\times\int_{-1}^{1}d\cos\theta'_1\ \sin^{-2\epsilon}\theta'_1\
\ F^{q\bar q\ (soft)}\left(s,\cos\theta'_1\right) \nonumber \\
\,
\end{eqnarray}
where
\begin{eqnarray}
\label{3.25}
F^{q\bar q\ (col\pm)}\left(s,x,\cos\theta'_1\right) &\equiv& F^{q\bar
q}\left(s,x,y=\pm 1,\cos\theta'_1,\cos\theta'_2\right)\nonumber \\
F^{q\bar q\ (soft)}\left(s,\cos\theta'_1\right) &\equiv& F^{q\bar
q}\left(s,x=1,y,\cos\theta'_1,\cos\theta'_2\right)\,.
\end{eqnarray}
To compute the quantities in  (\ref{3.25}) we first note that the following
relations hold for $y=1$:
\begin{eqnarray}
\label{3.26}
p_1-k &=&  xp_1 \nonumber \\
\sla{k} &=&  (1-x)\sla{p}_1\,.
\end{eqnarray}
Using these relations in  (\ref{3.16})-(\ref{3.17}) and noting that in the
limit $y\rightarrow 1$ ($-1$) only
$\left|M^{q\bar q}_{Ia}\right|^2$ ($\left|M^{q\bar q}_{Ib}\right|^2$)
 contributes to $F^{q\bar q}$ in  (\ref{3.19}) we obtain the covariant
expression:
\begin{eqnarray}
\label{3.27}
F^{q\bar q\ (col+)}\left(s,x,\cos\theta'_1\right) &=& 8sg^2_SC_F\frac{ \left( 1
+ x^2 -\epsilon(1-x)^2 \right) }{x}\sum\left|M^{q\bar q\rightarrow
W\gamma}(xs,xb^{+})\right|^2\,. \nonumber \\
\,
\end{eqnarray}
We implicitly include in (\ref{3.27}) an overall factor of
$C(Q_1,Q_2,(1-x)p_1)$.
An analogous computation yields:
\begin{eqnarray}
\label{3.28}
F^{q\bar q\ (col-)}\left(s,x,\cos\theta'_1\right) &=& 8sg^2_SC_F\frac{ \left( 1
+ x^2 -\epsilon(1-x)^2 \right) }{x}\sum\left|M^{q\bar q \rightarrow
W\gamma}(xs,b^{-})\right|^2\,. \nonumber \\
\,
\end{eqnarray}
In the latter we implicitly include an overall factor of $C(Q_1,Q_2,(1-x)p_2)$.
The collinear limits of $F^{q\bar q}$ have thus been reduced to $2$ to $2$ body
squared matrix elements. We checked that the same expressions for the collinear
limits of $F^{q\bar q}$ are obtained when the sum of gluon polarizations is
taken in the axial gauge.

Now we obtain the soft residues of each of the contributing terms in the
squared matrix element:
\begin{eqnarray}
\label{3.30b}
\lim_{x\rightarrow 1} 4(p_1\cdot k)(p_2\cdot k)\sum\left|M^{q\bar
q}_{Ia}\right|^2 &=&4sg^2_SC_F(1+y)^2\sum\left|M^{q\bar q \rightarrow
W\gamma}(s,b^{soft})\right|^2 \nonumber \\
\lim_{x\rightarrow 1} 4(p_1\cdot k)(p_2\cdot k)\sum\left|M^{q\bar
q}_{Ib}\right|^2 &=&4sg^2_SC_F(1-y)^2\sum\left|M^{q\bar q \rightarrow
W\gamma}(s,b^{soft})\right|^2 \nonumber \\
\lim_{x\rightarrow 1} 4(p_1\cdot k)(p_2\cdot k)\left( \sum M^{q\bar
q}_{Ia}M^{\star q\bar q}_{Ib} + c.c.
\right)&=&8sg^2_SC_F(1-y^2)\sum\left|M^{q\bar q \rightarrow
W\gamma}(s,b^{soft})\right|^2\,. \nonumber \\
\,
\end{eqnarray}
After summing the above three contributions we obtain:
\begin{eqnarray}
\label{3.29}
F^{q\bar q\ (soft)}\left(s,\cos\theta'_1\right) &=&
16sg^2_SC_F\sum\left|M^{q\bar q \rightarrow W\gamma}(s,b^{soft})\right|^2\,.
\end{eqnarray}
This contribution contains an implicit factor of $C(Q_1,Q_2,0)$. Note that the
dependence on $y$ cancels after all terms in (\ref{3.30b}) are added together.
We have checked that the same result is obtained if the gluon polarization sum
is taken in the axial gauge. In  (\ref{3.27}), (\ref{3.28}) and (\ref{3.29}) we
define:
\begin{eqnarray}
\label{3.30}
b^{+}&\equiv& b(s,x,y=1,\cos\theta'_1) =
\frac{s}{2}\beta(sx)\left(1+\cos\theta'_1\right) \nonumber \\
b^{-}&\equiv& b(s,x,y=-1,\cos\theta'_1)=
\frac{sx}{2}\beta(sx)\left(1+\cos\theta'_1\right) \nonumber \\
b^{soft}&\equiv& b(s,x=1,y,\cos\theta'_1)=
\frac{s}{2}\beta(s)\left(1+\cos\theta'_1\right)\,.
\end{eqnarray}
Note that in the soft limit the variable $\cos\theta'_1$ is equivalent to
$\cos\theta_1$ of a $2$ to $2$ body kinematics.

Noting that the squared matrix elements in  (\ref{3.27})-(\ref{3.29}) are of
the $2$ to $2$ body type we can now rewrite the soft and collinear terms in a
more convenient way:
\begin{eqnarray}
\label{3.31}
\sigma^P_{q\bar q\ (soft)} &=&\frac{\alpha_S}{\pi}C_F\left(-2V -
\frac{1}{\overline{\epsilon}}\left[\frac{3}{2} + 2\ln(1-x_0)\right] +
\frac{3}{2}\ln\frac{s}{\mu^2} + 2\ln^2(1-x_0)\right.\nonumber \\
& &\left.+ 2\ln(1-x_0)\ln\frac{s}{\mu^2} + 2\zeta(2) -
4\right)\sigma^{(0)}\left[q(p_1)\bar q(p_2)\rightarrow W\gamma\right] \nonumber
\\
\sigma^P_{q\bar q\ (col+)}&=&
-\frac{\alpha_S}{2\pi\overline{\epsilon}}\int_{\rho(s)}^{1}dx\left[C_F\left(1+x^2\right)\plusdist{1}{1-x}{x_0} + \epsilon C_F\left(\ln\left(\frac{2\mu^2}{sy_0}\right)\left(1+x^2\right)\right.\right.\nonumber \\
& &\hspace{.9in}\times\left.\left.\plusdist{1}{1-x}{x_0} -
2\left(1+x^2\right)\plusdist{\ln(1-x)}{1-x}{x_0} + x - 1
\right)\right]\nonumber \\
& &\hspace{.9in}\times\sigma^{(0)}\left[q(xp_1)\bar q(p_2)\rightarrow
W\gamma\right] \nonumber \\
\sigma^P_{q\bar q\ (col-)}&=&
-\frac{\alpha_S}{2\pi\overline{\epsilon}}\int_{\rho(s)}^{1}dx\left[C_F\left(1+x^2\right)\plusdist{1}{1-x}{x_0} + \epsilon C_F\left(\ln\left(\frac{2\mu^2}{sy_0}\right)\left(1+x^2\right)\right.\right.\nonumber \\
& &\hspace{.9in}\times\left.\left.\plusdist{1}{1-x}{x_0} -
2\left(1+x^2\right)\plusdist{\ln(1-x)}{1-x}{x_0} + x - 1
\right)\right]\nonumber \\
& &\hspace{.9in}\times\sigma^{(0)}\left[q(p_1)\bar q(xp_2)\rightarrow
W\gamma\right]\,. \nonumber \\
\,
\end{eqnarray}
In the previous formulae we neglected terms of $O(\epsilon)$ and we used
\begin{eqnarray}
\label{3.32}
V\equiv -e^{-\left(\gamma_E -
\ln(4\pi)\right)\epsilon}\left(\frac{1}{2\epsilon^2} -
\frac{1}{2\epsilon}\left(\ln\frac{s}{\mu^2} - \frac{3}{2}\right) +
\frac{1}{4}\ln^2\frac{s}{\mu^2} - \frac{3}{4}\ln\frac{s}{\mu^2} -
\frac{7}{4}\zeta(2) + 2 \right)\,. \nonumber \\
\,
\end{eqnarray}
We also made the replacement $g^2_S = 4\pi\alpha_S\mu^{2\epsilon}$.

The $O(\alpha_S)$ corrections to the $2$ to $2$ body partonic cross section for
$q\bar q\rightarrow W\gamma$ were computed in
\cite{STN} and they are given by:
\begin{eqnarray}
\label{3.33}
\sigma^{P(1)}\left[q(p_1)\bar q(p_2)\rightarrow W\gamma\right]
&=&\frac{\alpha_S}{\pi}C_F \sigma^{(0)}\left[q(p_1)\bar q(p_2)\rightarrow
W\gamma\right] 2V\nonumber \\
& & +
\frac{1}{4N^2_c}\frac{1}{2s}\beta(s)C^2_W\frac{\alpha}{9}\frac{\alpha_S}{\pi}N_cC_F\nonumber \\
& &\times\int_{-1}^{1}d\cos\theta_1 \frac{\left( 2\left(s-M^2_W-b\right) -
b\right )\left( 2F_1(s,b) - F_2(s,b)\right)}{s-M^2_W}\nonumber \\
& &\times C(Q_1,Q_2,0)\nonumber \\
\,
\end{eqnarray}
where
\begin{eqnarray}
\label{3.34}
F_1(s,b) &\equiv& F(-b,-(s-M^2_W-b),s,M^2_W) \nonumber \\
F_2(s,b) &\equiv& F(-(s-M^2_W-b),-b,s,M^2_W) \nonumber \\
C_W &\equiv& M_W\sqrt{\frac{G_F}{\sqrt{2}}} \,
\end{eqnarray}
and $G_F$ is the Fermi coupling constant, $b$ was defined in  (\ref{3.6}) and
$F(t_1,t_2,s,M^2_W)$ was defined in  (\ref{3.7}) of \cite{STN}.

Using  (\ref{3.23}), (\ref{3.24}), (\ref{3.31}) and (\ref{3.33}) in the RHS of
(\ref{2.18}) we can write the $O(\alpha_S)$ hard scattering cross section on
the LHS of (\ref{2.18}) after performing an integration over $Q_1$ and $Q_2$:
\begin{eqnarray}
\label{3.35}
\sigma^{(1)}\left[ q(p_1)\bar q(p_2) \rightarrow
W\gamma\right]+\sigma^{(1)}\left[ q(p_1)\bar q(p_2) \rightarrow W\gamma
g\right]&=&\nonumber \\
& &\hspace{-3.8in}\sigma^P_{q\bar q\ (finite)} + \sigma^P_{q\bar q\ (soft)} +
\sigma^{P(1)}\left[q(p_1)\bar q(p_2)\rightarrow W\gamma\right]\nonumber \\
& &\hspace{-3.8in} + \sigma^P_{q\bar q\ (col+)} +
\frac{\alpha_S}{2\pi\overline{\epsilon}}\int_{0}^{1}dv\
\overline{P}_{qq}(v)\sigma^{(0)}\left[ q(vp_1)\bar q(p_2) \rightarrow
W\gamma\right]\nonumber \\
& &\hspace{-3.8in} + \sigma^P_{q\bar q\ (col-)} +
\frac{\alpha_S}{2\pi\overline{\epsilon}}\int_{0}^{1}dv\ \overline{P}_{\bar q
\bar q}(v)\sigma^{(0)}\left[ q(p_1)\bar q(vp_2) \rightarrow
W\gamma\right]\nonumber \\
& &\hspace{-3.8in} = \sigma^P_{q\bar q\ (finite)} + \sigma^P_{q\bar q\ (SV)}
\nonumber \\
& &\hspace{-3.8in} -
\frac{\alpha_S}{2\pi}C_F\int_{\rho(s)}^{1}dx\left[\left(1+x^2\right)\ln\left(\frac{2\mu^2}{sy_0}\right)\plusdist{1}{1-x}{x_0} - 2\left(1+x^2\right)\plusdist{\ln(1-x)}{1-x}{x_0} + x - 1\right]\nonumber \\
& &\hspace{-3.8in}\times\left(\sigma^{(0)}\left[ q(xp_1)\bar q(p_2) \rightarrow
W\gamma\right] + \sigma^{(0)}\left[ q(p_1)\bar q(xp_2) \rightarrow
W\gamma\right]\right)\nonumber \\
\,
\end{eqnarray}
where we have defined the soft-plus-virtual contributions:
\begin{eqnarray}
\label{3.35a}
\sigma^P_{q\bar q\ (SV)} &\equiv&\sigma^P_{q\bar q\ (soft)} +
\sigma^{P(1)}\left[q(p_1)\bar q(p_2)\rightarrow W\gamma\right] =\nonumber \\
& &\frac{\alpha_S}{\pi}C_F\left(\frac{3}{2}\ln\frac{s}{\mu^2} + 2\ln^2(1-x_0) +
2\ln(1-x_0)\ln\frac{s}{\mu^2} + 2\zeta(2) - 4\right)\nonumber \\
& &\times\sigma^{(0)}\left[q(p_1)\bar q(p_2)\rightarrow W\gamma\right]\ +\
\frac{1}{4N^2_c}\frac{1}{2s}\beta(s)C^2_W\frac{\alpha}{9}\frac{\alpha_S}{\pi}N_cC_F\nonumber \\
& &\times\int_{-1}^{1}d\cos\theta_1 \frac{\left( 2\left(s-M^2_W-b\right) - b
\right)\left( 2F_1(s,b) - F_2(s,b) \right)}{s-M^2_W} C(Q_1,Q_2,0)\nonumber \\
\,
\end{eqnarray}

Summarizing all the contributions in the incoming $q\bar q$ partonic channel,
we have
\begin{eqnarray}
\label{3.35b}
\int DQ_1DQ_2\sum_{X} \left( \frac{D^2\sigma^H}{DQ_1DQ_2} \right)^{q \bar
q}\left[ p(P_1)\ \bar p(P_2) \rightarrow W(Q_1)\ \gamma(Q_2) \ X \right]& =&
\nonumber \\
& &\hspace{-4.4in}\int_{0}^{1}d\tau_1\int_{0}^{1}d\tau_2 \bigg\{
f_{qp}(\tau_1)f_{\bar q\bar p}(\tau_2) \bigg( \sigma^{(0)}\left[ q(p_1)\bar
q(p_2) \rightarrow W\gamma\right] \nonumber \\
& &\hspace{-3.4in} \left.\left. +\ \sigma^{(1)}\left[ q(p_1)\bar q(p_2)
\rightarrow W\gamma\right] + \sigma^{(1)}\left[ q(p_1)\bar q(p_2) \rightarrow
W\gamma g\right] \right.\right.\nonumber \\
& &\hspace{-3.4in}\left.\left. + \int_{0}^{1} d\tau f_{\gamma g}(\tau)\int
DQ_1Dq_2\frac{D^2\sigma^{(1)}}{DQ_1Dq_2}\left[ q(p_1)\bar q(p_2) \rightarrow
W(Q_1)g(q_2)\right] \right)\right. \nonumber \\
& &\hspace{-3.6in} + (q \leftrightarrow \bar q) \bigg\} \nonumber \\
\,
\end{eqnarray}
with $p_1=\tau_1P_1$, $p_2=\tau_2P_2$ and $q_2=Q_2/\tau$.
Since there are no singularities left all the necessary squared matrix elements
in (\ref{3.35b})
can now be safely evaluated in $n=4$ dimensions. On the RHS of (\ref{3.35b}) we
still need the following squared matrix elements:
\begin{eqnarray}
\label{3.36b}
\sum\left|M^{q\bar q\rightarrow W\gamma}(s,b)\right|^{2(0)} &=&
\frac{2^7}{9}\pi N_cC^2_W\alpha\left( 2(s-M^2_W-b)-b \right)^2\nonumber \\
& &\times\frac{ sM^2_W - b(s-M^2_W-b) + \frac{1}{2}(s-M^2_W)^2 }{
b(s-M^2_W-b)(s-M^2_W)^2 } \nonumber \\
\sum\left|M^{q\bar q\rightarrow Wg}(s,\hat{b})\right|^{2(1)} &=& 2^6\pi
N_cC_FC^2_W\alpha_S\left( \frac{ \hat{b}^2 + (s-M^2_W-\hat{b})^2 + 2sM^2_W }{
\hat{b}(s-M^2_W-\hat{b}) } \right)\nonumber \\
\,
\end{eqnarray}
where $\hat{b}\equiv b/\tau$. The integrand of the last term in (\ref{3.35b})
carries an implicit factor of $C(Q_1,\tau q_2,(1-\tau)q_2)$. In (\ref{3.35a})
and (\ref{3.36b}) the invariant $b$ is evaluated in the unprimed frame
as defined in (\ref{3.6}).

The expression for $\sum\left|M^{q\bar q\rightarrow W\gamma g}\right|^2$ in
$n=4$ needed when evaluating $\sigma^{P}_{q\bar q\ (finite)}$ is too long to be
presented here, but it may be obtained upon request. We note here that
$\gamma_5$ is
never needed in $n\neq 4$ dimensions: we obtained the cancellation of
singularities before fully
computing any squared matrix element where we had to explicitly evaluate
$\gamma_5$ and whatever remained after this cancellation could be safely
computed in $n=4$ dimensions.

Since we have integral expressions of all
quantities on the RHS of  (\ref{3.35b}) in terms of the variables $x, y,
\theta'_1$ and $\theta'_2$
which define all the independent invariants of the system we can compute these
integrals using numerical Montecarlo techniques
and histogram any physical variable of interest that can be expressed in terms
of these invariants.
We will reexamine these issues in more detail in sections IV and V.

\subsection{\sc The $qg$ channel.}

In this channel the singularities are not both of initial state (type $I$), as
in
the case of $q\bar q$ channel, but we have now one piece in the initial state
(type $I$) and another in the final state (type $II$),
as shown in Fig.2. In this case it is thus more convenient to integrate each
term separately and write for the total partonic cross section in this channel:
\def\sigtilde{{\tilde\sigma}}
\begin{eqnarray}
\label{3.38}
\sigma^P\left[q(p_1)g(p_2)\rightarrow W\gamma q\right] &=&
\sigma^{P,I}\left[q(p_1)g(p_2)\rightarrow W\gamma q\right] +
\sigma^{P,II}\left[q(p_1)g(p_2)\rightarrow W\gamma q\right]\nonumber \\
&+& \sigtilde^{P}\left[q(p_1)g(p_2)\rightarrow W\gamma q\right] \nonumber \\
\,
\end{eqnarray}
where the first two terms on the RHS of  (\ref{3.38}) contain only the partial
squared matrix elements $\sum\left|M^{qg}_{Ib}\right|^2$
and $\sum\left|M^{qg}_{II}\right|^2$ respectively, while the third term
contains $\sum\left|M^{qg}_{III}\right|^2$ and all interference
terms of the squared matrix element. If the gluon and the photon are summed
over physical polarizations only then the third term on the RHS of (\ref{3.38})
is free of singularities while the first two terms contain only collinear
singularities.

If we define the same kinematics in $\sigma^{P,I}_{qg\rightarrow W\gamma q}$ as
we did with the $q\bar q$ channel, the collinear pieces of integrand contain a
factor $(1+y)^{-1}$. To isolate the singularity in this term it is therefore
enough to define the non-singular function:
\begin{eqnarray}
\label{3.39}
F^{qg}_I\left(s,x,y,\cos\theta'_1,\theta'_2\right) &\equiv& -2p_2\cdot
k\sum\left|M^{qg}_{Ib}\right|^2 \,
\end{eqnarray}
so that the first term on the RHS of  (\ref{3.38}) may be written:
\begin{eqnarray}
\label{3.40}
\sigma^{P,I}\left[q(p_1)g(p_2)\rightarrow W\gamma q\right] &=&\nonumber \\
& &\hspace{-1.7in}
-\frac{1}{8N^2_cC_F(1-\epsilon)}\frac{1}{2s}\left(4\pi\right)^{\epsilon-2}\frac{\Gamma(1-\epsilon)}{\Gamma(1-2\epsilon)}\frac{s^{-\epsilon}}{\pi}\int_{\rho(s)}^{1}dx\ \Phi(sx)(1-x)^{-2\epsilon}\nonumber \\
&
&\hspace{-1.7in}\times\int_{-1}^{1}dy(1-y)^{-\epsilon}(1+y)^{-1-\epsilon}\int_{-1}^{1}d\cos\theta'_1\ \sin^{-2\epsilon}\theta'_1\int_{0}^{\pi}d\theta'_2\ \sin^{-2\epsilon}\theta'_2\nonumber \\
& &\hspace{-1.7in}\times F^{qg}_I\left(s,x,y,\cos\theta'_1,\theta'_2\right)\,.
\nonumber \\
\,
\end{eqnarray}
We now rewrite the factor $(1+y)^{-1-\epsilon}$ as a distribution:
\begin{eqnarray}
\label{3.41}
(1+y)^{-1-\epsilon} &\sim& -\ \frac{y_0^{-\epsilon}}{\epsilon}\delta(1+y)\ +
\plusdist{1}{1+y}{y_0} +\ O(\epsilon)\,.
\end{eqnarray}
Using  (\ref{3.41}) in  (\ref{3.40}) we can write the corresponding
contribution to the $O(\alpha_S)$ 2 to 3 body partonic cross section as
follows:
\begin{eqnarray}
\label{3.42}
\sigma^{P,I(1)}\left[q(p_1)g(p_2)\rightarrow W\gamma q\right]
=\sigma^{P,I}_{qg,finite} + \sigma^{P,I}_{qg,col-} + O(\epsilon)\,,
\end{eqnarray}
where
\begin{eqnarray}
\label{3.43}
\sigma^{P,I}_{qg,finite} &=&
-\frac{1}{8N^2_cC_F}\frac{1}{2s}\left(4\pi\right)^{-4}\int_{\rho(s)}^{1}dx\
\beta(sx)\int_{-1}^{1}dy\plusdist{1}{1+y}{y_0}\int_{-1}^{1}d\cos\theta'_1\nonumber \\
& &\times\int_{0}^{\pi}d\theta'_2\
F^{qg}_I\left(s,x,y,\cos\theta'_1,\theta'_2\right) \nonumber \\
\sigma^{P,I}_{qg,col-}
&=&\frac{1}{8N^2_cC_F(1-\epsilon)}\frac{1}{2s}(4\pi)^{\epsilon-2}\frac{1}{\Gamma(1-\epsilon)}\frac{s^{-\epsilon}}{\pi}\frac{\pi}{\epsilon}\left(\frac{2}{y_0}\right)^{\epsilon}\int_{\rho(s)}^{1}dx\ \Phi(sx)\nonumber \\
& &\times (1-x)^{-2\epsilon}\int_{-1}^{1}d\cos\theta'_1\
\sin^{-2\epsilon}\theta'_1\ F^{qg\ (col-)}_I\left(s,x,\cos\theta'_1\right)
\nonumber \\
\,
\end{eqnarray}
and
\begin{eqnarray}
\label{3.44}
F^{qg\ (col-)}_I\left(s,x,\cos\theta'_1\right) =
F^{qg}_I\left(s,x,y=-1,\cos\theta'_1,\theta'_2\right)\,.
\end{eqnarray}
To compute the limit on the RHS of  (\ref{3.44}) we write, as we did in
(\ref{3.16}) and (\ref{3.16.b}):
\begin{eqnarray}
\label{3.45}
\sum\left|M^{qg}_{Ib}\right|^2 &=& \frac{g^2_S}{\left(2p_2\cdot k\right)^2}C_F
R_{qg,I}^{\alpha\alpha'}\sum M\left[q\left(p_1\right)\bar
q\left(p_2-k,\alpha\right) \rightarrow W(Q_1)\gamma(Q_2)\right]\nonumber \\
& &\hspace{1.2in}\times M^{\ast}\left[q\left(p_1\right)\bar
q\left(p_2-k,\alpha'\right) \rightarrow W(Q_1)\gamma(Q_2)\right]\,, \nonumber
\\
\,
\end{eqnarray}
where now in the $qg$ center of mass frame:
\begin{eqnarray}
\label{3.46}
R_{qg,I}^{\alpha\alpha'} &=& -4p_2\cdot k \left[\gamma_0 \left\{ \left(
\frac{2-n}{2} + \frac{(1-x)(1-y)}{2} \right) \sla{p}_2 - \left((1-x)(1-y)-1
\right) \sla{k} \right.\right.\nonumber \\
& &\hspace{0.8in} -(1-x)(1+y)\sla{p}_1 \bigg\} \bigg]^{\alpha'\alpha}\,.
\nonumber \\
\,
\end{eqnarray}
{}From the above equation it is now clear that there will be no soft quark
singularity coming from the type $Ia$ squared matrix element.
Using  (\ref{3.46}), (\ref{3.44}) and (\ref{3.39}) we obtain:
\begin{eqnarray}
\label{3.47}
F^{qg\ (col-)}_I\left(s,x,\cos\theta'_1\right) &=& 2g^2_SC_F\frac{ \left( 2x(1
- x) - 1 +\epsilon \right) }{x}\sum\left|M^{q\bar q \rightarrow
W\gamma}(xs,b^{-})\right|^2\,. \nonumber \\
\,
\end{eqnarray}
The latter expression contains an implicit factor of $C(Q_1,Q_2,(1-x)p_2)$ to
account for experimental cuts.
Using (\ref{3.47}) in (\ref{3.43}) we can rewrite the collinear contribution
(neglecting terms of $O(\epsilon)$):
\begin{eqnarray}
\label{3.48}
\sigma^{P,I}_{qg,col-}
&=&\frac{\alpha_S}{2\pi\overline{\epsilon}}\int_{\rho(s)}^{1}dx\left[x(1-x)-\frac{1}{2} + \epsilon\left\{\ln\left(\frac{2\mu^2}{sy_0}\right)\left(x(1-x)-\frac{1}{2}\right) + \ln(1-x)\right.\right.\nonumber \\
& &\hspace{0.8in}+x(1-x)\left[1-2\ln(1-x)\right]\ \bigg\}
\bigg]\sigma^{(0)}\left[q(p_1)\bar q(xp_2)\rightarrow W\gamma\right]\,.
\nonumber \\
\,
\end{eqnarray}

To treat the type $II$ term on the RHS of (\ref{3.38}) it is convenient to
rotate the reference frame so that in the $W\gamma$ center of mass frame we
have:
\begin{eqnarray}
\label{3.50}
p'_1 &=&p'_{1,0}\left(1,0,\ldots,0,\sin\psi',\cos\psi'\right) \nonumber \\
p'_2 &=&p'_{2,0}\left(1,0,\ldots,0,-\sin\chi',\cos\chi'\right) \nonumber \\
k' &=&k'_{0}\left(1,0,\ldots,0,0,1\right) \nonumber \\
Q'_1
&=&\left|\vec{Q}'_1\right|\left(\frac{Q'_{1,0}}{\left|\vec{Q}'_1\right|},\ldots,-\sin\phi'_1\sin\phi'_2\cos\phi'_3,
-\sin\phi'_1\cos\phi'_2,-\cos\phi'_1\right) \nonumber \\
Q'_2
&=&\left|\vec{Q}'_1\right|\left(1,\ldots,\sin\phi'_1\sin\phi'_2\cos\phi'_3,\sin\phi'_1\cos\phi'_2,\cos\phi'_1\right)\,. \nonumber \\
\,
\end{eqnarray}
We can thus write for the $2$ to $3$ body integral:
\begin{eqnarray}
\label{3.51}
\int DQ_1DQ_2\frac{D^2\sigma^{P,II}}{DQ_1DQ_2}\left[q(p_1)g(p_2)\rightarrow
W(Q_1)\gamma(Q_2)q\right] &=& \nonumber \\
&
&\hspace{-3.5in}\frac{1}{8N^2_cC_F(1-\epsilon)}\frac{1}{2s}\left(4\pi\right)^{\epsilon-2}\frac{\Gamma(1-\epsilon)}{\Gamma(1-2\epsilon)}\frac{s^{1-\epsilon}}{2\pi}\beta^{2-2\epsilon}(s)\int_{0}^{1}dz\ \Phi\left(s[1+\beta(s)(z-1)]\right)\nonumber \\
& &\hspace{-3.5in}\times
(1-z)^{1-2\epsilon}\int_{-1}^{1}dy\left(1-y^2\right)^{-\epsilon}\int_{-1}^{1}dv
\left(1-v^2\right)^{-\epsilon}\int_{0}^{\pi}d\phi'_2\
\sin^{-2\epsilon}\phi'_2\nonumber \\
& &\hspace{-3.5in}\times\sum\left|M^{qg}_{II}(s,a,b,c,d)\right|^2 \nonumber \\
\,
\end{eqnarray}
with
\begin{eqnarray}
\label{3.52}
\cos\chi' &=& \frac{1-x-y(1+x)}{1+x-y(1-x)}\nonumber \\
v &\equiv& \cos\phi'_1 \nonumber \\
z &\equiv& 1 + \frac{x-1}{\beta(s)}\,.
\end{eqnarray}
The explicit form of the invariants in the squared matrix element in terms of
the new integration angles is given as follows:
\begin{eqnarray}
\label{3.53}
b&\equiv& 2p_1\cdot Q_2 = 2p'_{1,0}\left|\vec{Q}'_1\right|\left(1 -
\sin\psi'\sin\phi'_1\cos\phi'_2 - \cos\psi'\cos\phi'_1\right)   \nonumber \\
c&\equiv& 2k\cdot Q_2 = 2k'_0\left|\vec{Q}'_1\right|\left( 1 - \cos\phi'_1
\right)\nonumber \\
d&\equiv& 2p_2\cdot Q_2 = 2p'_{2,0}\left|\vec{Q}'_1\right|\left(1 +
\sin\chi'\sin\phi'_1\cos\phi'_2 - \cos\chi'\cos\phi'_1\right)\,. \nonumber \\
\,
\end{eqnarray}
The rest of the invariants and variables remain as defined in
(\ref{3.11})-(\ref{3.13}).
To isolate the singularity in the type $II$ squared matrix element it is enough
to define:
\begin{eqnarray}
\label{3.54}
F^{qg}_{II}\left(s,z,y,v,\phi'_2\right) &\equiv& -2k\cdot
Q_2\sum\left|M^{qg}_{II}\right|^2 \,
\end{eqnarray}
so that now the second term on the RHS of (\ref{3.38}) may be written:
\begin{eqnarray}
\label{3.55}
\sigma^{P,II}\left[q(p_1)g(p_2)\rightarrow W\gamma q\right] &=&\nonumber \\
& &\hspace{-0.9in}
-\frac{1}{8N^2_cC_F(1-\epsilon)}\frac{1}{2s}\left(4\pi\right)^{\epsilon-2}\frac{\Gamma(1-\epsilon)}{\Gamma(1-2\epsilon)}\frac{s^{-\epsilon}}{\pi}\frac{2^{2\epsilon}}{\Gamma(1-\epsilon)}\frac{1}{16\pi}\beta^{1-4\epsilon}(s)\left(\frac{4\pi}{s}\right)^{\epsilon}\nonumber \\
&
&\hspace{-0.9in}\times\int_{0}^{1}dz[z(1-z)]^{-2\epsilon}\left[1+(z-1)\beta(s)\right]^{\epsilon}\int_{-1}^{1}dy\left(1-y^2\right)^{-\epsilon}\nonumber \\
&
&\hspace{-0.9in}\times\int_{-1}^{1}dv(1+v)^{-\epsilon}(1-v)^{-1-\epsilon}\int_{0}^{\pi}d\phi'_2\ \sin^{-2\epsilon}\phi'_2\ F^{qg}_{II}\left(s,z,y,v,\phi'_2\right)\,. \nonumber \\
\,
\end{eqnarray}
We rewrite the factor $(1-v)^{-1-\epsilon}$ as a distribution:
\begin{eqnarray}
\label{3.56}
(1-v)^{-1-\epsilon} &\sim& -\ \frac{v_0^{-\epsilon}}{\epsilon}\delta(1-v) +
\plusdist{1}{1-v}{v_0}\ +\ O(\epsilon) \,
\end{eqnarray}
with $0<v_0\leq 2$, so that (\ref{3.55}) may be rewritten as follows:
\begin{eqnarray}
\label{3.57}
\sigma^{P,II(1)}\left[q(p_1)g(p_2)\rightarrow W\gamma q\right]
=\sigma^{P,II}_{qg,finite} + \sigma^{P,II}_{qg,col+} + O(\epsilon) \,
\end{eqnarray}
where
\begin{eqnarray}
\label{3.58}
\sigma^{P,II}_{qg,finite} &=&
-\frac{1}{8N^2_cC_F}\frac{1}{2s}\left(4\pi\right)^{-4}\beta(s)\int_{0}^{1}dz\int_{-1}^{1}dy\int_{-1}^{1}dv\plusdist{1}{1-v}{v_0}\nonumber \\
& &\times\int_{0}^{\pi}d\phi'_2\ F^{qg}_{II}\left(s,z,y,v,\phi'_2\right)
\nonumber \\
\sigma^{P,II}_{qg,col+}
&=&\frac{1}{8N^2_cC_F(1-\epsilon)}\frac{1}{2s}(4\pi)^{\epsilon-2}\frac{1}{\Gamma(1-\epsilon)}\frac{s^{-\epsilon}}{\pi}\frac{\pi}{\epsilon}\left(\frac{2}{v_0}\right)^{\epsilon}\Phi(s)\beta^{-2\epsilon}(s)\nonumber \\
&
&\times\int_{0}^{1}dz[z(1-z)]^{-2\epsilon}\left[1+(z-1)\beta(s)\right]^{\epsilon}\int_{-1}^{1}dy\left(1-y^2\right)^{-\epsilon}\nonumber \\
& &\times F^{qg\ (col+)}_{II}(s,z,y) \nonumber \\
\,
\end{eqnarray}
and
\begin{eqnarray}
\label{3.59}
F^{qg\ (col+)}_{II}(s,z,y) = F^{qg}_{II}\left(s,z,y,v=1,\phi'_2\right)\,.
\end{eqnarray}
The type $II$ squared matrix element is given by:
\begin{eqnarray}
\label{3.60}
\sum\left|M^{qg}_{II}\right|^2 &=& \frac{e^2_q}{\left(2k\cdot Q_2\right)^2}
R_{qg,II}^{\alpha\alpha'}\sum M\left[q(p_1)g(p_2) \rightarrow
W(Q_1)q(Q_2+k,\alpha)\right]\nonumber \\
& &\hspace{1.2in}\times M^{\ast}\left[q(p_1)g(p_2) \rightarrow
W(Q_1)q(Q_2+k,\alpha')\right] \nonumber \\
\,
\end{eqnarray}
with $e_q$ the charge of the outgoing quark. In the $qg$ center of mass frame
we have:
\begin{eqnarray}
\label{3.61}
R_{qg,II}^{\alpha\alpha'} &=& -4k\cdot Q_2 \left[\gamma_0 \left\{ \left(
\frac{2-n}{2} - \frac{(1-z)(1+v)}{2z} \right)\sla{Q}_2 - \left(
1+\frac{(1-z)(1+v)}{z} \right)\sla{k} \right. \right. \nonumber \\
& &\left. \left. \hspace{1.0in}+
\left(\frac{(1-z)(1-v)}{2z}\right)\sla{\overline{Q}}_2 \right\}
\right]^{\alpha'\alpha}\,, \nonumber \\
\,
\end{eqnarray}
where we have summed over physical polarizations of the outgoing photon in the
covariant gauge.
Again we note that there will be no singularities in the soft quark limit, that
is when $z\rightarrow 1$.
For the collinear limit of $F^{qg}_{II}$ we obtain:
\begin{eqnarray}
\label{3.62}
F^{qg\ (col+)}_{II}(s,z,y) = -2e^2_q\frac{ \left( 1+(1-z)^2 -\epsilon z^2
\right) }{z}\sum\left|M^{qg\rightarrow
Wq}\left(s,\frac{b^+_{II}}{z}\right)\right|^2 \nonumber \\
\,
\end{eqnarray}
with
\begin{eqnarray}
\label{3.63}
b^+_{II} &\equiv& b(s,z,y,v=1) = \frac{s}{2}\beta(s)z(1-y)\,.
\end{eqnarray}
Remember that (\ref{3.62}) has an implicit factor of $C(Q_1,Q_2,(1-z)Q_2/z)$.
Using (\ref{3.62}) in  (\ref{3.58}) we can rewrite the collinear contribution
(neglecting terms of $O(\epsilon)$:)
\begin{eqnarray}
\label{3.64}
\sigma^{P,II}_{qg,col+} &=& -
\frac{\alpha}{2\pi\overline{\epsilon}}\int_{0}^{1}dz\left[\hat{e}^2_q\frac{1+(1-z)^2}{z} + \epsilon \hat{e}^2_q\left\{ \frac{1+(1-z)^2}{z}\left( \ln\left(\frac{2\mu^2}{sv_0}\right) \right.\right.\right.\nonumber \\
&
&\left.\left.\left.\hspace{0.9in}+\ln\left(\frac{1+(z-1)\beta(s)}{z^2(1-z)^2\beta^2(s)}\right) \right) -z \right\}\right]\nonumber \\
& &\times\int DQ_1DQ_2\frac{D^2\sigma^{(1)}}{DQ_1DQ_2}\left[
q(p_1)g(p_2)\rightarrow W(Q_1)q\left(\frac{Q_2}{z}\right) \right]\,. \nonumber
\\
\,
\end{eqnarray}
In (\ref{3.64}) we made the replacement $e^2 = 4\pi\alpha\mu^{2\epsilon}$.

Using (\ref{3.38}), (\ref{3.42}),(\ref{3.43}), (\ref{3.48}),
(\ref{3.57}),(\ref{3.58}) and (\ref{3.64}) in
(\ref{2.26}) we obtain for the $O(\alpha_S)$ hard scattering cross section:
\begin{eqnarray}
\label{3.65}
\sigma^{(1)}\left[ q(p_1)g(p_2) \rightarrow W\gamma q\right] &=& \nonumber \\
& & \hspace{-2.in} \sigma^{P,I}_{qg,finite} +
\frac{\alpha_S}{2\pi}\int_{\rho(s)}^{1}dx\left\{ \frac{1}{2} + \left(x^2 +
(1-x)^2\right) \left[ \ln(1-x) - \frac{1}{2}\left( 1+\ln\left(
\frac{2\mu^2}{sy_0} \right) \right) \right] \right\}\nonumber \\
& & \hspace{-1.3in}\times\sigma^{(0)}\left[q(p_1)\bar q(xp_2)\rightarrow
W\gamma\right] \nonumber \\
& & \hspace{-2.1in}  + \sigma^{P,II}_{qg,finite} +
\frac{\alpha}{2\pi}\hat{e}^2_q\int_{0}^{1}dz\left\{ z -
\left(\frac{1+(1-z)^2}{z} \right)\left[\ln\left(\frac{2\mu^2}{sv_0}\right) +
\ln\left(\frac{1+(z-1)\beta(s)}{z^2(1-z)^2\beta^2(s)}\right) \right] \right\}
\nonumber \\
& &\hspace{-1.3in} \times\int DQ_1Dq_2\frac{D^2\sigma^{(1)}}{DQ_1Dq_2}\left[
q(p_1)g(p_2)\rightarrow W(Q_1)q(q_2) \right] \nonumber \\
& & \hspace{-2.1in} + \sigtilde^{P}\left[q(p_1)g(p_2)\rightarrow W\gamma
q\right] \nonumber \\
\,
\end{eqnarray}
with $q_2=Q_2/\tau$.
We have thus cancelled all singularities and we can now summarize for all the
contributions in the incoming $qg$ partonic channel:
\begin{eqnarray}
\label{3.66}
\int DQ_1DQ_2\sum_{X} \left( \frac{D^2\sigma^H}{DQ_1DQ_2} \right)^{qg}\left[
p(P_1)\ \bar p(P_2) \rightarrow W(Q_1)\ \gamma(Q_2) \ X \right]& =&  \nonumber
\\
& &\hspace{-4.4in}\int_{0}^{1}d\tau_1\int_{0}^{1}d\tau_2 \left\{
f_{qp}(\tau_1)f_{g\bar p}(\tau_2) \left( \sigma^{(1)}\left[ q(p_1)g(p_2)
\rightarrow W\gamma q\right] \right.\right.\nonumber \\
& &\hspace{-3.4in}\left.\left. + \int_{0}^{1}d\tau f_{\gamma q}(\tau)\int
DQ_1Dq_2\frac{D^2\sigma^{(1)}}{DQ_1Dq_2}\left[ q(p_1)g(p_2) \rightarrow
W(Q_1)q(q_2)\right] \right)\right. \nonumber \\
& &\hspace{-3.6in}\left. + (p \leftrightarrow \bar p, \tau_1 \leftrightarrow
\tau_2, p_1 \leftrightarrow p_2)\ \ \right\} \nonumber \\
\,
\end{eqnarray}
with $p_1=\tau_1P_1$, $p_2=\tau_2P_2$.
In (\ref{3.65}) and (\ref{3.66}) we need the following squared matrix element:
\begin{eqnarray}
\label{3.67}
\sum\left|M^{qg\rightarrow Wq}(s,\hat{b})\right|^{2(1)} &=& 2^6\pi
N_cC_FC^2_W\alpha_S\left( \frac{ s^2 + (s-M^2_W-\hat{b})^2 - 2\hat{b}M^2_W }{
s(s-M^2_W-\hat{b}) } \right)\nonumber \\
\,
\end{eqnarray}
where $\hat{b} = b/\tau$. (\ref{3.67}) carries an implicit factor of
$C(Q_1,\tau q_2,(1-\tau)q_2)$.
The pieces of squared matrix element needed in the last term in (\ref{3.65})
are too long to be presented here but they may be obtained upon request.

\subsection{\sc The $g \bar q$ channel.}

The treatment of this channel is analogous to the $qg$ channel. We can again
decompose the
partonic cross section as follows:
\begin{eqnarray}
\label{3.68}
\sigma^P\left[g(p_1)\bar q(p_2)\rightarrow W\gamma \bar q\right] &=&
\sigma^{P,I}\left[g(p_1)\bar q(p_2)\rightarrow W\gamma \bar q\right] +
\sigma^{P,II}\left[g(p_1)\bar q(p_2)\rightarrow W\gamma \bar q\right]\nonumber
\\
&+& \sigtilde^{P}\left[g(p_1)\bar q(p_2)\rightarrow W\gamma \bar q\right]
\nonumber \\
\,
\end{eqnarray}
where the non-singular term $\sigtilde^{P}\left[g(p_1)\bar q(p_2)\rightarrow
W\gamma \bar q\right]$ contains all interference pieces of the squared matrix
element and also $\left|M^{g\bar q}_{III}\right|^2$. The other terms in
(\ref{3.68}) are decomposed as follows:
\begin{eqnarray}
\label{3.69}
\sigma^{P,I(1)}\left[g(p_1)\bar q(p_2)\rightarrow W\gamma \bar q\right]
=\sigma^{P,I}_{g\bar q,finite} + \sigma^{P,I}_{g\bar q,col+} + O(\epsilon)
\nonumber \\
\sigma^{P,II(1)}\left[g(p_1)\bar q(p_2)\rightarrow W\gamma \bar q\right]
=\sigma^{P,II}_{g\bar q,finite} + \sigma^{P,II}_{g\bar q,col+} + O(\epsilon) \,
\end{eqnarray}
with
\begin{eqnarray}
\label{3.70}
\sigma^{P,I}_{g\bar q,finite} &=&
-\frac{1}{8N^2_cC_F}\frac{1}{2s}\left(4\pi\right)^{-4}\int_{\rho(s)}^{1}dx\
\beta(sx)\int_{-1}^{1}dy\plusdist{1}{1-y}{y_0}\int_{-1}^{1}d\cos\theta'_1\nonumber \\
& &\times\int_{0}^{\pi}d\theta'_2\ F^{g\bar
q}_I\left(s,x,y,\cos\theta'_1,\theta'_2\right) \nonumber \\
\sigma^{P,I}_{g\bar q,col+}
&=&\frac{\alpha_S}{2\pi\overline{\epsilon}}\int_{\rho(s)}^{1}dx\left[x(1-x)-\frac{1}{2} + \epsilon\left\{\ln\left(\frac{2\mu^2}{sy_0}\right)\left(x(1-x)-\frac{1}{2}\right) + \ln(1-x)\right.\right.\nonumber \\
& &\hspace{0.8in}+x(1-x)\left[1-2\ln(1-x)\right]\ \bigg\}
\bigg]\sigma^{(0)}\left[q(xp_1)\bar q(p_2)\rightarrow W\gamma\right] \nonumber
\\
\sigma^{P,II}_{g\bar q,finite} &=&
-\frac{1}{8N^2_cC_F}\frac{1}{2s}\left(4\pi\right)^{-4}\beta(s)\int_{0}^{1}dz\int_{-1}^{1}dy\int_{-1}^{1}dv\plusdist{1}{1-v}{v_0}\nonumber \\
& &\times\int_{0}^{\pi}d\phi'_2\ F^{g\bar q}_{II}\left(s,z,y,v,\phi'_2\right)
\nonumber \\
\sigma^{P,II}_{g\bar q,col+} &=& -
\frac{\alpha}{2\pi\overline{\epsilon}}\int_{0}^{1}dz\left[\hat{e}^2_{\bar
q}\frac{1+(1-z)^2}{z} + \epsilon \hat{e}^2_{\bar q}\left\{
\frac{1+(1-z)^2}{z}\left(
\ln\left(\frac{2\mu^2}{sv_0}\right)\right.\right.\right.\nonumber \\
& &\left.\left.\left.\hspace{1.in} +
\ln\left(\frac{1+(z-1)\beta(s)}{z^2(1-z)^2\beta^2(s)}\right) \right) -z
\right\}\right]\nonumber \\
& &\times\int DQ_1DQ_2\frac{D^2\sigma^{(1)}}{DQ_1DQ_2}\left[ g(p_1)\bar
q(p_2)\rightarrow W(Q_1)\bar q\left(\frac{Q_2}{z}\right) \right]\,. \nonumber
\\
\,
\end{eqnarray}
In (\ref{3.70}) we used the following non-singular functions:
\begin{eqnarray}
\label{3.70b}
F^{g\bar q}_I\left(s,x,y,\cos\theta'_1,\theta'_2\right) &\equiv& -2p_1\cdot
k\sum\left|M^{g\bar q}_{Ia}\right|^2 \nonumber \\
F^{g\bar q}_{II}\left(s,z,y,v,\phi'_2\right) &\equiv& -2k\cdot
Q_2\sum\left|M^{g\bar q}_{II}\right|^2 \,
\end{eqnarray}
Using (\ref{3.68})-(\ref{3.70}) in  (\ref{2.30}) we obtain the cancellation of
all singularities in this channel:
\begin{eqnarray}
\label{3.71}
\sigma^{(1)}\left[ g(p_1)\bar q(p_2) \rightarrow W\gamma \bar q\right] &=&
\nonumber \\
& & \hspace{-2.in} \sigma^{P,I}_{g\bar q,finite} +
\frac{\alpha_S}{2\pi}\int_{\rho(s)}^{1}dx\left\{ \frac{1}{2} + \left(x^2 +
(1-x)^2\right) \left[ \ln(1-x) - \frac{1}{2}\left( 1+\ln\left(
\frac{2\mu^2}{sy_0} \right) \right) \right] \right\}\nonumber \\
& & \hspace{-1.4in}\times\sigma^{(0)}\left[q(xp_1)\bar q(p_2)\rightarrow
W\gamma\right] \nonumber \\
& & \hspace{-2.1in}  + \sigma^{P,II}_{g\bar q,finite} +
\frac{\alpha}{2\pi}\hat{e}^2_{\bar q}\int_{0}^{1}dz\left\{ z -
\left(\frac{1+(1-z)^2}{z} \right)\left[\ln\left(\frac{2\mu^2}{sv_0}\right) +
\ln\left(\frac{1+(z-1)\beta(s)}{z^2(1-z)^2\beta^2(s)}\right) \right] \right\}
\nonumber \\
& &\hspace{-1.4in} \times\int DQ_1Dq_2\frac{D^2\sigma^{(1)}}{DQ_1Dq_2}\left[
g(p_1)\bar q(p_2)\rightarrow W(Q_1)\bar q(q_2) \right] \nonumber \\
& & \hspace{-2.1in} + \sigtilde^{P}\left[g(p_1)\bar q(p_2)\rightarrow W\gamma
\bar q\right]\,. \nonumber \\
\,
\end{eqnarray}
We can now summarize all contributions in the incoming $g\bar q$ partonic
channel:
\begin{eqnarray}
\label{3.72}
\int DQ_1DQ_2\sum_{X} \left( \frac{D^2\sigma^H}{DQ_1DQ_2} \right)^{g\bar
q}\left[ p(P_1)\ \bar p(P_2) \rightarrow W(Q_1)\ \gamma(Q_2) \ X \right]& =&
\nonumber \\
& &\hspace{-4.4in}\int_{0}^{1}d\tau_1\int_{0}^{1}d\tau_2 \bigg\{
f_{gp}(\tau_1)f_{\bar q\bar p}(\tau_2) \bigg( \sigma^{(1)}\left[ g(p_1)\bar
q(p_2) \rightarrow W\gamma \bar q\right] \nonumber \\
& &\hspace{-3.4in}\left.\left. + \int_{0}^{1} d\tau f_{\gamma \bar q}(\tau)\int
DQ_1Dq_2\frac{D^2\sigma^{(1)}}{DQ_1Dq_2}\left[ g(p_1)\bar q(p_2) \rightarrow
W(Q_1)\bar q(q_2)\right] \right)\right. \nonumber \\
& &\hspace{-3.4in}+ (p \leftrightarrow \bar p, \tau_1 \leftrightarrow \tau_2,
p_1 \leftrightarrow p_2)\ \ \bigg\} \nonumber \\
\,
\end{eqnarray}
with $p_1=\tau_1P_1$, $p_2=\tau_2P_2$ and $q_2 = Q_2/\tau$. In (\ref{3.71}) and
(\ref{3.72}) we still need the squared matrix element:
\begin{eqnarray}
\label{3.73}
\sum\left|M^{g\bar q\rightarrow W\bar q}(s,\hat{b})\right|^{2(1)} &=& 2^6\pi
N_cC_FC^2_W\alpha_S\left( \frac{ s^2 + \hat{b}^2 - 2(s-M^2_W-\hat{b})M^2_W }{
s\hat{b} } \right)\nonumber \\
\,
\end{eqnarray}
with $\hat{b} = b/\tau$. The comments after (\ref{3.67}) apply here too.

\newpage

\pagestyle{myheadings}
\mysection{\bf THE THREE SCENARIOS AND THE EXPERIMENTAL CUTS.}
\label{4}

\subsection{\sc $2$ body inclusive production of $W^+$ and $\gamma$.}

In this scenario (``$2$ body inclusive scenario'') one does not tag the
outgoing jet, so it will include events with zero and $1$ outgoing jet.
We may define this scenario requiring the following conditions for the outgoing
particles
\begin{eqnarray}
\label{4.1}
|\cos\theta_{\gamma}|, |\cos\theta_{W}| &<& \cos(\theta^-) \nonumber \\
Pt_{\gamma}, Pt_{W}  &>& Pt^- \nonumber \\
R_{W,\gamma} &>& R^- \nonumber \\
( R_{jet,\gamma} < R^- )&\Longrightarrow& (s_{(jet,\gamma)} < s^-) \nonumber \\
( R_{jet,W} < R^- )&\Longrightarrow& (s_{(jet,W)} < s^-) \,
\end{eqnarray}
where we call $\theta_i$ the angle between the incoming proton axis and the
axis of the outgoing particle $i$; $Pt_{i}$ is the transverse momentum of
particle $i$. $R_{i,j}$ is the cone size between a pair of outgoing particles:
$R_{i,j}=\sqrt{(\Delta_{i,j}\eta^{\ast})^2 + (\Delta_{i,j}\phi)^2}$ with the
pseudorapidity $\eta^{\ast}\equiv (1/2)\ln[(1+\cos\theta)/(1-\cos\theta)]$
and $\phi$ the azimuthal angle; $s_{(jet,W)}=E_{jet}/E_W$ is the ``shadowing
ratio'' between the untagged
jet and the $W$ boson. The last two conditions in (\ref{4.1}) discard events
where the jet being too
close to the $W$ or the photon is at the same time of comparable energy so that
it would ``shadow'' one of the two
tagged particles making it undetectable. For this purpose we check the cone
size $R_{jet,\gamma}$ ($R_{jet,W}$) and if this is less than $R^-$ we keep the
event only when $s_{(jet,\gamma)}$ ($s_{(jet,W)}$) is less than $s^-$, setting
the differential cross section to zero otherwise. The quantities
$\theta^-,Pt^-,R^-,s^-$ are constants related to the acceptance and resolution
of the detector. All the quantities are defined in the proton-antiproton center
of mass frame.

\subsection{\sc Production of $W^+$ and $\gamma$ with $1$ jet.}

Here one detects three outgoing particles, namely $W^+$, $\gamma$ and $1$ jet.
We call this the ``$1$ jet scenario'' and we define it by imposing the
following conditions:
\begin{eqnarray}
\label{4.3}
|\cos\theta_{\gamma}|, |\cos\theta_{W}|,|\cos\theta_{jet}|  &<& \cos(\theta^-)
\nonumber \\
Pt_{\gamma}, Pt_{W}, Pt_{jet}  &>& Pt^-  \nonumber \\
R_{W,\gamma} &>&  R^-  \nonumber \\
R_{jet,\gamma} &>&  R^-  \nonumber \\
R_{jet,W} &>&  R^-  \,.
\end{eqnarray}

\subsection{\sc Production of $W^+$ and $\gamma$ with $0$ jets.}

In this scenario (``$0$ jet scenario'') we select events where the $W^+$ and
$\gamma$ are detected but no outgoing jet is detected. This includes
$2$ to $2$ body events and $2$ to $3$ body events where the outgoing jet has a
small angle with respect to
the beam, a small transverse momentum or it is ``shadowed'' by the photon or
the $W$ so that it remains undetected. We may define this scenario requiring
the following conditions for the outgoing particles
\begin{eqnarray}
\label{4.2}
|\cos\theta_{\gamma}|, |\cos\theta_{W}| &<& \cos(\theta^-) \nonumber \\
Pt_{\gamma}, Pt_{W}  &>& Pt^- \nonumber \\
R_{W,\gamma} &>& R^-  \nonumber \\
( R_{jet,\gamma} < R^-)&\Longrightarrow& (s_{(jet,\gamma)} < s^-) \nonumber \\
( R_{jet,W} < R^- )&\Longrightarrow& (s_{(jet,W)} < s^-) \nonumber \\
( |\cos\theta_{jet}| > \cos(\theta^-) )& \rm{or} & (Pt_{jet} < Pt^- )\,.
\end{eqnarray}

\subsection{\sc General remarks.}
\label{4.D}

We note that the second and third scenarios are complementary, in the sense
that an event in the first scenario falls
in either of the last two. In other words, we may obtain the histograms of the
$0$ jet production scenario by subtracting the histograms of the $1$ jet
scenario from the corresponding histograms for the $2$ body inclusive scenario.

To implement the three experimental cut functions $C(Q_1,Q_2,k)$ which define
each of the scenarios in {\sc A,B } and {\sc C}
all quantities involved in the above conditions have to be defined in terms of
the partonic invariants
that are used in the integrands of the corresponding cross section formulae.
In the  proton-antiproton center of mass frame we have:
\begin{eqnarray}
\label{4.4}
E_{\gamma} &=& \frac{ P_1\cdot Q_2 + P_2\cdot Q_2 }{\sqrt{S}} \nonumber \\
\cos\theta_{\gamma} &=& -\frac{ P_1\cdot Q_2 - P_2\cdot Q_2 }{ P_1\cdot Q_2 +
P_2\cdot Q_2 } \nonumber \\
E_{W} &=& \frac{ P_1\cdot Q_1 + P_2\cdot Q_1 }{\sqrt{S}} \nonumber \\
\cos\theta_{W} &=& -\frac{ P_1\cdot Q_1 - P_2\cdot Q_1 }{ \sqrt{ (P_1\cdot Q_1
+ P_2\cdot Q_1)^2 -SM^2_W } } \nonumber \\
E_{jet} &=& \frac{ P_1\cdot k + P_2\cdot k }{\sqrt{S}} \nonumber \\
\cos\theta_{jet} &=& -\frac{ P_1\cdot k - P_2\cdot k }{ P_1\cdot k + P_2\cdot k
} \,.
\end{eqnarray}
$P_1$ and $P_2$ represent the proton and antiproton momenta respectively; they
must be appropriately expressed in terms of the incoming parton momenta
$p_1,p_2$ and their momentum fractions $\tau_1,\tau_2$ in all the cross section
formulae. $\sqrt{S}=2P_1\cdot P_2$ is the proton-antiproton center of mass
energy. $Q_1, Q_2$ and $k$ are the momenta of the $W$ boson, the photon and the
jet respectively. The rest of the quantities needed can be computed using the
ones in (\ref{4.4}).

When we replaced the divergent factors $(1-x)^{-1-2\epsilon}$, $(1 \pm
y)^{-1-\epsilon}$ and $(1-v)^{-1-\epsilon}$ in section III with distributions
the resulting equations remained valid as long as the variables $x, y$ and $v$
were integrated over their whole
range. The energy of the outgoing jet in the incoming parton-parton center of
mass frame is linearly related to the variable $x$ (see (\ref{3.10}),) so it is
in principle not a physical quantity unless $x<x_0$, in which case the symbol
$\sim $ can be replaced by $=$ in the corresponding distribution in
(\ref{3.21}). Similarly, the angle between the outgoing jet and the beam
in the parton-parton frame is related to $y$ and the angle between the outgoing
jet and the photon is related to $v$, so
these quantities are not physical either, unless the variables $y$ and $v$ fall
inside the ranges where we can replace
$\sim $ for $=$ in the corresponding distributions.

According to the above observations we shoudn't have any trouble in the $2$
body inclusive and in the $0$ jet scenarios, since in these cases the outgoing
jet is not being tagged so the unphysical variables are not ``observed'', but
they are rather integrated over their whole range. However, in the $1$ jet
scenario the energy and angles of the jet are observed and these are directly
related to the variables $x,y$ and $v$. According to the way we defined the $1$
jet scenario in (\ref{4.3}) the outgoing jet is never allowed to be soft or
collinear to the beams or the outgoing photon so the subtraction of divergences
will never be active. With this in mind we can easilychoose the parameters
$x_0, y_0$ and $v_0$ in our Monte Carlo in such a way that the sampled ranges
of $x,y$ and $v$ always fall inside the regions where $\sim $ may be replaced
for $=$ in (\ref{3.21}), (\ref{3.41}) and (\ref{3.56}). To accomplish this we
can just take $x_0 = 1$ and $y_0=v_0=0$. The experimental cut function
$C(Q_1,Q_2,k)$ will a

ccordingly set to zero all the terms containing ill defined logarithms.

\newpage

\pagestyle{myheadings}
\mysection{\bf The numerical implementation.}

When numerically implementing the ``generalized plus'' distributions defined in
(\ref{3.22}) to compute total cross
sections the second terms on the RHS of (\ref{3.22}) are finite when the soft
or collinear limits are approached.
However, when we produce histograms of single or double differential cross
sections it is necessary to split the second terms on the RHS of these
definitions into two parts, as we will explain next. For the case of the $x$
integration we have:
\begin{eqnarray}
\label{5.1}
\int_{x_0}^{1}dx \frac{f(x)-f(1)}{1-x} = \int_{x_0}^{1}dx \frac{f(x)}{1-x} -
\int_{x_0}^{1}dx \frac{f(1)}{1-x}\,.
\end{eqnarray}
The first term on the RHS of (\ref{5.1}) is naturally histogrammed using $2$ to
$3$ body kinematics. The soft pieces that resulted from
the expansion in (\ref{3.21}) were added to other $2$ to $2$ body contributions
in order to cancel singularities so the remaining
pieces are naturally histogrammed using $2$ to $2$ body kinematics. This means
that in order to keep consistency in our computation we have to
histogram the second term on the RHS of (\ref{5.1}) -which is the term that
compensates for the soft singular terms in (\ref{3.21})-
using $2$ to $2$ body kinematics as well. It is thus clear that a consistent
histogramming cannot be achieved in a simple way without
splitting the LHS of (\ref{5.1}).
In doing so we introduce logarithmic singularities in each of the terms on the
RHS of (\ref{5.1}) that cancel each other only
after summing both contributions bin by bin. To control the numerical
cancellations we introduce small adimensional cuts $\Delta_x,\Delta_y$ and
$\Delta_v$ in the lower or upper limits of the corresponding integrals.
A first order estimate of the error
introduced by the cuts along with the requirement of good numerical convergence
will help us find the best values for these parameters.

In what follows we will rewrite the partonic hard scattering cross sections for
each channel taking into account the $\Delta$ parameters introduced above.
The contribution of each of these terms to the hadronic cross section is
obtained after multiplying by the corresponding experimental cut function,
convoluting with parton densities (see section II) and adding the corresponding
``inverted channels'' (i.e. the ones obtained by interchange of the incoming
partons.)
Numerical results for each of these hadronic contributions are presented in the
following paper \cite{MS}.

For the $q\bar q$ hard scattering channel cross section needed in (\ref{3.35b})
we have:
\begin{eqnarray}
\label{5.2}
\sigma_{q\bar q}&=&\sigma_{q\bar q}^{Born} + \sigma^P_{q\bar q\ (SV)} +
\sigma_{Ia} + \sigma_{Ib} + \sigma_{I,4} + \sigma_{q\bar q\ (finite)}^{P} +
\sigma_{q\bar q\ (Brems)}  + \sigma_{q\bar q\ (error)} \nonumber \\
\,
\end{eqnarray}
where
\begin{eqnarray}
\label{5.3}
\sigma_{Ia} &=& \sigma_{Ia,1} + \sigma_{Ia,2} + \sigma_{Ia,3} \nonumber \\
\sigma_{Ib} &=& \sigma_{Ib,1} + \sigma_{Ib,2} + \sigma_{Ib,3} \nonumber \\
\sigma_{q\bar q\ (finite)}^{P} &=& \sigma_{f,1,1,a} + \sigma_{f,1,2,a} +
\sigma_{f,1,3,a} + \sigma_{f,1,1,b} + \sigma_{f,1,2,b} + \sigma_{f,1,3,b}
\nonumber \\
& &+\sigma_{f,2,1,a} + \sigma_{f,2,2,a} + \sigma_{f,2,3,a} + \sigma_{f,2,1,b} +
\sigma_{f,2,2,b} + \sigma_{f,2,3,b} + \sigma_{f,3} \nonumber \\
\sigma_{q\bar q\ (error)} &=&\sigma_{Ia,error} + \sigma_{Ib,error} +
\sigma_{f,1,error,a} + \sigma_{f,1,error,b} +  \sigma_{f,2,error,a}\nonumber \\
& & + \sigma_{f,2,error,b} + \sigma_{f,error} \nonumber \\
\,
\end{eqnarray}
and
\begin{eqnarray}
\label{5.4}
\sigma_{q\bar q}^{Born} &\equiv& \sigma^{(0)}\left[ q(p_1)\bar q(p_2)
\rightarrow W\gamma\right] \nonumber \\
                        &=& C_{q\bar q,3}\ \beta(s)\int_{0}^{1}d\cos\theta_1
\sum\left|M^{q\bar q\rightarrow W\gamma}(s,b)\right|^{2(0)}\nonumber \\
\sigma_{Ia,1} &\equiv& \frac{C_{q\bar q,1}}{2}\int_{\rho(s)}^{1}dx
\frac{\beta(sx)}{x}(1-x)\int_{-1}^{1}d\cos\theta'_1\sum\left|M^{q\bar
q\rightarrow W\gamma}(xs,xb^{+})\right|^{2(0)} \nonumber \\
\sigma_{Ia,2} &\equiv& -\frac{C_{q\bar q,1}}{2}\int_{\rho(s)}^{x_0}dx
\frac{\beta(sx)}{x}\left(\frac{1+x^2}{1-x}\right)\left[\ln\left(\frac{2\mu^2}{sy_0}\right)-2\ln(1-x)\right]\nonumber \\
              & &\times\int_{-1}^{1}d\cos\theta'_1\sum\left|M^{q\bar
q\rightarrow W\gamma}(xs,xb^{+})\right|^{2(0)} \nonumber \\
\sigma_{Ia,3} &\equiv& -\frac{C_{q\bar q,1}}{2}\int_{x_0}^{1-\Delta_x}dx
\frac{\beta(sx)}{x}\left(\frac{1+x^2}{1-x}\right)\left[\ln\left(\frac{2\mu^2}{sy_0}\right)-2\ln(1-x)\right]\nonumber \\
              & &\times\int_{-1}^{1}d\cos\theta'_1\sum\left|M^{q\bar
q\rightarrow W\gamma}(xs,xb^{+})\right|^{2(0)} \nonumber \\
\sigma_{Ia,error} &\equiv&\frac{C_{q\bar q,1}}{2}\Delta_x\left(
\ln\left(\frac{2\mu^2}{sy_0}\right) + 2 -2\ln\Delta_x \right)\nonumber \\
& &\times\left.\frac{\partial}{\partial x}\left[
\frac{\beta(sx)}{x}(1+x^2)\int_{-1}^{1}d\cos\theta'_1\sum\left|M^{q\bar
q\rightarrow W\gamma}(xs,xb^{+})\right|^{2(0)}\right]\right|_{x=1}\nonumber \\
& & + O(\Delta^2_x\ln\Delta_x) \nonumber \\
\sigma_{Ib,1} &\equiv& \frac{C_{q\bar q,1}}{2}\int_{\rho(s)}^{1}dx
\frac{\beta(sx)}{x}(1-x)\int_{-1}^{1}d\cos\theta'_1\sum\left|M^{q\bar
q\rightarrow W\gamma}(xs,b^{-})\right|^{2(0)} \nonumber \\
\sigma_{Ib,2} &\equiv& -\frac{C_{q\bar q,1}}{2}\int_{\rho(s)}^{x_0}dx
\frac{\beta(sx)}{x}\left(\frac{1+x^2}{1-x}\right)\left[\ln\left(\frac{2\mu^2}{sy_0}\right)-2\ln(1-x)\right]\nonumber \\
              & &\times\int_{-1}^{1}d\cos\theta'_1\sum\left|M^{q\bar
q\rightarrow W\gamma}(xs,b^{-})\right|^{2(0)} \nonumber \\
\sigma_{Ib,3} &\equiv& -\frac{C_{q\bar q,1}}{2}\int_{x_0}^{1-\Delta_x}dx
\frac{\beta(sx)}{x}\left(\frac{1+x^2}{1-x}\right)\left[\ln\left(\frac{2\mu^2}{sy_0}\right)-2\ln(1-x)\right]\nonumber \\
              & &\times\int_{-1}^{1}d\cos\theta'_1\sum\left|M^{q\bar
q\rightarrow W\gamma}(xs,b^{-})\right|^{2(0)} \nonumber \\
\sigma_{Ib,error} &\equiv&\frac{C_{q\bar q,1}}{2}\Delta_x\left(
\ln\left(\frac{2\mu^2}{sy_0}\right) + 2 -2\ln\Delta_x \right)\nonumber \\
                  & &\times\left.\frac{\partial}{\partial x}\left[
\frac{\beta(sx)}{x}(1+x^2)\int_{-1}^{1}d\cos\theta'_1\sum\left|M^{q\bar
q\rightarrow W\gamma}(xs,b^{-})\right|^{2(0)}\right]\right|_{x=1}\nonumber \\
& & + O(\Delta^2_x\ln\Delta_x) \nonumber \\
\sigma_{I,4} &\equiv& 2C_{q\bar q,1}\
\beta(s)\ln\left(\frac{1-x_0}{\Delta_x}\right)\left[\ln\left(\frac{2\mu^2}{sy_0}\right)-\ln\left[(1-x_0)\Delta_x\right]\right]\nonumber \\
             & &\times\int_{-1}^{1}d\cos\theta'_1\sum\left|M^{q\bar
q\rightarrow W\gamma}(s,b^{soft})\right|^{2(0)} \nonumber \\
\sigma_{f,1,1,a} &\equiv& \frac{C_{q\bar q,2}}{s}\int_{\rho(s)}^{x_0}dx
\frac{\beta(sx)}{1-x}\int_{-1}^{1-y_0}dy
\frac{1}{1-y}\int_{-1}^{1}d\cos\theta'_1\int_{0}^{\pi}d\theta'_2\ F^{q\bar
q}(s,x,y,\cos\theta'_1,\theta'_2)\nonumber \\
\sigma_{f,1,2,a} &\equiv& \frac{C_{q\bar q,2}}{s}\int_{\rho(s)}^{x_0}dx
\frac{\beta(sx)}{1-x}\int_{1-y_0}^{1-\Delta_y}dy
\frac{1}{1-y}\int_{-1}^{1}d\cos\theta'_1\int_{0}^{\pi}d\theta'_2\ F^{q\bar
q}(s,x,y,\cos\theta'_1,\theta'_2)\nonumber \\
\sigma_{f,1,3,a} &\equiv&-\frac{C_{q\bar
q,1}}{4}\ln\left(\frac{y_0}{\Delta_y}\right)\int_{\rho(s)}^{x_0}dx
\frac{\beta(sx)}{1-x}\left(\frac{1+x^2}{x}\right)\nonumber \\
& &\times\int_{-1}^{1}d\cos\theta'_1\sum\left|M^{q\bar q\rightarrow
W\gamma}(xs,xb^{+})\right|^{2(0)} \nonumber \\
\sigma_{f,1,error,a} &\equiv&-\frac{C_{q\bar
q,2}}{s}\Delta_y\int_{\rho(s)}^{x_0}dx
\frac{\beta(sx)}{1-x}\int_{-1}^{1}d\cos\theta'_1\int_{0}^{\pi}d\theta'_2\left.\left(\frac{\partial F^{q\bar q}(s,x,y,\cos\theta'_1,\theta'_2)}{\partial y}\right)\right|_{y=1}\nonumber \\
& & + O(\Delta^2_y) \nonumber \\
\sigma_{f,1,1,b} &\equiv& \frac{C_{q\bar q,2}}{s}\int_{\rho(s)}^{x_0}dx
\frac{\beta(sx)}{1-x}\int_{-1+y_0}^{1}dy
\frac{1}{1+y}\int_{-1}^{1}d\cos\theta'_1\int_{0}^{\pi}d\theta'_2\ F^{q\bar
q}(s,x,y,\cos\theta'_1,\theta'_2)\nonumber \\
\sigma_{f,1,2,b} &\equiv& \frac{C_{q\bar q,2}}{s}\int_{\rho(s)}^{x_0}dx
\frac{\beta(sx)}{1-x}\int_{-1+\Delta_y}^{-1+y_0}dy
\frac{1}{1+y}\int_{-1}^{1}d\cos\theta'_1\int_{0}^{\pi}d\theta'_2\ F^{q\bar
q}(s,x,y,\cos\theta'_1,\theta'_2)\nonumber \\
\sigma_{f,1,3,b} &\equiv&-\frac{C_{q\bar
q,1}}{4}\ln\left(\frac{y_0}{\Delta_y}\right)\int_{\rho(s)}^{x_0}dx
\frac{\beta(sx)}{1-x}\left(\frac{1+x^2}{x}\right)\nonumber \\
& &\times\int_{-1}^{1}d\cos\theta'_1\sum\left|M^{q\bar q\rightarrow
W\gamma}(xs,b^{-})\right|^{2(0)} \nonumber \\
\sigma_{f,1,error,b} &\equiv&\frac{C_{q\bar
q,2}}{s}\Delta_y\int_{\rho(s)}^{x_0}dx
\frac{\beta(sx)}{1-x}\int_{-1}^{1}d\cos\theta'_1\int_{0}^{\pi}d\theta'_2\left.\left(\frac{\partial F^{q\bar q}(s,x,y,\cos\theta'_1,\theta'_2)}{\partial y}\right)\right|_{y=-1}\nonumber \\
& & + O(\Delta^2_y) \nonumber \\
\sigma_{f,2,1,a} &\equiv& \frac{C_{q\bar q,2}}{s}\int_{x_0}^{1-\Delta_x}dx
\frac{\beta(sx)}{1-x}\int_{-1}^{1-y_0}dy
\frac{1}{1-y}\int_{-1}^{1}d\cos\theta'_1\int_{0}^{\pi}d\theta'_2\ F^{q\bar
q}(s,x,y,\cos\theta'_1,\theta'_2)\nonumber \\
\sigma_{f,2,2,a} &\equiv& \frac{C_{q\bar q,2}}{s}\int_{x_0}^{1-\Delta_x}dx
\frac{\beta(sx)}{1-x}\int_{1-y_0}^{1-\Delta_y}dy
\frac{1}{1-y}\int_{-1}^{1}d\cos\theta'_1\int_{0}^{\pi}d\theta'_2\ F^{q\bar
q}(s,x,y,\cos\theta'_1,\theta'_2)\nonumber \\
\sigma_{f,2,3,a} &\equiv&-\frac{C_{q\bar
q,1}}{4}\ln\left(\frac{y_0}{\Delta_y}\right)\int_{x_0}^{1-\Delta_x}dx
\frac{\beta(sx)}{1-x}\left(\frac{1+x^2}{x}\right)\nonumber \\
& &\times\int_{-1}^{1}d\cos\theta'_1\sum\left|M^{q\bar q\rightarrow
W\gamma}(xs,xb^{+})\right|^{2(0)} \nonumber \\
\sigma_{f,2,error,a} &\equiv&-\frac{C_{q\bar
q,2}}{s}\Delta_y\int_{x_0}^{1-\Delta_x}dx
\frac{\beta(sx)}{1-x}\int_{-1}^{1}d\cos\theta'_1\int_{0}^{\pi}d\theta'_2\left.\left(\frac{\partial F^{q\bar q}(s,x,y,\cos\theta'_1,\theta'_2)}{\partial y}\right)\right|_{y=1}\nonumber \\
                     & & + O(\Delta^2_y) \nonumber \\
\sigma_{f,2,1,b} &\equiv& \frac{C_{q\bar q,2}}{s}\int_{x_0}^{1-\Delta_x}dx
\frac{\beta(sx)}{1-x}\int_{-1+y_0}^{1}dy
\frac{1}{1+y}\int_{-1}^{1}d\cos\theta'_1\int_{0}^{\pi}d\theta'_2\ F^{q\bar
q}(s,x,y,\cos\theta'_1,\theta'_2)\nonumber \\
\sigma_{f,2,2,b} &\equiv& \frac{C_{q\bar q,2}}{s}\int_{x_0}^{1-\Delta_x}dx
\frac{\beta(sx)}{1-x}\int_{-1+\Delta_y}^{-1+y_0}dy
\frac{1}{1+y}\int_{-1}^{1}d\cos\theta'_1\int_{0}^{\pi}d\theta'_2\ F^{q\bar
q}(s,x,y,\cos\theta'_1,\theta'_2)\nonumber \\
\sigma_{f,2,3,b} &\equiv&-\frac{C_{q\bar
q,1}}{4}\ln\left(\frac{y_0}{\Delta_y}\right)\int_{x_0}^{1-\Delta_x}dx
\frac{\beta(sx)}{1-x}\left(\frac{1+x^2}{x}\right)\nonumber \\
& &\times\int_{-1}^{1}d\cos\theta'_1\sum\left|M^{q\bar q\rightarrow
W\gamma}(xs,b^{-})\right|^{2(0)} \nonumber \\
\sigma_{f,2,error,b} &\equiv&\frac{C_{q\bar
q,2}}{s}\Delta_y\int_{x_0}^{1-\Delta_x}dx
\frac{\beta(sx)}{1-x}\int_{-1}^{1}d\cos\theta'_1\int_{0}^{\pi}d\theta'_2\left.\left(\frac{\partial F^{q\bar q}(s,x,y,\cos\theta'_1,\theta'_2)}{\partial y}\right)\right|_{y=-1}\nonumber \\
                     & & + O(\Delta^2_y) \nonumber \\
\sigma_{f,3} &\equiv& -2C_{q\bar q,1}\
\beta(s)\ln\left(\frac{1-x_0}{\Delta_x}\right)\ln\left(\frac{2}{y_0}\right)\int_{-1}^{1}d\cos\theta'_1\sum\left|M^{q\bar q\rightarrow W\gamma}(s,b^{soft})\right|^{2(0)} \nonumber \\
\sigma_{f,error} &\equiv&-\frac{C_{q\bar
q,2}}{s}\Delta_x\int_{-1}^{1}dy\left[\plusdist{1}{1-y}{y_0} +
\plusdist{1}{1+y}{y_0}\right]\nonumber \\
                 &
&\times\int_{-1}^{1}d\cos\theta'_1\int_{0}^{\pi}d\theta'_2\left.\frac{\partial}{\partial x}\left[\beta(sx)F^{q\bar q}(s,x,y,\cos\theta'_1,\theta'_2)\right]\right|_{x=1} + O(\Delta^2_x) \nonumber \\
\sigma_{q\bar q\ (Brems)} &\equiv& \int_{0}^{1} d\tau f_{\gamma g}(\tau)\int
DQ_1Dq_2\frac{D^2\sigma^{(1)}}{DQ_1Dq_2}\left[ q(p_1)\bar q(p_2) \rightarrow
W(Q_1)g(q_2)\right] \nonumber \\
                           &=& C_{q\bar q,3}\ \beta(s)\int_{0}^{1} d\tau
f_{\gamma g}(\tau)\int_{0}^{1}d\cos\theta_1 \sum\left|M^{q\bar q\rightarrow
Wg}(s,\frac{b}{\tau})\right|^{2(1)}\,. \nonumber \\
\,
\end{eqnarray}
The invariant $b$ in the unprimed frame was defined in (\ref{3.6}). In
(\ref{5.4}) appropriate experimental cut functions are implicit in each of the
corresponding integrands. We have introduced the constants:
\begin{eqnarray}
\label{5.5}
C_{q\bar q,1} &\equiv& \frac{1}{4N^2_c}\frac{1}{2s}\frac{1}{16\pi^2}\alpha_S
C_F \nonumber \\
C_{q\bar q,2} &\equiv& \frac{1}{4N^2_c}\frac{1}{2s}\frac{1}{2^{10}\pi^4}
\nonumber \\
C_{q\bar q,3} &\equiv& \frac{1}{4N^2_c}\frac{1}{2s}\frac{1}{16\pi} \,.
\end{eqnarray}

Now we rewrite the hard scattering cross section for the $qg$ channel needed in
(\ref{3.66}):
\def\sigtilde{{\tilde\sigma}}
\begin{eqnarray}
\label{5.6}
\sigma_{qg}&=&\sigma_{qg,finite}^{P,I} + \sigma_{qg}^{I,col} +
\sigma_{qg,finite}^{P,II} + \sigma_{qg}^{II,col} + \sigtilde^{P}_{qg} +
\sigma_{qg(Brems)} + \sigma_{qg(error)} \nonumber \\
\,
\end{eqnarray}
where
\begin{eqnarray}
\label{5.7}
\sigma_{qg,finite}^{P,I} &=& \sigma_{qg,f,1}^{I} + \sigma_{qg,f,2}^{I} +
\sigma_{qg,f,3}^{I}  \nonumber \\
\sigma_{qg,finite}^{P,II} &=& \sigma_{qg,f,1}^{II} + \sigma_{qg,f,2}^{II} +
\sigma_{qg,f,3}^{II} \nonumber \\
\sigma_{qg(error)} &=& \sigma_{qg,error}^{I} + \sigma_{qg,error}^{II} \,
\end{eqnarray}
and
\begin{eqnarray}
\label{5.8}
\sigma_{qg,f,1}^{I} &\equiv& -C_{qg,3}\int_{\rho(s)}^{1}dx\
\beta(sx)\int_{-1+y_0}^{1}dy\frac{1}{1+y}\int_{-1}^{1}d\cos\theta'_1\int_{0}^{\pi}d\theta'_2\ F^{qg}_I\left(s,x,y,\cos\theta'_1,\theta'_2\right) \nonumber \\
\sigma_{qg,f,2}^{I} &\equiv& -C_{qg,3}\int_{\rho(s)}^{1}dx\
\beta(sx)\int_{-1+\Delta_y}^{-1+y_0}dy\frac{1}{1+y}\int_{-1}^{1}d\cos\theta'_1\int_{0}^{\pi}d\theta'_2\ F^{qg}_I\left(s,x,y,\cos\theta'_1,\theta'_2\right) \nonumber \\
\sigma_{qg,f,3}^{I} &\equiv&
-\frac{C_{qg,1}}{2}\ln\left(\frac{y_0}{\Delta_y}\right)\int_{\rho(s)}^{1}dx
\frac{\beta(sx)}{x}\left( x^2 + (1-x)^2\right)\nonumber \\
& &\times\int_{-1}^{1}d\cos\theta'_1\sum\left|M^{q\bar q\rightarrow
W\gamma}(xs,b^{-})\right|^{2(0)} \nonumber \\
\sigma_{qg,error}^{I} &\equiv& C_{qg,3}\ \Delta_y\int_{\rho(s)}^{1}dx\
\beta(sx)\int_{-1}^{1}d\cos\theta'_1\int_{0}^{\pi}d\theta'_2\left.\left(\frac{\partial F^{qg}_{I}(s,x,y,\cos\theta'_1,\theta'_2)}{\partial y}\right)\right|_{y=-1}\nonumber \\
& & + O(\Delta^2_y) \nonumber \\
\sigma_{qg}^{I,col} &\equiv& C_{qg,1}\int_{\rho(s)}^{1}dx
\frac{\beta(sx)}{x}\bigg\{\frac{1}{2} + \left(x^2 + (1-x)^2\right) \bigg[
\ln(1-x)\nonumber \\
& &\hspace{1.3in}\left.\left. - \frac{1}{2} - \frac{1}{2}\ln\left(
\frac{2\mu^2}{sy_0} \right) \right]
\right\}\int_{-1}^{1}d\cos\theta'_1\sum\left|M^{q\bar q\rightarrow
W\gamma}(xs,b^{-})\right|^{2(0)} \nonumber \\
\sigma_{qg,f,1}^{II} &\equiv& -C_{qg,3}\
\beta(s)\int_{0}^{1}dz\int_{-1}^{1}dy\int_{-1}^{1-v_0}dv\frac{1}{1-v}\int_{0}^{\pi}d\phi'_2\ F^{qg}_{II}\left(s,z,y,v,\phi'_2\right) \nonumber \\
\sigma_{qg,f,2}^{II} &\equiv& -C_{qg,3}\
\beta(s)\int_{0}^{1}dz\int_{-1}^{1}dy\int_{1-v_0}^{1-\Delta_v}dv\frac{1}{1-v}\int_{0}^{\pi}d\phi'_2\ F^{qg}_{II}\left(s,z,y,v,\phi'_2\right) \nonumber \\
\sigma_{qg,f,3}^{II} &\equiv& -C_{qg,2}\
\hat{e}^2_{q}\ln\left(\frac{v_0}{\Delta_v}\right)\beta(s)\int_{0}^{1}dz
\left(\frac{ 1+(1-z)^2 }{z}\right)\int_{-1}^{1}dy\sum\left|M^{qg\rightarrow
Wq}\left(s,\frac{b^+_{II}}{z}\right)\right|^{2(1)} \nonumber \\
\sigma_{qg,error}^{II} &\equiv& C_{qg,3}\ \Delta_v\ \beta(s)\int_{0}^{1}dz
\int_{-1}^{1}dy\int_{0}^{\pi}d\phi'_2\left.\left(\frac{\partial
F^{qg}_{II}(s,z,y,v,\phi'_2)}{\partial v}\right)\right|_{v=1} + O(\Delta^2_v)
\nonumber \\
\sigma_{qg}^{II,col} &\equiv& C_{qg,2}\ \hat{e}^2_{q}\ \beta(s)\nonumber \\
& &\times\int_{0}^{1}dz \left\{ z - \left(\frac{ 1+(1-z)^2 }{z}\right)\left[
\ln\left(\frac{2\mu^2}{sv_0}\right) +
\ln\left(\frac{1+(z-1)\beta(s)}{z^2(1-z)^2\beta^2(s)}\right) \right] \right\}
\nonumber \\
& &\times\int_{-1}^{1}dy\sum\left|M^{qg\rightarrow
Wq}\left(s,\frac{b^+_{II}}{z}\right)\right|^{2(1)} \nonumber \\
\sigtilde^{P}_{qg} &\equiv& \sigtilde^{P}\left[q(p_1)g(p_2)\rightarrow W\gamma
q\right]\nonumber \\
\sigma_{qg(Brems)} &\equiv& \int_{0}^{1} d\tau f_{\gamma q}(\tau)\int
DQ_1Dq_2\frac{D^2\sigma^{(1)}}{DQ_1Dq_2}\left[ q(p_1)g(p_2) \rightarrow
W(Q_1)q(q_2)\right] \nonumber \\
                           &=& C_{qg,4}\ \beta(s)\int_{0}^{1} d\tau f_{\gamma
q}(\tau)\int_{0}^{1}d\cos\theta_1 \sum\left|M^{qg\rightarrow
Wq}(s,\frac{b}{\tau})\right|^{2(1)}\,. \nonumber \\
\,
\end{eqnarray}
The comments after (\ref{5.4}) are also valid here. In (\ref{5.8}) we have
introduced the constants:
\begin{eqnarray}
\label{5.9}
C_{qg,1} &\equiv& \frac{1}{8N^2_c}\frac{1}{2s}\frac{1}{16\pi^2}\alpha_S
\nonumber \\
C_{qg,2} &\equiv& \frac{1}{8N^2_cC_F}\frac{1}{2s}\frac{1}{32\pi^2}\alpha
\nonumber \\
C_{qg,3} &\equiv& \frac{1}{8N^2_cC_F}\frac{1}{2s}\frac{1}{2^8\pi^4} \nonumber
\\
C_{qg,4} &\equiv& \frac{1}{8N^2_cC_F}\frac{1}{2s}\frac{1}{16\pi} \,.
\end{eqnarray}

Finally, the hard scattering cross section in the $g\bar q$ channel needed in
(\ref{3.72}) may be rewritten:
\def\sigtilde{{\tilde\sigma}}
\begin{eqnarray}
\label{5.10}
\sigma_{g\bar q}&=&\sigma_{g\bar q,finite}^{P,I} + \sigma_{g\bar q}^{I,col} +
\sigma_{g\bar q,finite}^{P,II} + \sigma_{g\bar q}^{II,col} +
\sigtilde^{P}_{g\bar q} + \sigma_{g\bar q(Brems)}  + \sigma_{g\bar q(error)}
\nonumber \\
\,
\end{eqnarray}
where
\begin{eqnarray}
\label{5.11}
\sigma_{g\bar q,finite}^{P,I} &=& \sigma_{g\bar q,f,1}^{I} + \sigma_{g\bar
q,f,2}^{I} + \sigma_{g\bar q,f,3}^{I} \nonumber \\
\sigma_{g\bar q,finite}^{P,II} &=& \sigma_{g\bar q,f,1}^{II} + \sigma_{g\bar
q,f,2}^{II} + \sigma_{g\bar q,f,3}^{II} \nonumber \\
\sigma_{g\bar q(error)} &=& \sigma_{g\bar q,error}^{I} + \sigma_{g\bar
q,error}^{II} \,
\end{eqnarray}
and
\begin{eqnarray}
\label{5.12}
\sigma_{g\bar q,f,1}^{I} &\equiv& -C_{qg,3}\int_{\rho(s)}^{1}dx\
\beta(sx)\int_{-1}^{1-y_0}dy\frac{1}{1-y}\int_{-1}^{1}d\cos\theta'_1\int_{0}^{\pi}d\theta'_2\ F^{g\bar q}_I\left(s,x,y,\cos\theta'_1,\theta'_2\right) \nonumber \\
\sigma_{g\bar q,f,2}^{I} &\equiv& -C_{qg,3}\int_{\rho(s)}^{1}dx\
\beta(sx)\int_{1-y_0}^{1-\Delta_y}dy\frac{1}{1-y}\int_{-1}^{1}d\cos\theta'_1\int_{0}^{\pi}d\theta'_2\ F^{g\bar q}_I\left(s,x,y,\cos\theta'_1,\theta'_2\right) \nonumber \\
\sigma_{g\bar q,f,3}^{I} &\equiv&
-\frac{C_{qg,1}}{2}\ln\left(\frac{y_0}{\Delta_y}\right)\int_{\rho(s)}^{1}dx
\frac{\beta(sx)}{x}\left( x^2 + (1-x)^2\right)\nonumber \\
                         & &\times\int_{-1}^{1}d\cos\theta'_1\sum\left|M^{q\bar
q\rightarrow W\gamma}(xs,xb^{+})\right|^{2(0)} \nonumber \\
\sigma_{g\bar q,error}^{I} &\equiv& C_{qg,3}\ \Delta_y\int_{\rho(s)}^{1}dx\
\beta(sx)\int_{-1}^{1}d\cos\theta'_1\int_{0}^{\pi}d\theta'_2\left.\left(\frac{\partial F^{g\bar q}_{I}(s,x,y,\cos\theta'_1,\theta'_2)}{\partial y}\right)\right|_{y=1}\nonumber \\
& & + O(\Delta^2_y) \nonumber \\
\sigma_{g\bar q}^{I,col} &\equiv& C_{qg,1}\int_{\rho(s)}^{1}dx
\frac{\beta(sx)}{x}\bigg\{\frac{1}{2} + \left(x^2 + (1-x)^2\right) \bigg[
\ln(1-x)\nonumber \\
& &\hspace{1.2in}\left.\left. - \frac{1}{2} - \frac{1}{2}\ln\left(
\frac{2\mu^2}{sy_0} \right)\right]
\right\}\int_{-1}^{1}d\cos\theta'_1\sum\left|M^{q\bar q\rightarrow
W\gamma}(xs,xb^{+})\right|^{2(0)} \nonumber \\
\sigma_{g\bar q,f,1}^{II} &\equiv& -C_{qg,3}\
\beta(s)\int_{0}^{1}dz\int_{-1}^{1}dy\int_{-1}^{1-v_0}dv\frac{1}{1-v}\int_{0}^{\pi}d\phi'_2\ F^{g\bar q}_{II}\left(s,z,y,v,\phi'_2\right) \nonumber \\
\sigma_{g\bar q,f,2}^{II} &\equiv& -C_{qg,3}\
\beta(s)\int_{0}^{1}dz\int_{-1}^{1}dy\int_{1-v_0}^{1-\Delta_v}dv\frac{1}{1-v}\int_{0}^{\pi}d\phi'_2\ F^{g\bar q}_{II}\left(s,z,y,v,\phi'_2\right) \nonumber \\
\sigma_{g\bar q,f,3}^{II} &\equiv& -C_{qg,2}\ \hat{e}^2_{\bar
q}\ln\left(\frac{v_0}{\Delta_v}\right)\beta(s)\int_{0}^{1}dz \left(\frac{
1+(1-z)^2 }{z}\right)\int_{-1}^{1}dy\sum\left|M^{g\bar q\rightarrow
Wq}\left(s,\frac{b^+_{II}}{z}\right)\right|^{2(1)} \nonumber \\
\sigma_{g\bar q,error}^{II} &\equiv& C_{qg,3}\ \Delta_v\ \beta(s)\int_{0}^{1}dz
\int_{-1}^{1}dy\int_{0}^{\pi}d\phi'_2\left.\left(\frac{\partial F^{g\bar
q}_{II}(s,z,y,v,\phi'_2)}{\partial v}\right)\right|_{v=1} + O(\Delta^2_v)
\nonumber \\
\sigma_{g\bar q}^{II,col} &\equiv& C_{qg,2}\ \hat{e}^2_{\bar q}\
\beta(s)\nonumber \\
& &\times\int_{0}^{1}dz \left\{ z - \left(\frac{ 1+(1-z)^2 }{z}\right)\left[
\ln\left(\frac{2\mu^2}{sv_0}\right) +
\ln\left(\frac{1+(z-1)\beta(s)}{z^2(1-z)^2\beta^2(s)}\right) \right]
\right\}\nonumber \\
& &\times\int_{-1}^{1}dy\sum\left|M^{g\bar q\rightarrow W\bar
q}\left(s,\frac{b^+_{II}}{z}\right)\right|^{2(1)} \nonumber \\
\sigtilde^{P}_{g\bar q} &\equiv& \sigtilde^{P}\left[g(p_1)\bar
q(p_2)\rightarrow W\gamma \bar q\right] \nonumber \\
\sigma_{g\bar q(Brems)} &\equiv& \int_{0}^{1} d\tau f_{\gamma \bar q}(\tau)\int
DQ_1Dq_2\frac{D^2\sigma^{(1)}}{DQ_1Dq_2}\left[ g(p_1)\bar q(p_2) \rightarrow
W(Q_1)\bar q(q_2)\right] \nonumber \\
                           &=& C_{qg,4}\ \beta(s)\int_{0}^{1} d\tau f_{\gamma
\bar q}(\tau)\int_{0}^{1}d\cos\theta_1 \sum\left|M^{g\bar q\rightarrow W\bar
q}(s,\frac{b}{\tau})\right|^{2(1)} \nonumber \\
\,
\end{eqnarray}
The comments after (\ref{5.4}) are also valid here.

All the above terms will contribute in the $2$ body inclusive scenario and in
the $0$ jet scenario. However, in the $1$ jet
scenario, as we mentioned in section \ref{4.D}, by setting $x_0 = 1$ and
$y_0=v_0=0$ we are left only with the following contributions
\begin{eqnarray}
\label{5.13}
\sigma_{q\bar q}&=& \sigma_{f,1,1,a} + \sigma_{f,1,1,b}  \nonumber \\
\sigma_{qg}&=&\sigma_{qg,f,1}^{I} + \sigma_{qg,f,1}^{II} + \sigtilde^{P}_{qg}
\nonumber \\
\sigma_{g\bar q}&=&\sigma_{g\bar q,f,1}^{I} + \sigma_{g\bar q,f,1}^{II} +
\sigtilde^{P}_{g\bar q} \,.
\end{eqnarray}

\bigskip
\centerline {\bf Acknowledgements}
\bigskip
S.M. would like to thank Prof. D.\ Soper for some clarifying discussions during
the CTEQ '93 summer school. The work in this paper was supported in part by the
contract NSF 9309888.

\newpage

\pagestyle{myheadings}
%

\end{document}